\documentclass{article}
\usepackage[T1]{fontenc}
\usepackage[latin9]{inputenc}
\usepackage{amsmath}
\usepackage{amssymb}
\usepackage{mathtools}
\usepackage{amsthm}
\usepackage[linesnumbered,ruled,vlined]{algorithm2e}
\usepackage{graphicx}
\usepackage{pgfplots}
\pgfplotsset{compat=1.18}
\usepackage{subcaption}
\usepackage{appendix}
\usepackage[colorlinks=true,citecolor=blue,linkcolor=blue,urlcolor=black]{hyperref}
\usepackage{xcolor}
\usepackage{colortbl}
\usepackage{comment}
\usepackage{thm-restate}
\usepackage[numbers,square,sort]{natbib}
\usepackage{fullpage}
\usepackage{array} 
\newcolumntype{C}[1]{>{\centering\arraybackslash}m{#1}}
\usepackage{tikz}
\usepackage{caption}
\usepackage{enumitem}
\usepackage{accents}
\usepackage{bbm}
\usepackage{hhline}
\usepackage{multirow}
\usepackage{bm}
\usetikzlibrary{positioning, fit, calc, backgrounds, shapes.geometric, patterns}
\usepackage{makecell} 
\usepackage{float}

\allowdisplaybreaks
\let\textstyle\relax

\newcommand\blfootnote[1]{%
  \begingroup
  \renewcommand\thefootnote{}%
  \NoHyper\footnote{#1}\endNoHyper
  \addtocounter{footnote}{-1}%
  \endgroup
}

\newtheorem{theorem}{Theorem}[section]

\newtheorem{proposition}[theorem]{Proposition}
\newtheorem{lemma}[theorem]{Lemma}

\newtheorem{definition}[theorem]{Definition}
\newtheorem{remark}[theorem]{Remark}
\newtheorem{observation}[theorem]{Observation}

\newtheorem*{theorem*}{Theorem}
\newtheorem{maintheorem}{Main Theorem}

\newcommand{\reals}{\mathbb{R}}
\newcommand{\indicator}{\textbf{1}}
\DeclarePairedDelimiter{\floor}{\lfloor}{\rfloor}

\newcommand{\eps}{\varepsilon}
\DeclareMathOperator*{\argmax}{arg\,max}

\newcommand{\cmmnt}[1]{}

\newcommand{\agents}{A}
\newcommand{\goodobj}{BEST}

\newcommand{\prob}{\mathbb{P}}
\newcommand{\alg}{\textsc{Alg}}

\newcommand{\pequalsnp}{$\textsc{P} = \textsc{NP}$}

\newcommand{\Seq}{S^\dagger}

\newcommand{\price}{price of equality}
\newcommand{\Price}{Price of Equality}
\newcommand{\POE}{\textsc{PoE}}
\newcommand{\eqstarset}[1]{S^\star|_{#1}}

\newcommand{\nash}{\mathsf{NE}}

\newcommand{\I}{\mathcal{I}}

\newcommand{\B}{\mathcal{B}}
\newcommand{\F}{\mathcal{F}}

\newcommand{\noindexcontract}{\boldsymbol{\alpha}}
\newcommand{\icontract}{\alpha_i}

\newcommand{\noindexsubcontract}[1]{\noindexcontract|_{#1}}
\newcommand{\noindexsubcontractstar}[1]{\noindexcontract^\star|_{#1}}

\newcommand{\contract}{\noindexcontract}

\newcommand{\constar}{{\contract^\star}}
\newcommand{\alpstar}{\alpha^\star}
\newcommand{\eqstar}{S^\star}

\newcommand{\multiInstance}{\langle \agents, \{T_i\}_{i \in \agents}, f, c \rangle}

\newcommand{\actions}{T}

\newcommand{\eqcontract}[2]{\alpha^{\text{eq}}\left({#1},{#2}\right)}

\title{Equal-Pay Contracts}

\author{ 
Michal Feldman$^\ast$
\quad
Yoav Gal-Tzur$^\dagger$
\quad 
Tomasz Ponitka$^\ddagger$
\quad
Maya Schlesinger$^\mathsection$
}
\date{\today}

\begin{document}

\maketitle
\blfootnote{
This project has been partially funded by the European Research Council (ERC) under the European Union's Horizon 2020 program (grant agreement No.~866132), by the European Union's Horizon Europe Program (grant agreement No.~101170373), by an Amazon Research Award, by the Israel Science Foundation Breakthrough Program (grant No.~2600/24), and by a grant from TAU Center for AI and Data Science (TAD), and by the NSF-BSF (grant number 2020788).}
\blfootnote{$^\ast$Tel Aviv University and Microsoft ILDC, Israel. Email: \texttt{mfeldman@tauex.tau.ac.il}}
\blfootnote{$^\dagger$Tel Aviv University, Israel. Email: \texttt{yoavgaltzur@mail.tau.ac.il}}
\blfootnote{$^\ddagger$Tel Aviv University, Israel. Email: \texttt{tomaszp@mail.tau.ac.il}}
\blfootnote{$^\mathsection$Tel Aviv University, Israel. Email: \texttt{mayas1@mail.tau.ac.il}}

\begin{abstract}
We study multi-agent contract design, where a principal incentivizes a team of agents to take costly actions that jointly determine the project success via a combinatorial reward function.
While prior work largely focuses on unconstrained contracts that allow heterogeneous payments across agents, many real-world environments impose strong fairness constraints that limit payment dispersion. 
Motivated by this, we study equal-pay contracts, where all hired agents receive identical payments. 
Our results also extend to nearly-equal-pay contracts where any two payments are identical up to a constant factor.

We provide both algorithmic and hardness results across a broad hierarchy of reward functions, under both binary and combinatorial action models. 
While our focus is on equal-pay contracts, our analysis also yields new insights into unconstrained contract design and resolves two important open problems in this regime.
On the positive side, we establish polynomial-time constant-factor approximation algorithms for (i)
submodular rewards under combinatorial actions, and (ii) XOS rewards under binary actions. 
We further show that these guarantees are tight by establishing the following hardness results:
We rule out the existence of (i) a PTAS for combinatorial actions, even for gross substitutes rewards (unless \pequalsnp), and (ii) any constant-factor approximation for XOS rewards with combinatorial actions. 
Crucially, these hardness results hold even in the absence of any fairness constraints, thereby settling the corresponding open problems for unconstrained contracts.
Moreover, our impossibility result for XOS rewards provides the first separation between binary and combinatorial actions for unconstrained contracts.

Finally, we quantify the loss induced by fairness via the \emph{price of equality}, defined as the worst-case ratio between the optimal principal's utility achievable by unconstrained contracts and that achievable by equal-pay contracts. 
We obtain a tight bound of $\Theta(\log n/ \log \log n)$ on the price of equality, where $n$ is the number of agents. This gap is tight in a strong sense: the upper bound applies even for XOS rewards with combinatorial actions, {while} the lower bound arises already for additive rewards with binary actions.
\end{abstract}

\newpage
\clearpage

\section{Introduction}
\label{sec:intro}
Contracts are performance-based payment schemes designed to incentivize agents to exert costly effort on behalf of a principal, even when such effort is unobservable.
Classic contract theory provides elegant characterizations of optimal contracts \cite{holmstrom1979moral, grossman-hart-1983}.
Modern contracting environments, such as online labor markets, healthcare systems, and AI delegation, are increasingly implemented on computational platforms, calling for an algorithmic approach to contract design. 
Indeed, the field of algorithmic contract design has grown rapidly in recent years, with a substantial body of work studying computational tractability and approximation guarantees in complex contracting environments \cite{BabaioffFN10,BabaioffFNW12,ho2014adaptive,dutting2019simple, DuttingFT24,Feldman25}.

A major focus within algorithmic contract design is the study of contracts in \emph{multi-agent combinatorial-actions} models \cite{babaioff2006combinatorial,dutting2022combinatorial, duetting2022multi,multimulti}.
In these settings, a principal incentivizes a \emph{team} of agents, where each agent may choose a (possibly empty) set of costly actions. The project's outcome depends on the \emph{joint} collection of actions taken by all agents.
This framework can capture rich interdependencies between actions, such as substitutes and complements, modeled through a combinatorial reward function.

Existing work on this model has predominantly focused on \emph{unconstrained} contracts, in which the principal can tailor payments to individual agents. 
Under this assumption, optimal contracts typically generate large pay gaps across agents, reflecting differences in costs and marginal contributions to the formed team.
In practice, however, such disparities often conflict with fairness desiderata and with institutional or legal constraints.

These constraints are especially visible in the public sector, including civil service systems, public education, and academia, where compensation is often determined by rigid pay scales tied to rank or seniority. This leads to identical base salaries for large groups of employees. Importantly, even when performance-based incentives exist in these environments, they are often implemented symmetrically. For instance, in many academic systems, faculty members who secure external research grants receive a salary supplement equal to a fixed fraction of their base salary, and this fraction is set uniformly across all professors.

From the perspective of contract design, these considerations motivate the study of \emph{equal-pay} contracts, in which all agents receive the same payments regardless of their characteristics. More broadly, many settings impose weaker restrictions on wage dispersion, including policies that regulate or penalize excessive pay gaps \cite[e.g.][]{highpaycentre_wagemark_2013}. Such settings can be modeled as \emph{nearly-equal-pay} contracts, where agents' payments may differ only by a bounded multiplicative factor.
Together, these considerations make equal-pay and nearly-equal-pay contracts a natural and economically relevant constraint on the principal's design space, and form the basis of our study in this paper.

This perspective gives rise to two fundamental questions:
\begin{enumerate}
[
noitemsep, 
topsep=1pt, 
label=(\textit{\arabic*})]
\item Can we efficiently compute approximately optimal equal-pay (or nearly-equal-pay) contracts?
\item How much profit is lost by restricting contracts to be equal-pay relative to unconstrained contracts?
\end{enumerate}

We address both questions within the general multi-agent combinatorial-actions model of \cite{multimulti} for a broad class of reward functions. In addition, we resolve several open problems related to this model, under \emph{unconstrained} contracts.

\paragraph{Multi-Agent Combinatorial-Actions Model.} 
Before presenting our contribution, we describe the multi-agent combinatorial-actions model of \cite{multimulti} in more detail.
In this model, a principal (she) assigns a binary-outcome project to a set of $n$ agents. 
The project yields some reward to the principal upon success (normalized to $1$) and $0$ otherwise.
Each agent $i \in [n]$ has a set $T_i$ of available actions and may select any subset $S_i \subseteq T_i$. 
Each action $j \in T_i$ incurs a non-negative cost $c_j$, and the total cost incurred by agent $i$ equals $\sum_{j \in S_i} c_j$. 
The probability of a project's success is given by a set function $f$, which maps the combined action profile $S = S_1 \sqcup \cdots \sqcup S_n$ to a probability $f(S) \in [0,1]$. 
Since the principal's reward is normalized to $1$, $f(S)$ also represents the expected reward generated by $S$; thus we also refer to $f$ as the reward function.
An important special case, studied in \cite{BabaioffFNW12,duetting2022multi}, is the multi-agent \emph{binary-action} setting, in which each agent chooses whether to exert effort or shirk, and the outcome depends solely on the set of agents who exert effort; this setting corresponds to the case where $|T_i| = 1$.

To provide incentives, the principal commits to a linear contract $\noindexcontract = (\alpha_1, \ldots, \alpha_n)$, where each $\alpha_i \in [0,1]$ specifies the fraction of the realized reward paid to agent $i$ in the event of success. 
Given a contract $\noindexcontract$, agent $i$'s utility under an action profile $S$ is
$\alpha_i f(S) - \sum_{j \in S_i} c_j$.
Any given contract induces a game among the agents, where an action profile $S$ constitutes a Nash equilibrium if no agent can increase her utility by unilaterally deviating to a different action set. 

The principal's goal is to induce an equilibrium action profile that maximizes her utility (also known as profit), defined as the expected reward minus the total expected payments to the agents, namely
$f(S)\cdot \bigl(1 - \sum_{i \in [n]} \alpha_i\bigr)$.
The task of computing a contract that maximizes the principal's utility is known as the \emph{optimal contract problem}.
This problem raises substantial algorithmic challenges, as the principal's utility depends on the combinatorial structure of the agents' action profiles. 
As is common in the literature, since the reward function $f$ has exponential representation size, we assume oracle access to $f$ via two query models: value queries (given $S$, return $f(S)$), and demand queries (given a price vector $\boldsymbol{p}$ over actions, return a set $S$ maximizing $f(S)-\sum_{j\in S}p_j$); see Section~\ref{sec:model} for details.

\subsection{Our Results}
\label{sec:our-results}

In this section, we present our algorithmic results, followed by tight bounds on the loss incurred due to the restriction to equal-pay contracts. 
In the algorithmic part, we characterize the computational complexity of finding (near-)optimal equal-pay contracts over standard families of combinatorial functions $f$: $\text{additive} \subsetneq \text{gross substitutes} \subsetneq \text{submodular} \subsetneq \text{XOS} \subsetneq \text{subadditive}$ (see Section~\ref{sec:model} for precise definitions).
Moreover, we also resolve two open problems regarding the tractability of optimal contracts in unconstrained settings. 
Our results regarding equal-pay contracts are summarized in Table~\ref{table:equal-pay-results:eqpay}, and our contributions to the regime of  unconstrained contracts (along with previously known results) are presented in Table~\ref{table:equal-pay-results:general}.

Notably, we observe that any result for equal-pay contracts extends to 
$\gamma$-equal-pay contracts, where payments may differ by at most a multiplicative factor $\gamma>1$. In particular, approximation guarantees degrade gracefully, incurring only an additional factor-$\gamma$ loss (see Theorem~\ref{thm:gamma_reduction}). This observation allows us  to restrict attention to equal-pay contracts for the remainder of the paper.

\paragraph{Algorithms and Hardness.} 

For additive rewards, it is not difficult to show that an optimal equal-pay contract can be computed in polynomial time (see Theorem~\ref{thm:opt_additive}). This stands in sharp contrast to the unconstrained setting, where computing an optimal contract is NP-hard even when agents have binary actions \cite{duetting2022multi}.

At first glance, this might suggest that restricting attention to equal-pay contracts leads to new tractability in regimes previously known to be hard. Our results, however, show that this intuition is misleading: In all other cases, the computational hardness persists even when the optimal contract itself is an equal-pay contract. Moreover, positive results from the unconstrained setting do not carry over automatically; achieving comparable guarantees for equal-pay contracts requires new ideas and techniques.

For submodular rewards, prior work shows that the unconstrained problem admits a polynomial-time constant-factor approximation, but no PTAS \cite{multimulti}. Interestingly, their hardness result applies also to the equal-pay regime, implying that the best one can hope for in our setting is a constant-factor approximation. Our next result shows that this guarantee is indeed achievable.

\begin{table}[!ht]
\centering
\renewcommand{\arraystretch}{1.3}
\begin{subtable}{\textwidth}
\centering
\begin{tabular}{|w{c}{4.2cm}||w{c}{4.2cm}|w{c}{4.2cm}|}
\hhline{|-||--|}
 \textbf{Equal-Pay Contracts} & Binary Actions & Combinatorial Actions \\
\hhline{|-||--|}
\noalign{\vskip 2pt}
\hhline{|-||--|}
Additive
 & \cellcolor{green!35}
   \begin{tabular}{@{}c@{}}
     P
   \end{tabular}
 & \cellcolor{green!35}
   \begin{tabular}{@{}c@{}}
     P
     {\footnotesize (Thm~\ref{thm:opt_additive})}
   \end{tabular}
\\
\hhline{|-||--|}
Gross Substitutes
 &
   \begin{tabular}{@{}c@{}}
     $O(1)$-apx \\
     ?
   \end{tabular}
 & \cellcolor{yellow!20}
   \begin{tabular}{@{}c@{}}
     $O(1)$-apx \\
     {\text{No PTAS}}
     {\footnotesize (Thm~\ref{thm:matroid_hardness})}
   \end{tabular}
\\
\hhline{|-||--|}
Submodular
 & \cellcolor{yellow!20}
   \begin{tabular}{@{}c@{}}
     $O(1)$-apx \\
     No PTAS
     {\footnotesize \cite{multimulti}}
   \end{tabular}
 & \cellcolor{yellow!20}
   \begin{tabular}{@{}c@{}}
     {\text{$O(1)$-apx}}
     {\footnotesize (Thm~\ref{thm:equal-pay_const_approx})} \\
     No PTAS
   \end{tabular}
\\
\hhline{|-||--|}
XOS
 & \cellcolor{yellow!20}
   \begin{tabular}{@{}c@{}}
     $O(1)$-apx
     {\footnotesize (Thm~\ref{thm:xos_binary_approx})} \\
     No PTAS
   \end{tabular}
 & \cellcolor{red!20}
   \begin{tabular}{@{}c@{}}
     {\text{No $O(1)$-apx}}
     {\footnotesize (Thm~\ref{thm:XOS_inapprox})}
   \end{tabular}
\\
\hhline{|-||--|}
Subadditive
 & \cellcolor{red!20}
   \begin{tabular}{@{}c@{}}
     No $O(1)$-apx
     {\footnotesize \cite{duetting2022multi}}
   \end{tabular}
 & \cellcolor{red!20}
   \begin{tabular}{@{}c@{}}
     No $O(1)$-apx
   \end{tabular}
\\
\hhline{|-||--|}
\end{tabular}
\caption{The computational landscape under value and demand oracle access for equal-pay contracts. All results for equal-pay contracts are new, except for the hardness results in the binary-action setting. }
\label{table:equal-pay-results:eqpay}
\end{subtable}

\vspace{0.5cm}

\begin{subtable}{\textwidth}
\centering
\begin{tabular}{|w{c}{4.2cm}||w{c}{4.2cm}|w{c}{4.2cm}|}
\hhline{|-||--|}
 \textbf{Unconstrained Contracts} & Binary Actions & Combinatorial Actions \\
\hhline{|-||--|}
\noalign{\vskip 2pt}
\hhline{|-||--|}
Additive
 & \cellcolor{green!20}
   \begin{tabular}{@{}c@{}}
     FPTAS   {\footnotesize \cite{duetting2022multi}} \\
     NP-hard
     {\footnotesize \cite{duetting2022multi}}
   \end{tabular}
 & \cellcolor{green!20}
   \begin{tabular}{@{}c@{}}
     FPTAS
     {\footnotesize \cite{oneactiontoomany}} \\
     NP-hard
   \end{tabular}
\\
\hhline{|-||--|}
Gross Substitutes
 &
   \begin{tabular}{@{}c@{}}
     $O(1)$-apx \\
     ?
   \end{tabular}
 & 
\cellcolor{yellow!20}
   \begin{tabular}{@{}c@{}}
 $O(1)$-apx \\
       No PTAS
          {\footnotesize (Thm~\ref{thm:matroid_hardness})} {\Large\textcolor{red}{$\star$}}
   \end{tabular}
\\
\hhline{|-||--|}
Submodular
 & \cellcolor{yellow!20}
   \begin{tabular}{@{}c@{}}
     $O(1)$-apx \\
     No PTAS
     {\footnotesize \cite{multimulti}}
   \end{tabular}
 & \cellcolor{yellow!20}
   \begin{tabular}{@{}c@{}}
     $O(1)$-apx
     {\footnotesize \cite{multimulti}} \\
     No PTAS
   \end{tabular}
\\
\hhline{|-||--|}
XOS
 & \cellcolor{yellow!20}
   \begin{tabular}{@{}c@{}}
     $O(1)$-apx
     {\footnotesize \cite{duetting2022multi}} \\
     No PTAS
   \end{tabular}
& 
\cellcolor{red!20} 
   \begin{tabular}{@{}c@{}} No ${O(1)}$-apx  {\footnotesize (Thm~\ref{thm:XOS_inapprox})}  {\Large \textcolor{red}{$\star$}}\end{tabular} \\
\hhline{|-||--|}
Subadditive
 & \cellcolor{red!20}
   \begin{tabular}{@{}c@{}}
     No $O(1)$-apx
     {\footnotesize \cite{duetting2022multi}}
   \end{tabular}
 & \cellcolor{red!20}
   \begin{tabular}{@{}c@{}}
     No $O(1)$-apx
   \end{tabular}
\\
\hhline{|-||--|}
\end{tabular}
\caption{The computational landscape under value and demand oracle access for unconstrained contracts. The cells marked with 
{\textcolor{red}{$\star$}} 
denote our new contributions to the literature on unconstrained contract design.}
\label{table:equal-pay-results:general}
\end{subtable}
\caption{A summary of algorithmic and hardness results for equal-pay and unconstrained contracts. In both tables, any positive result immediately extends to all cells to its north and west, while any negative result propagates to all cells to its south and east. Colors indicate the tractability spectrum, ranging from green (tractable) through yellow (intermediate) to red (intractable); white indicates remaining open problems.}
\label{table:equal-pay-results}
\end{table}

\begin{maintheorem}[Constant Approximation for Submodular Rewards; Theorem~\ref{thm:equal-pay_const_approx}]\label{thm:constant_apx_submodular}
For submodular reward functions, there exists a polynomial-time algorithm that computes an equal-pay contract achieving a constant-factor approximation to the optimal equal-pay contract.
The algorithm uses demand oracle access to the reward function.
In binary-action settings, value oracles suffice.
\end{maintheorem}

As we show below, Theorem~\ref{thm:constant_apx_submodular} is tight in two ways: First, one cannot hope for a PTAS for this problem, even for \emph{gross substitutes} (GS) rewards --- a strict subclass of submodular functions (see Theorem~\ref{thm:no-ptas-gs}).
Second, the constant-factor approximation guarantee for submodular rewards \emph{cannot} be extended to XOS rewards --- a strict superclass of submodular functions (see Theorem~\ref{thm:no_constant_xos}).
Interestingly, these two impossibility results apply to both equal-pay and unconstrained contracts, thereby resolving two important open problems in the literature.
We next elaborate on these results. 

We start with gross substitutes (GS) rewards. 
GS functions are a subclass of submodular that has attracted significant attention in the literature on algorithmic mechanism and contract design (e.g., \cite{leme2017gross,dutting2022combinatorial}).
A fundamental open problem in multi-agent settings has been whether GS rewards admit a PTAS/FPTAS for the optimal contract problem. 
We resolve this question in the negative, for combinatorial-action settings.
Specifically, we show that even for matroid rank functions --- a strict subclass of GS --- the problem admits no PTAS.
This inapproximability result holds for both equal-pay and unconstrained contracts, thereby resolving the open problem for unconstrained contracts.

\begin{maintheorem}
[No PTAS for Gross Substitutes Rewards; Theorem~\ref{thm:matroid_hardness}]
\label{thm:no-ptas-gs}
For matroid rank function rewards, it is NP-hard to compute a $0.7$-approximation to the optimal contract.  
This result holds for both unconstrained contracts and equal-pay contracts.
\end{maintheorem}
Two remarks are in order: First, for GS functions demand and value oracles are essentially equivalent, thus the above theorem holds under both oracle models.
Second, the existence of a PTAS/FPTAS for GS rewards in the binary actions model remains an open problem.

We next turn to XOS rewards, for which we show that the optimal contract problem admits no constant-factor approximation.
As in the case of GS rewards, this impossibility result holds for both equal-pay and unconstrained contracts. Consequently, our result resolves another important open problem in multi-agent contract design, as posed for example in \cite{Feldman25}.

\begin{maintheorem}[No Constant-Factor Approximation for XOS Rewards; Theorem~\ref{thm:XOS_inapprox}]\label{thm:no_constant_xos}
For XOS reward functions, no algorithm that makes a polynomial number of value and demand queries can achieve a constant-factor approximation to the optimal contract, with or without the equal-pay restriction.
\end{maintheorem}

Notably, prior work on unconstrained contract design has given significant attention to the tractability of the optimal contract problem along two dimensions: binary versus combinatorial actions, and submodular versus XOS rewards. Constant-factor approximations were established for three of the four resulting cases --- submodular rewards under both action models, and XOS rewards under binary actions (see Table~\ref{table:equal-pay-results:general}) --- while the case of XOS rewards with combinatorial actions remained open. Our result closes this gap by showing that, unlike the other cases, this remaining problem is inapproximable.

More broadly, our findings suggest that the equal-pay model, beyond its intrinsic significance as a fairness constraint, also provides a useful lens for understanding unconstrained contracts.
Indeed, all current hardness results beyond additive rewards (including all previous hardness results \cite{duetting2022multi,ezra2023Inapproximability, multimulti}, and the two established here) arise in scenarios where optimal contracts turn out to be equal-pay contracts.

Finally, we also show that in the case of binary actions with XOS $f$, there exists a polynomial-time algorithm that yields a constant-factor approximation to the optimal equal-pay contract (see Theorem~\ref{thm:xos_binary_approx}). This result shows that Theorem~\ref{thm:no_constant_xos} does not apply to binary actions.

\paragraph{Price of Equality.} 

We next study the \emph{price of equality} (PoE), defined as the ratio between the optimal profit achievable by an unconstrained contract and that achievable by an equal-pay contract.

Our main result here is a tight bound on the PoE for XOS rewards under combinatorial actions. Remarkably, this bound is tight in a strong sense: no better bound is possible even for additive rewards and binary actions.

\begin{maintheorem}[Price of Equality; Proposition~\ref{prop:pof_add} and Theorem~\ref{thm:pof_xos}]\label{thm:price}
For XOS reward functions and combinatorial actions, the price of equality is $\Theta(\log n / \log \log n)$. This bound is tight, even for additive rewards and binary actions.
\end{maintheorem}

We further show that this bound does not extend to the broader class of subadditive rewards. In particular, we establish a lower bound of $\Omega(\sqrt{n})$ on the price of equality under subadditive rewards, even in binary-action settings (see Proposition~\ref{prop:pof_subadditive}).

\paragraph{Beyond Profit Maximization.}
\cite{feldman2025budget,oneactiontoomany} introduce a broader class of objectives, termed BEST objectives (beyond standard), which include profit, social welfare, and reward maximization, among others. 
In Appendix~\ref{app:best} we show that some of our results (for profit maximization) carry over to any BEST objective. 
Most importantly, for GS rewards with combinatorial actions, both our constant-factor approximation and our tight bound on the price of equality extend to any BEST objective.

\subsection{Our Techniques}
\label{sec:techniques}

Before presenting our techniques, we emphasize a fundamental challenge in multi-agent settings with combinatorial actions, which we describe as demand queries being \emph{agent-agnostic}.
We then explain how our positive results overcome the difficulties imposed by this property, and how we leverage it to construct hard instances of the problem.

\paragraph{Demand Queries Are Agent-Agnostic.}
A demand query takes as input a price vector $\boldsymbol{p}$ that assigns a price $p_j$ to each action $j$, and returns a set of actions $S \subseteq T$ that maximizes
$f(S) - \sum_{j \in S} p_j$.
A central challenge in applying demand queries in the multi-agent combinatorial-actions setting is that they are inherently \emph{agent-agnostic}: they ignore the ownership structure of actions.
In particular, a demand query does not depend on which actions are controlled by the same agent, and treats all actions as independent items.

This limitation is illustrated by the following example.
Consider two agents $A=\{1,2\}$, each controlling two actions, with $T_1=\{1,2\}$ and $T_2=\{3, 4\}$.
The costs of actions are $c_1 = 1/8 - \epsilon$, $c_2 = 1/8$, $c_3 = 1/8$, and $c_4 = 1/8 + \epsilon$.
The success probability is additive, given by $f(S) = (1/4)\cdot |S|$.
The optimal contract is $\contract = (1/2, 0)$, which induces the action profile $S = \{1,2\}$.
Under this contract, agent~1 takes both of her actions, yielding a principal's utility of
$u_P = (1 - 1/2)\cdot (1/2) = 1/4$.

By contrast, even though actions $2$ and $3$ are identical from the perspective of the reward function $f$ and incur identical costs, the action profile $S' = \{1,3\}$ can only be incentivized with the contract $\contract' = (1/2-4\epsilon, 1/2)$, which yields a negligible utility to the principal.
Thus, the optimal contract crucially depends on the ownership of actions, a distinction that agent-agnostic demand queries cannot capture without explicitly encoding ownership information in the price vector.

We next present our techniques for the main results.

\paragraph{Constant-Approximation for Submodular Rewards, for Equal-Pay Contract (Theorem~\ref{thm:constant_apx_submodular}).}  
Equal-pay contracts are defined by two parameters: a per-agent payment $t\in \reals_{\ge 0}$, and a subset of agents who receive this payment, with all remaining agents receiving zero. Our algorithm proceeds by ``guessing" the optimal value of $t$, reducing the problem to identifying an action set that is both (i) rewarding --- one with a high value of $f$, and (ii) agent-sparse --- involving only a small number of agents, {so as to keep total payments low}. {A} standard approach is to design prices such that demand queries find such a rewarding action set. {However}, since demand queries are {inherently} agent-agnostic (see the discussion above), {there is no guarantee that the resulting set will be agent-sparse.}

To address this difficulty, we develop an approximate demand-query procedure for submodular rewards subject to a cardinality constraint on the participating agents\footnote{It returns a set of actions $S\subseteq T$ with at most $k$ participating agents, such that for some constant $\beta \in (0,1)$, for any $S'\subseteq T$ with at most $k$ participating agents, $f(S) - \sum_{i\in S} p_i \ge \beta \cdot f(S') - \sum_{i\in S'} p_i$.}. 
Such constrained submodular demand queries are concerned with two spaces: the action space in which the objective is defined, and the agent space in which the constraint is defined. This interdependency of the action and agent spaces requires new ideas.
Roughly speaking, our procedure uses the \emph{distorted greedy} algorithm of \cite{Harshaw2019SubmodularMB} recursively at two levels:  at the agent level, and for each agent at the action level.
Applying our procedure with the ``right'' prices (for the actions) and cardinality constraints (on the agents) yields a set of actions which is both rewarding and agent sparse.
Finally, given such a desirable action set, we adapt machinery from \cite{multimulti} to convert it into an equal-pay contract with similar guarantees.

\paragraph{Hardness of Approximation for XOS Rewards, for Equal-Pay and Unconstrained Contracts (Theorem~\ref{thm:no_constant_xos}).}
As in the case of submodular rewards, a demand query to $f$, even with carefully chosen prices, cannot guarantee an action set that is both rewarding and agent-sparse. Unfortunately, the approach used for submodular rewards, namely, optimizing a surrogate submodular objective, does not extend to the XOS case, since demand queries for XOS functions cannot be approximated using only polynomially many value queries.

Indeed, we show that the agent-agnostic nature of demand queries leads to an inapproximability of the optimal contract within any constant factor.
Fix a hardness parameter $\ell \in \mathbb{N}$, and an instance with $\ell^7$ agents, each with $\ell^3$ actions.
We construct a family of reward functions parametrized by a subset of agents $G$ of size $\ell^{2}$.
For each action set $S$, the reward function is defined as the maximum of three terms: (i) the number of agents that perform at least one action in $S$, (ii) the number of actions taken by the agents in $G$, divided by $\ell$, and (iii) the constant $\ell^3$. 
The intuition behind this construction is that agents in $G$ are ``special'': if sufficiently many of them are incentivized to take many actions, term~(ii) yields a large reward.
However, because demand queries are agent-agnostic, they typically return solutions that maximize term~(i), namely involving many agents each taking only one action, rather than concentrating effort on the agents in $G$.

Action costs are chosen so that at most $O(\ell^3)$ agents can be incentivized simultaneously. 
As a result, the only way to obtain an $o(\ell)$-approximation to the optimal reward is to incentivize sufficiently many agents in $G$ to perform all their actions. In particular, any such approximation requires (roughly) identifying $G$.

However, terms (i) and (iii) in the reward function ``obfuscate'' the identity of the special set $G$:
We show that if a price vector $\boldsymbol{p}$ reveals information about $G$ (i.e., if the value of its demand set is achieved by term (ii)), then the total number of agents that have a cheap action (with price at most $1/\ell$) must be small (at most $2\ell^4$), while the number of agents in $G$ with a cheap action must be at least $\ell$.
Since $G$ is chosen uniformly at random, the probability that a demand query is informative is exponentially small in $\ell=n^{1/7}$. The inapproximability  then follows from  Yao's principle.

\paragraph{No PTAS for GS Rewards, for Equal-Pay and Unconstrained Contracts (Theorem~\ref{thm:no-ptas-gs}).}
Our result builds on the hardness result of \cite{ezra2023Inapproximability}, which proves that there is no PTAS in the multi-agent binary-action model, where the reward function is a coverage function defined over sets of agents.
We exploit the richness of the multi-agent combinatorial-action model to turn the coverage structure over agents into a matroid structure over actions, while maintaining the binary behavior of agents.
To this end, given a matroid-rank function $f$, we assign costs to actions such that agents exhibit a binary behavior relative to a payment threshold $t$: an agent performs all actions with positive marginal contribution if his payment is at least $t$, and shirks (i.e., performs no actions) otherwise.

Notably, the hardness in \cite{ezra2023Inapproximability} applies when only value-query access is given, whereas our result holds under both value and demand queries, since they are essentially equivalent for matroid rank functions (as a subclass of GS).

\paragraph{Price of Equality (Theorem~\ref{thm:price}).}
In the proof, we bound the loss incurred by restricting attention to equal-pay contracts, relative to an optimal unconstrained contract 
$\contract^\star$.

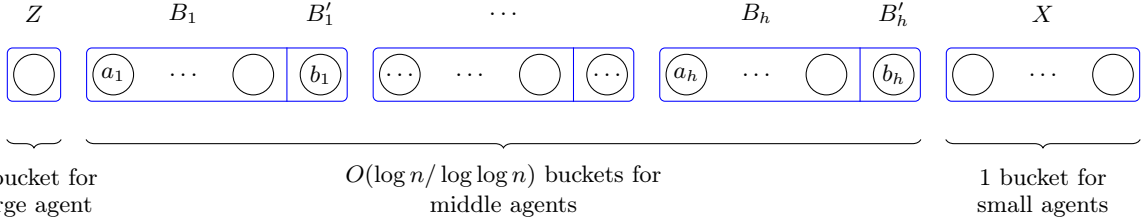
\begin{figure}[t]
\centering
\begin{tikzpicture}[
  node distance=3.2mm,
  circ/.style={circle, draw, minimum size=5.4mm, inner sep=0.6pt},
  box/.style={draw=blue, rectangle, rounded corners=2pt, inner sep=2.2pt},
  every node/.style={font=\small},
]

\def\k{0.7} 

\node[circ] (s) {};
\node[box, fit=(s)] (Sbox) {};

\node[circ, right=\k*5mm of s] (a1) {$a_1$};
\node[right=\k*3.4mm of a1, draw=none] (adots) {$\ldots$};
\node[circ, right=\k*3.4mm of adots] (a2) {};
\node[circ, right=\k*3.4mm of a2] (a3) {$b_1$};
\node[box, fit=(a1) (adots) (a2) (a3)] (A) {};

\node[circ, right=\k*5mm of a3] (b1) {$\ldots$};
\node[right=\k*3.4mm of b1, draw=none] (bdots) {$\ldots$};
\node[circ, right=\k*3.4mm of bdots] (b2) {};
\node[circ, right=\k*3.4mm of b2] (b3) {$\ldots$};
\node[box, fit=(b1) (bdots) (b2) (b3)] (B) {};

\node[circ, right=\k*5mm of b3] (c1) {$a_h$};
\node[right=\k*3.4mm of c1, draw=none] (cdots) {$\ldots$};
\node[circ, right=\k*3.4mm of cdots] (c2) {};
\node[circ, right=\k*3.4mm of c2] (c3) {$b_h$};
\node[box, fit=(c1) (cdots) (c2) (c3)] (C) {};

\node[circ, right=\k*5mm of c3] (d1) {};
\node[right=\k*3.4mm of d1, draw=none] (ddots) {$\ldots$};
\node[circ, right=\k*3.4mm of ddots] (d2) {};
\node[box, fit=(d1) (ddots) (d2)] (D) {};

\coordinate (labYtop) at ($(A.north)+(0,\k*4.5mm)$);
\node[draw=none] (Zlab) at (s     |- labYtop) {$Z$};
\node[draw=none] at (adots |- labYtop) {$B_1$};
\node[draw=none] at (a3    |- labYtop) {$B'_1$};
\node[draw=none] at (cdots |- labYtop) {$B_h$};
\node[draw=none] at (c3    |- labYtop) {$B'_h$};
\node[draw=none] (Xlab) at (ddots |- labYtop) {$X$};

\coordinate (labYbot) at ($(A.south)+(0,-\k*4.5mm)$);
\node[draw=none] (Zbase) at (s     |- labYbot) {};
\node[draw=none] (Xbase) at (ddots |- labYbot) {};

\node[draw=none] (MidEll) at (B.center |- labYbot) {};

\coordinate (AlabYbot) at ($(A.north)+(0,\k*4.5mm)$);
\node[draw=none] (AMidEll) at (B.center |- AlabYbot) {$\ldots$};

\node[draw=none] (Zc)  at ($(Zbase.south)+(0,-\k*4.2mm)$) {1 bucket for};
\node[draw=none] (Zcc) at ($(Zbase.south)+(0,-\k*8.2mm)$) {large agent};
\node[draw=none] (Mc)  at ($(MidEll.south)+(0,-\k*4.2mm)$) {$O(\log n / \log \log n)$ buckets for};
\node[draw=none] (Mcc) at ($(MidEll.south)+(0,-\k*8.2mm)$) {middle agents};
\node[draw=none] (Xc)  at ($(Xbase.south)+(0,-\k*4.2mm)$) {1 bucket for};
\node[draw=none] (Xcc) at ($(Xbase.south)+(0,-\k*8.2mm)$) {small agents};

\coordinate (braceY) at ($(Zc.north)+(0,\k*2.6mm)$);

\draw[decorate, decoration={brace, mirror, amplitude=\k*3.2pt}]
  (Sbox.south west |- braceY) -- (Sbox.south east |- braceY);

\draw[decorate, decoration={brace, mirror, amplitude=\k*3.2pt}]
  (A.south west |- braceY) -- (C.south east |- braceY);

\draw[decorate, decoration={brace, mirror, amplitude=\k*3.2pt}]
  (D.south west |- braceY) -- (D.south east |- braceY);

\coordinate (midA) at ($(a2)!0.5!(a3)$);
\draw[blue] (midA |- A.north) -- (midA |- A.south);

\coordinate (midB) at ($(b2)!0.5!(b3)$);
\draw[blue] (midB |- B.north) -- (midB |- B.south);

\coordinate (midC) at ($(c2)!0.5!(c3)$);
\draw[blue] (midC |- C.north) -- (midC |- C.south);

\end{tikzpicture}
\caption{The structure of the buckets used in Algorithm~\ref{alg:pof_xos}. Nodes represent agents ordered from left to right by non-increasing $\alpha_i^\star$. The text inside each node denotes the label assigned to its index. Blue boxes indicate the buckets, and the text above each box shows the label assigned to that bucket by the algorithm.}
\label{fig:bucketing}
\end{figure}

We first rename the agents such that 
$\alpha_1^\star \geq \alpha_2^\star \geq \cdots \geq \alpha_n^\star$.
If $\alpha_1^\star > 1/2$ and incentivizing agent 1 alone in $\contract^\star$ gives sufficiently high utility, then we form a contract that only incentivizes agent 1.
Otherwise, we decompose all agents with $\alpha_i^\star \leq 1/2$ into buckets, as follows:
We group all agents $i$ such that $\alpha_i^\star \leq \frac{1}{2n}$ into a single bucket, and partition the remaining (middle) agents greedily, adding agents to a bucket as long as the total payment does not exceed $1/2$ (see Figure~\ref{fig:bucketing} for an illustration).

Using the scaling-for-existence lemma from \cite{dutting2025black}, we show that, for each of these buckets, we can compute an equal-pay contract whose reward is at least  a constant fraction of the bucket's reward in the original equilibrium under $\contract^\star$. 
Additionally, the total payment of this equal-pay contract is at most $3/4$, implying that the principal's utility is at least $1/4$ of the bucket's reward. 
Finally, using the arithmetic-geometric (AM--GM) inequality alongside additional properties, we show that our bucketing procedure results in at most $O(\log n / \log \log n)$ buckets.
Thus, by the subadditivity of $f$, taking the highest-reward bucket yields the desired $O(\log n / \log \log n)$-approximation guarantee.

\subsection{Related Work}\label{subsec:RelatedWork}

\paragraph{Combinatorial Team Contracts.}

A combinatorial model for contracting a team of agents was introduced in \cite{babaioff2006combinatorial,BabaioffFNW12}, 
where each agent can either work or shirk, and the principal's reward for a subset of working agents is given by a Boolean function $f$. Follow-up works introduce mixed strategies to the model \cite{babaioff2006mixed}, and analyzed free-riding \cite{babaioff2009free}.

The work of \cite{duetting2022multi} generalized this model by allowing $f$ to be any monotone set function, while focusing on the complements-free hierarchy of \cite{lehmann2001combinatorial}.
For XOS $f$, they give a poly-time algorithm that achieves a constant-factor approximation to the optimal contract with demand oracle access. For submodular $f$ they show that value queries are sufficient to achieve the same guarantees.
Subsequent works considered the same model with additional budget constraints, and objectives beyond the principal's utility (profit)  \cite{aharoni2025welfare,feldman2025budget}. In particular, \cite{feldman2025budget} have introduced the class of BEST objectives, (which includes, among other objectives, profit, reward, and social welfare), and showed that any function from this class can be reduced to the unbudgeted profit maximization problem of \cite{duetting2022multi}, while only losing a constant factor to the approximation. 
\cite{alon2025multi} considered a multi-project setting, where each project corresponds to a multi-agent binary-action instance as in \cite{duetting2022multi}, and any agent can participate in at most a single project. For XOS $f$, they provide an efficient approximation algorithm for the optimal contract and allocation that achieves a constant-factor approximation, given access to a demand oracle.

Going beyond binary-actions, \cite{multimulti} introduced a model in which any agent can perform any subset of an individual collection of actions. This generalizes the binary-action model of \cite{duetting2022multi}, where each agent has one individual action.
For submodular $f$, they show how to achieve a constant-factor approximation to the optimal contract in poly-time with demand queries, and that no PTAS exists in this setting, even for binary actions.
\cite{oneactiontoomany} have shown that introducing budgets to this setting changes the computational landscape: when $f$ is submodular, exponentially-many demand queries are required to achieve any finite approximation, but a constant-factor approximation is achievable when $f$ is GS.
\cite{dutting2025black} examined more general equilibrium notions than pure Nash equilibrium (PNE), the most permissive being coarse correlated equilibrium (CCE). They consider profit maximization and show that, under XOS $f$, there exists a PNE that only loses a constant-factor compared to the best CCE.
For submodular $f$, they show how to extend guarantees made for a specific PNE to all CCEs of a contract, exemplifying the robustness of PNEs.

\paragraph{Fair Contracts.}
A recent line of work has considered fairness in contract design.
\cite{castiglioni2025fair} consider the multi-project model of \cite{alon2025multi}, but required that the allocation and contract are envy-free. They prove that achieving a constant factor approximation to the optimal envy-free contract and allocation is NP-hard, and show positive results when the number of projects is a constant. They also consider well-known relaxations of envy-freeness such as $\eps$-EF and EF1.

Closer to our setting, in an independent and concurrent work, \cite{castiglioni2025fairteam} consider the multi-agent binary-action model of \cite{duetting2022multi} with submodular rewards and study envy-free contracts; agent $i$ envies agent $j$ if swapping their contracts leads to a new equilibrium in which $i$ is strictly better-off.
Clearly, equal-pay contracts are envy-free in the above sense, and it is shown that the best equal-pay contract yields constant-factor approximation to  the optimal envy-free contract.
Thus, to approximate the optimal envy-free contract, they give a poly-time algorithm that computes an $O(1)$-approximation to the optimal equal-pay contract, using value and demand queries.
In comparison, for this setting our algorithm also achieves an $O(1)$-approximation, but for the more general XOS rewards.

In another concurrent and independent work, \cite{ding2026multi} study the equal-pay contracts, and specifically the price of equality (which they term price of non-discrimination), in the multi-agent binary-action setting of \cite{duetting2022multi}.
Their main result is an (almost) tight \POE: For submodular rewards and binary-actions the \POE\ is at most $O(\log n)$. Additionally, the \POE\ is at least $\Omega(\log n / \log \log n)$ for additive rewards.
While the lower bound on the \POE\ matches ours, our upper bound is stronger in three ways: It closes the $\log \log n$ factor gap,  holds for XOS $f$, and applies under combinatorial actions.
They also study $\gamma$-equal-pay and provide a near-tight bounds on the price of equality, for any possible $\gamma$, even if it depends on the number of agents.

Finally, the work of \cite{Feng2024priceof} consider a multi-agent environment in which every agent has combinatorial actions, as in \cite{dutting2022combinatorial}, and an observable individual outcome. 
They study the price of equality in that context, showing that for arbitrary rewards and costs the \POE\ can be as large as $n$, the number of agents. They prove refined bounds when agents have gross-substitutes rewards.

\paragraph{Other Combinatorial Models.}
A contracting model in which a single agent can perform any subset of $n$ actions was introduced by \cite{dutting2022combinatorial}. They showed that for gross substitutes $f$, the optimal contract can be computed in poly-time with value queries, while it is NP-hard to do so when $f$ is submodular. A series of follow-up works studied the tractability frontier of this problem \cite{contractsBeyondGS,deo2024supermodular,dutting2024query,feldman2025ultraefficient,ezra2023Inapproximability}. The work of
\cite{contractsSequential} studied a single agent which makes sequential, rather than simultaneous, actions and observe intermediate outcomes. 
Combinatorial inspection was introduced by \cite{contractsInspection}, where the agent takes a single action and the principal recommends an action and picks a set of actions to inspect. The principal withholds payment if the agent caught not performing the recommended actions, and pays the cost of inspection, given by a combinatorial function.

\paragraph{Linear Contracts.}
An important class of contracts is \emph{linear} contracts, in which the agent's payment is a fixed fraction of the principal's realized reward.
In the case of binary outcomes, as considered in this work, the optimal contract is linear.
Moreover, linear contracts exhibit robustness under uncertainty, formalized via the max-min optimality criterion.
This robustness is established by \cite{carroll2015robustness} under uncertainty about the feasible action set, and by \cite{dutting2019simple} under uncertainty about the outcome distribution; the former is extended to randomized contracts in \cite{peng2024optimal}.

A separate line of work considers settings in which the principal may strategically introduce ambiguity to increase her profit. In this context, \cite{dutting2024ambiguous,duetting2025succinct} show that linear contracts are ambiguity-proof: they are robust to such manipulations and cannot be improved by deliberately adding ambiguity.

\paragraph{Other Contractual Models.}
Non-combinatorial multi-agent settings with agent externalities are studied in \cite{segal1999contracting,segal2003coordination,bernstein2012contracting}. 
More recently, \cite{cacciamani2024multi,castiglioni2023multi} have studied multi-agent contracts beyond binary outcomes. 
Contracts where agents have private types are studied in \cite{alon2021contracts, alon2022bayesian, castiglioni2025reduction, CastiglioniM021, GuruganeshSW023, guruganesh2021contracts, castiglioni2022designing}. 
Several recent works explore the intersection of contracts and learning \cite{ZhuBYWJJ23, cohen2023learning, ho2014adaptive, BacchiocchiC0024,chen2024boundedcontractslearnableapproximately, duetting2025pseudodimensioncontracts,hogsgaard2026optimalsamplecomplexitylinear}.
\section{Model and Preliminaries}\label{sec:model}

In this section, we review the multi-agent combinatorial-actions model introduced in \cite{multimulti} (see Section~\ref{subsec:model}), and present several useful lemmas (see Section~\ref{apx:toolbox}).

\subsection{The Multi-Agent Combinatorial-Actions Setting}\label{subsec:model}

We study a contractual environment consisting of a single principal and a finite set $A=[n]$ of $n$ agents.
The principal owns a project whose outcome is binary: it either succeeds or fails.
A successful outcome yields a payoff of $1$ to the principal, while failure yields $0$.
The probability of success depends on the agents' (costly) actions, as detailed below.
The principal does not observe the agents' actions, only the realized outcome, and can condition payments upon it.

To incentivize the agents, the principal offers each agent $i \in A$ a \emph{linear contract} $\icontract \in [0,1]$, which specifies the payment made to agent $i$ upon project success; no payment is made if the project fails.
We denote the contract offered to all agents with $\noindexcontract \in [0,1]^A$.
Importantly, considering only linear contracts is without loss of generality in this binary-outcome model.\footnote{
Equivalently, one may view the model as a multi-outcome project in which the principal observes only the realized reward and is restricted to linear payment schemes.
}

Each agent $i \in A$ is associated with a set of available actions $T_i$, from which the agent may select any subset.
The special case $T_i=\{j\}$ corresponds to the binary-action model studied in \cite{duetting2022multi, feldman2025budget}, as the agent may choose either $j$ or $\emptyset$.
We assume that action sets are disjoint across agents, that is, $T_i \cap T_{i'} = \emptyset$ for all $i \neq i'$.
Each action $j \in T_i$ incurs a nonnegative cost $c_j \ge 0$, and costs are additive: for any $S_i \subseteq T_i$, we define
$c(S_i)=\sum_{j \in S_i} c_j$, with $c(\emptyset)=0$.
Let $T = \bigsqcup_{i \in A} T_i$ denote the disjoint union of all actions.

For any action profile $S \subseteq T$, we write $S_i = S \cap T_i$ for the actions chosen by agent $i$, and
$S_{-i} = \bigsqcup_{i' \in A \setminus \{i\}} S_{i'}$ for the actions selected by all other agents.

The project's success probability is determined by a monotone set function
$f : 2^T \to [0,1]$, which maps each action profile
$S = \bigsqcup_{i \in A} S_i$ to the probability that the project succeeds.
Since the principal's payoff upon success is normalized to $1$, $f(S)$ also represents her expected reward, and we refer to $f$ as the reward function.
We assume $f(\emptyset)=0$.

An instance of the multi-agent combinatorial-actions model is given by a tuple
$\multiInstance$, where $A$ is the set of agents, $\{T_i\}_{i \in A}$ are the agents' action sets,
$f$ is the success probability function, and $c=\{c_j\}_{j \in T}$ specifies action costs.

\paragraph{Utilities and Equilibrium.}
Given a contract $\noindexcontract$ and an action profile $S=(S_1,\ldots,S_n)$,
agent $i$'s expected utility is
\[
u_i(\noindexcontract,S)
=
\icontract \cdot f(S_i \sqcup S_{-i}) - c(S_i),
\]
namely, it is the expected payment minus the incurred cost.
We say that $\noindexcontract$ \emph{induces} an action profile $S$ if $S$ constitutes a (pure) Nash equilibrium under $\noindexcontract$.
That is, for every agent $i \in A$ and every alternative choice $S'_i \subseteq T_i$,
\[
\icontract \cdot f(S_i \sqcup S_{-i}) - c(S_i)
\;\ge\;
\icontract \cdot f(S'_i \sqcup S_{-i}) - c(S'_i).
\]

It follows from \cite{multimulti} that every contract $\noindexcontract$ admits at least one pure Nash equilibrium.
A contract may generally support multiple equilibria; we denote the set of all equilibria induced by $\noindexcontract$ by $\nash(\noindexcontract)$.

For a contract $\noindexcontract$ and an equilibrium $S \in \nash(\noindexcontract)$,
the principal's utility (or \emph{profit}) is defined as the expected reward minus payments:
\[
u_P(\noindexcontract,S)
=
\textstyle \left(1 - \sum_{i \in A} \icontract\right) f(S).
\]

We next define the notion of an equal-pay contract.

\begin{definition}[Equal-Pay Contract]
    A contract $\noindexcontract$ is an equal-pay contract if $\icontract = \alpha_j$ for any two agents $i,j$ with non-zero payments.
\end{definition}
For a scalar $t \in [0,1]$ and a subset of agents $G \subseteq \agents$, we denote by $\eqcontract{t}{G}$ the equal-pay contract which pays $t$ to every agent $i \in G$, and zero to all others.

\paragraph{Notation.}
We find it useful to introduce the following notation.
Given an action profile $S$ and a set of agents $G$, let $S|_G$ denote the action profile $S_i$ for each agent $i \in G$, and the empty action set for any agent $i \notin G$.
We also define $\agents(S) = \{i \in \agents \mid S_i \ne \emptyset\}$.
Observe that $S|_{A(S)} = S$.
Similarly, for a contract 
$\contract$ and a set of agents 
$G$, we write $\contract|_G$
for the projection of $\contract$ onto $G$, defined by
$\contract|_G = \contract \cdot \indicator[i \in G]$. 
When $G=\{i\}$ is a singleton, we often omit the brackets and write $\noindexsubcontract{i}$.

\paragraph{Reward Functions and Access Oracles}
Throughout this paper, we restrict attention to monotone reward functions $f$.
For sets $S,S'$, we write $f(S' \mid S)$ to denote the marginal gain of adding $S'$ to $S$, defined as $f(S' \mid S) = f(S' \cup S) - f(S)$.
We also use the following notation to denote marginal contribution of action $i$ to a set $S$ (which may include $i$): $f_S(i)=f({i} \mid S \setminus \{i\})$.

We focus on reward functions drawn from one of the following families. 

\begin{itemize}
    \item Additive: $f(S) = \sum_{a \in S} f(\{a\})$ for any $S \subseteq T$.
    \item Matroid Rank: For any set $S \subseteq T$, $f(S)$ is the rank of the set $S$ with respect to some matroid $\mathcal{M}=(T,I)$ over the ground set $T$. Namely, $f(S)$ is the cardinality of the largest independent set in $S$.
    \item Gross-Substitutes (GS): For any two price vectors $p,q \in \reals^{|T|}$, such that $p \le q$ coordinate-wise, and any $S^\star \in \argmax_{S \subseteq T} \{ f(S) - \sum_{a \in S} p_a \}$, there exists a \textit{demand bundle} with respect to $q$, i.e., $S' \in \argmax_{S \subseteq T} \{ f(S) - \sum_{a \in S} q_a \}$, such that $S^\star \cap \{j \mid p_j = q_j\} \subseteq S'$.
    \item Coverage: There exists a universe $U$, a weight function $w:U \to \reals_+$, and a mapping $h:T \to 2^U$ such that $f(S) = \sum_{u \in h(S)}w(u)$, where $h(S) = \bigcup_{i \in S} h(\{i\})$.
    \item Submodular: For any $S' \subseteq S \subseteq T$ and any $a \notin S$, it holds that
    $f(a \mid S') \ge f(a \mid S)$.
    \item XOS: There is a finite collection of additive functions $\mathcal{L}$ such that for any $S \subseteq T$, $f(S)=\max_{\ell \in \mathcal{L}} \ell(S)$.
    \item Subadditive: For any $S', S \subseteq T$, it holds that $f(S) + f(S') \ge f(S \cup S')$.
\end{itemize}
It is well-known that,
$\text{additive} \subsetneq \text{gross-substitutes} \subsetneq \text{submodular} \subsetneq \text{subadditive}$ \cite{lehmann2001combinatorial}.
Matroid rank functions are a strict subclass of GS, and coverage is a strict subclass of submodular functions, incomparable to GS.
As $f$ may have an exponential representation in $n$ (and $|T|$) we assume {algorithms} access {$f$} via two standard models:
\begin{itemize}
    \item Value oracle: accepts a set $S \subseteq T$ and returns $f(S)$.
    \item Demand oracle: accepts a vector $p \in \reals^{|T|}$ and returns  $S^\star \in \argmax_{S \subseteq T} \{ f(S) - \sum_{a \in S} p_a \}$.
\end{itemize}
Generally, demand oracles are strictly stronger than value oracles \cite{blumrosen2005computational}. However, for gross-substitutes $f$ a demand query can be computed in poly-time with value oracle access \cite{PaesLeme17}.

\subsection{Toolbox}\label{apx:toolbox}
In this section we provide several lemmas which we utilize throughout this work.

\paragraph{Subset Stability and the Doubling Lemma.}
Two key concepts in our analysis are \emph{subset stability} and the \emph{doubling lemma}. Subset stability weakens the standard Nash equilibrium requirement: an action profile $S = \bigsqcup_{i \in \agents} S_i$ is said to be subset-stable under a contract $\noindexcontract$ if no agent $i \in \agents$ can strictly improve her utility by switching to a subset of her chosen actions $S_i$.

\begin{definition}[Subset Stability, Definition 3.2 of \cite{multimulti}]\label{def:subsetstable}
    A set of actions $S$ is subset-stable with respect to contract $\noindexcontract$, if for every agent $i$, every subset of his actions $S'_i \subseteq S_i$ satisfies
    $$
    \icontract \cdot f(S_i \sqcup S_{-i}) - c(S_i) \ge \icontract \cdot f(S'_i \sqcup S_{-i})- c(S'_i).
    $$
\end{definition}

The doubling lemma, introduced in \cite{multimulti}, shows that any subset-stable action profile 
$S$ under a contract 
$\noindexcontract$ can be leveraged to induce a Nash equilibrium whose expected reward is at least half that of $S$. We present a closely related variant tailored to our equal-pay setting. The proof of the following lemma closely follows the original argument and is therefore deferred to Appendix~\ref{apx:modified_doubling}.

\begin{restatable}[Modified Doubling Lemma]{lemma}{modifiedDoublingLemma}\label{lemma:modified_doubling}
Suppose $f$ is submodular.  Let $S$ be a subset-stable action set with respect to a contract $\contract$, such that $S_i = \emptyset$ for all $i$ with $\alpha_i=0$.
Then any equilibrium $\Seq{}$ with respect to $2 \contract$ fulfills $f(\Seq{}) \geq \frac{1}{2} f(S)$.
\end{restatable}

When $f$ is submodular, restricted contracts maintain subset stability.
\begin{lemma}[\cite{multimulti}]\label{lem:subStableDownwards}
    Let $\multiInstance$ be an instance with submodular $f$.
    Let $S=\bigsqcup_{i \in \agents} S_i$ be a subset-stable profile of actions with respect to contract $\contract$.
    For any subset of agents $G \subseteq A$, 
    it holds that 
    $S|_G$
    is subset stable with respect to the contract $\noindexsubcontract{G}$. 
\end{lemma}

\paragraph{Dropout Stability and the Scaling-for-Existence Lemma.}
We next introduce the notion of dropout stability and the scaling-for-existence lemma.

\begin{definition}[Dropout Stability \cite{dutting2025black}]\label{def:dropout_stabl;e}
    A profile of actions $S$ is dropout stable with respect to $\contract$ if for every agent $i$, under $\alpha_i$, $S_i$ is a better response than shirking. Namely, 
    $\alpha_if(S_i\cup S_{-i}) -c(S_i) \ge \alpha_if(S_{-i})$.
\end{definition}

\begin{lemma}[Scaling-for-Existence Lemma \cite{dutting2025black}]\label{lem:scaling_for_exist}
    Let $f$ be any XOS function. For any contract $\contract$, and any dropout stable profile $S$ with respect to $\contract$.
    For any $\gamma >1$, and any $\agents' \subseteq \agents$ there exists an equilibrium $S' \in \nash(\contract')$ such that 
    $f(S') \ge (1-1/\gamma)f(S|_{A'})$,
    where $\contract' = \gamma \cdot \noindexsubcontract{A'}$.
    Moreover, $S'$ can be computed via a demand query to $f$ with prices $p_j = c_j/\alpha'_i$ for any $j \in T_i$, where $0/0=0$ and $c/0=\infty$ for $c>0$.
\end{lemma}

Lastly, we also utilize this useful lemma, due to \cite{duetting2022multi}.
\begin{lemma}[\cite{duetting2022multi}]\label{lem:xos_margs}
    For any XOS function $f$, and any sets $S \subseteq T$, it holds that 
    $
    \sum_{i \in S} f_T(i) \le f(S)
    $.
\end{lemma}
\section{Constant-Factor Approximation of Optimal Equal-Pay Contract}\label{sec:const_approx}

In this section we devise an efficient constant-factor approximation algorithm to the optimal equal-pay contract for submodular rewards with combinatorial actions.
\begin{theorem}\label{thm:equal-pay_const_approx}
    When $f$ is submodular, there exists a poly-time algorithm that computes a constant-factor approximation to the optimal equal-pay contract.
\end{theorem}

To achieve this result we utilize the reduction of \cite{multimulti}, which implies that a constant-factor approximation to the optimal contract can be achieved by a contract-equilibrium pair in which either no agent is large, or only one agent is paid (see statement below).
We observe that this reduction preserves equal-pay, and thus considering these two cases suffices when approximating the optimal equal-pay contract.

\begin{theorem}[Reduction to no agent is large / single agent (equal-pay), \cite{multimulti}] \label{thm:modified_defk_cases}
    Consider any submodular success probability function $f$. Let $\contract^\star$ be any equal-pay contract and $S^\star$ be an equilibrium of $\contract^\star$. For any $0 \leq \phi \leq 1$, there exists an equal-pay contract $\contract'$ and an equilibrium $S'$ of $\contract'$
    fulfilling
    $(1-\sum_i \alpha'_i) \cdot f(S') \geq ({\phi}/{1120}) \cdot (1-\sum_i \alpha^\star_i) \cdot f(S^\star)$ and
    \begin{enumerate}
        \item \textbf{(no large agent)} $f(S'_i) \leq \phi \cdot f(S')$ for all $i \in \agents$, or
        \item \textbf{(single agent)} 
        $\alpha'_i = 0$ and $S_i = \emptyset$ for all but one $i \in \agents$.
    \end{enumerate}
\end{theorem}

We also utilize the FPTAS of \cite{multimulti} for single-agent contracts:
\begin{theorem}[\cite{multimulti}]\label{thm:single-fptas}
    There exists an FPTAS for the single-agent contract design problem with general $f$ and demand oracle access.
\end{theorem}

Observe that, by combining Theorem~\ref{thm:modified_defk_cases} and Theorem~\ref{thm:single-fptas}, to prove Theorem~\ref{thm:equal-pay_const_approx} it suffices to show the following {theorem, which handles the case of ``no large agent".}

\begin{restatable}{theorem}{ConstApproxNoSmall} \label{thm:const_approx_no_small}
    When $f$ is submodular, Algorithm~\ref{alg:approx_equal_pay} runs in polynomial time with value oracle access to $f$, and returns an equal-pay contract $\alpha$ such that in any equilibrium, the principal's utility is at least $\Lambda$ with the following guarantee.
    For any equal-pay contract $\constar$ and any equilibrium $S^\star$ of $\constar$ such that $f(S^\star_i) \le (1/32) \cdot f(S^\star)$ for all $i\in \agents$, it holds that 
    \[ \textstyle
    \Lambda \ge \Omega(1) \left(1-\sum_{i \in A}\alpha^\star_i\right)f(S^\star).
    \]
\end{restatable}

\begin{remark}
    In the case of binary actions and submodular $f$, the single-agent problem is tractable with just value queries. This obviates the need for Theorem~\ref{thm:single-fptas}, and implies that, in this setting, the guarantees of Theorem~\ref{thm:equal-pay_const_approx} hold using value queries alone.
\end{remark}

The algorithm we provide for Theorem~\ref{thm:const_approx_no_small} relies on approximate demand queries to find approximately optimal action sets, augmented with a cardinality constraint to limit the number of participating agents and, as a result, the total payments. 
Consequently, in Section~\ref{sec:approx-demand}, we demonstrate how to approximate demand queries subject to a cardinality constraint on the number of participating agents, and in Section~\ref{sec:proof-gs}, we provide the full proof of Theorem~\ref{thm:const_approx_no_small}.

\subsection{Approximate Demand Queries with Agent Cardinality Constraints}
\label{sec:approx-demand}

In this section we address the correctness of Algorithm~\ref{alg:agent_constrained_approx_demand}, which is our main tool to overcome the fact that deamnd queries are agent-agnostic (see discussion in Section~\ref{sec:intro}).

\begin{restatable}[Approximate Demand Query With Agent Cardinality Constraints]{lemma}{approximateDemandQueryWtihConstraint}\label{lem:agent_constrained_approx_demand}
    When $f$ is submodular, Algorithm~\ref{alg:agent_constrained_approx_demand} runs in polynomial time and outputs a set {$S \subseteq T$} such that 
    \[
    \textstyle 
    |A(S)| \leq k \quad\text{and
}\quad f(S)-\sum_{i\in S} p_i \ge  \max_{U\subseteq [m]: |A(U)|\le k}\left\{(1-1/e)^2 \cdot f(U)-\sum_{i\in U} p_i\right\}.
    \]
\end{restatable}

The algorithm implements a two-level extension of the distorted greedy method of \cite{Harshaw2019SubmodularMB}. 
The upper level applies distorted greedy selection over agents: in iteration $i$ it estimates each agent's contribution relative to the current aggregate action set $S_i$ under a distorted objective where the value function $f$ is scaled by $(1-1/k)^{k-(i+1)}$. 
To estimate an agent's contribution, the lower level runs the original Distorted Greedy algorithm restricted to that agent's actions. 
The algorithm then selects the agent with the highest contribution, adds its action bundle to $S_{i+1}$, and repeats this process $k$ times. The full proof of Lemma~\ref{lem:agent_constrained_approx_demand} is deferred to Lemma~\ref{sec:proof_of_agent_constrained_approx_demand}.

\begin{algorithm}[t]
    \caption{An $(1-1/e)^2$-approximate demand query subject to agent cardinality constraints.}\label{alg:agent_constrained_approx_demand}
    \KwIn{A partition of actions $[m]=T_1\uplus \dots \uplus T_n$, a number $k\in \mathbb{N}^+$, a price vector $p\in \reals^m$, and value oracle access to a monotone submodular set function $f:2^{[m]}\rightarrow \reals_{\ge 0}$.}
    \KwOut{A set $S\subseteq [m]$ such that $|A(S)|\le k$ and $f(S)-\sum_{i\in S} p_i \ge \max_{T\subseteq [m]: |A(T)|\le k}\left\{ (1-1/e)^2 \cdot f(T)-\sum_{i\in T} p_i \right\}$.}
    {$S_0 \gets \emptyset$\;}
    \For{$i \gets 0, 1, \ldots, k-1$}{
        \For{$j \in [n]$}{
            Define $h_{i,j}(A)\coloneqq (1-1/k)^{k-(i+1)} \cdot  f(A\mid S_i)-  p(A)$ for $A\subseteq T_j$ \; 
                Use the Distorted Greedy algorithm  \cite[Theorem 3]{Harshaw2019SubmodularMB}  with objective $h_{i,j}$  to compute $A_{i,j} \subseteq T_j$ such that $h_{i,j}(A_{i,j}) \geq (1-1/e) \cdot (1-1/k)^{k-(i+1)} \cdot  f(A\mid S_i)-p(A)$ for all $A \subseteq T_j$\; \label{line:greedy}
            $v_{i,j}\gets \max\{0,\,h_{i,j}(A_{i,j})\}$ \;
        }
            Choose $j_i\in\arg\max_{j\in [n]} v_{i,j}$ (break ties arbitrarily) \;
    \lIf{$v_{i,j_i} > 0$}{$S_{i+1}\gets S_i\cup A_{i,j_i}$ \label{line:if_positiev}} 
     \lIf{$v_{i,j_i} = 0$}{$S_{i+1}\gets S_i$  \label{line:if_negative}}
    }
    \Return $S_k$\;\label{line:return_action_set}
\end{algorithm}

\subsection{Proof of Theorem~\ref{thm:const_approx_no_small}}
\label{sec:proof-gs}

\begin{algorithm}[t]
\caption{An $O(1)$-approximation for the optimal equal-pay contract under small agents}\label{alg:approx_equal_pay}
\KwIn{A partition of actions $[m]=T_1\uplus \dots \uplus T_n$, costs $c_1,\dots,c_m \ge 0$, and value oracle access to a submodular set function $f:2^{[m]}\rightarrow \reals_{\ge 0}$}
\KwOut{A contract $\alpha$ satisfying the guarantees of Theorem~\ref{thm:const_approx_no_small}.}
\For{$k\gets 16,17,\dots,n$}{
$p_j\gets {c_jk}/{4}$ for all $j\in [m]$\;
Let $Q\subseteq T$ be the output of Algorithm~\ref{alg:agent_constrained_approx_demand} with input $k,p,f$\; \label{line:demand_Q}
\If{$|A(Q)|\le k/16$} {
$S^k \gets Q, \contract^k \gets \eqcontract{(8/k)}{A(Q)}$\;\label{line:small_Q} 
}
\Else{
$V\gets \emptyset$\;
\For{$\ell\gets 1,\dots,\lfloor k/16\rfloor$}{
$i^\star\gets \argmax_{i\in \agents} f(V\cup Q_i)$\;
$V\gets V\cup Q_{i^\star}$\;
}
$S^k \gets V, \contract^k \gets\eqcontract{(8/k)}{A(V)}$ \label{line:large_Q}
}
}
$k^\star \gets \arg\max_{k\in \{16,\dots,n\}} f(S^k)$\;
\Return $\alpha^{k^\star}$\;
\end{algorithm}

The proof of Theorem~\ref{thm:const_approx_no_small} relies on two key lemmas. The first establishes a performance guarantee on the reward achieved by Algorithm~\ref{alg:approx_equal_pay}.
\begin{lemma} \label{lem:f_approx}
    Consider the sequence $S^{16},\dots S^n$ generated by Algorithm~\ref{alg:approx_equal_pay}. When $f$ is submodular,
    for any equal-pay contract $\constar$ such that $\sum_{i \in A} \alpha^\star_i \le 1$ and any equilibrium $S^\star$ of $\constar$ such that $f(S^\star_i) \le ({1}/{32}) \cdot f(S^\star)$ for all $i\in \agents$, 
    there exists $k$ such that 
    \[
    f(S^k) \ge ({1}/{620}) \left(1/e\right) f(S^\star)
    \]
\end{lemma}
\begin{proof}
    Let $\constar$ be an equal-pay contract and $S^\star$ be an equilibrium of $\constar$ such that $f(S^\star_i)\le (1/32) \cdot f(S^\star)$.
    Define $A_0 = \{i\in A(S^\star)\mid \alpha^\star_i = 0\}$ and $A_1 = A(S^\star)\setminus A_0$.
    
    We construct a set $B$ as follows:
    if $f(S^{\star}_{A_0}) \ge (1/2) \cdot {f(S^\star)}$ then $B = A_0$, and otherwise $B = A_1$. Note that by subadditivity, $f(S^\star_B) \geq (1/2) \cdot {f(S^\star)}$.
    
    We claim that $S^\star_B$ is subset-stable with respect to the equal-pay contract $\eqcontract{(1/|B|)}{B}$.
    Indeed, if $B=A_0$, the claim holds since all actions in $S^\star_B$ have zero cost.  If $B=A_1$, then by Lemma~\ref{lem:subStableDownwards}, $S^{\star}_B$ is subset-stable with respect to the contract $\constar$. Since $\constar$ has $|B|$ equal positive entries, $\alpha^{\star}_i \le {1}/{|B|}$, and increasing the payments  maintains subset stability. 
    
    Furthermore, for any $i\in B$, it holds that $f(S^\star_i) \le ({1}/{32}) \cdot f(S^\star) \le (1/16) \cdot f(S^\star_B)$, and thus, by subadditivity $|B| \ge 16$. Hence there is an iteration with $k=|B|$. 
    Consider the set $Q\subseteq T$, and the vector $\boldsymbol{p} \in \reals_{\ge 0}^m$ in that iteration, and observe that:
    \begin{align}\label{eq:f_Q_bound}
       f(Q) &\ge
        f(Q)-\textstyle  \sum_{j\in Q} p_j \nonumber && (\text{since $p_j \geq 0$}) \\
        &\ge(1-1/e)^2 \cdot f(S^\star_B) -\textstyle  \sum_{j\in S^\star_B} \frac{c_jk}{4}   && (\text{by Lemma~\ref{lem:agent_constrained_approx_demand}}) \nonumber\\
        &\ge (1-1/e)^2 \cdot f(S^\star_B) -\textstyle  \sum_{j\in S^\star_B} \frac14f_{S^\star_B}(\{j\})  && (\text{since $S^\star_B$ is subset-stable w.r.t $\eqcontract{(1/k)}{B}$}) \nonumber \\
        &\ge \left( (1-1/e)^2 - 1/4 \right) \cdot f(S^\star_B) && (\text{by Lemma~\ref{lem:xos_margs}})
        \nonumber\\
        &\ge (1/2) \cdot \left( (1-1/e)^2 - 1/4 \right) \cdot f(S^\star) && (\text{by the choice of $B$}) \nonumber \\
        &\geq (1/20) \cdot f(S^\star) \nonumber.
\end{align}

First, if $S^k$ is determined in Line~\ref{line:small_Q}, then $f(S^k)=f(Q) \geq (1/20) \cdot f(S^\star)$, as needed.

Otherwise, if $S^k$ is determined in Line~\ref{line:large_Q}, observe that lines 8-11 perform greedy maximization of the  function
$g : 2^{A} \to \mathbb{R}_{\geq 0}$ given by $g(S) = f(Q|_{S})$. Since $f$ is submodular, $g$ is submodular, and
thus, by the analysis of greedy maximization of \cite{submodularmaximization} (see also \cite[Lemma 2.2]{bazzi2016submodular}), 
\begin{align*}
f(S^k)&=f(V) = f(Q|_{A(V)}) =g(A(V)) \ge (1-e^{ - \lfloor k/16 \rfloor /|A(Q)|}) f(Q|_{A(Q)}) 
\end{align*}
Using inequalities $1-e^{-x}\ge x/e$ for all $x\in[0,1]$ and $\frac{\lfloor k/16\rfloor}{k}\ge \frac1{31}$ for all $k\ge16$, we obtain:
\begin{equation*}
\begin{aligned}
\label{eq:fv_lower}
f(S^k)
&\ge \left(\frac{ \lfloor k/16 \rfloor }{|A(Q)| \cdot e}\right) f(Q|_{A(Q)}) 
\ge \left(\frac{\lfloor k/16 \rfloor}{k \cdot e}\right) f(Q|_{A(Q)}) = \left(\frac{\lfloor k/16 \rfloor}{k \cdot e}\right) f(Q) \ge \left(\frac{1}{31 \cdot 20 \cdot e}\right) f(S^\star),
\end{aligned}
\end{equation*}
as needed.
\end{proof}

The next lemma shows that the principal's utility from the contract returned by Algorithm~\ref{alg:approx_equal_pay} is at least a constant fraction of the reward.

\begin{restatable}{lemma}{WorstCaseEquilibriumAlphaK} \label{lem:worst_case_equilibrium_alphak}
    Consider the sequences $\alpha^{16},\dots,\alpha^n$ and $S^{16},\dots, S^n$ generated by Algorithm~\ref{alg:approx_equal_pay}. When $f$ is submodular, for every $k\in \{16, \ldots, n\}$, for every equilibrium $S$ of $\alpha^k$ it holds that
    \[
    \textstyle \left(1-\sum_{i \in A}\alpha^k_i\right)f(S) \ge (1/4) \cdot f(S^k)
    \]
\end{restatable}

\begin{proof}
\newcommand{\betacon}{\boldsymbol{\beta}}
    Let $k\in \{16,\dots, n\}$.
    Recall the set profile of actions $Q$, which is the ``approximate'' demand bundle computed in Line~\ref{line:demand_Q}.
    First, observe that it is without loss of generality to assume that $f_Q(j) \ge p_j$ for any action $j \in Q$. Otherwise, we remove such $j$ from $Q$, and improve its guarantee.
    
    Let $\betacon = \eqcontract{(4/k)}{A(Q)}$ be the equal-pay contract which pays $(4/k)$ to the agents in $A(Q)$, and zero to others. We claim that $Q$ is subset-stable with respect to $\betacon$.
    Fix $ i \in A(Q)$, and let $Q'_i \subseteq Q_i$.
    Indeed, we show that under the contract $\betacon$, the agent's utility when performing $Q_i$ is at least as high as when performing $Q'_i$:
    \begin{align*}
    &\beta_i f(Q_i, Q_{-i})-c(Q_i) -\left(\beta_i f(Q_i', Q_{-i})-c(Q_i')\right) &&\\
    &= \frac{4}{k}\left(f(Q_i, Q_{-i})-f(Q_i',Q_{-i})\right) - c(Q_i\setminus Q_i') \\
    &\ge \frac{4}{k} \sum_{j\in Q_i \setminus Q_i'} f(j\mid Q\setminus \{j\})-\sum_{j\in Q_i \setminus Q_i'} c_j && \text{(by submodularity of $f$)} \\
    &=\sum_{j\in Q_i\setminus Q_i'}\frac{4}{k} f(j\mid Q\setminus \{j\}) - c_j && \\
    &\ge \sum_{j\in Q_i\setminus Q_i'} 0. && (\text{since $f(j\mid Q\setminus \{j\}) \geq  p_j ={k c_j}/{4}$})
    \end{align*}
       
     If Algorithm~\ref{alg:approx_equal_pay} determines $\alpha^k$ in Line~\ref{line:small_Q}, then $\alpha^k = 2\betacon$ and $S^k=Q$. Thus, since $Q$ is subset-stable with respect to the contract $\betacon$, by Lemma~\ref{lemma:modified_doubling} we have that for every equilibrium $S$ of $\alpha^k$, $f(S)\ge \frac{1}{2}f(Q)=\frac{1}{2}f(S^k)$.
     
     Otherwise, 
     $\contract^k=\eqcontract{(8/k)}{A(V)}$, where $V \subseteq Q$ is constructed in the for-loop in lines 8-10 of the algorithm. 
     Since $V$ is exactly the actions taken by agents $A(V)$ in the profile $Q$, Lemma~\ref{lem:subStableDownwards} implies that $V$ is subset stable with respect to $\eqcontract{(4/k)}{A(V)}$. By Lemma~\ref{lemma:modified_doubling} every equilibrium $S$ of $\alpha^k$,
     $f(S) \ge \frac{1}{2}f(V)=\frac{1}{2}f(S^k)$.
     Finally, observe that it always holds that $|A(S_k)|\le \lfloor \frac{k}{16}\rfloor$. Thus, the total payment of the equal-pay contract $\contract^k$ is at most $1/2$: $ \sum_{i \in A} \alpha_i^k = \frac{8}{k}|A(S^k)| \le \frac{1}{2}$.
     Putting everything together, we get
     \[
     \left(1-\sum_{i \in A} \alpha_i^k\right) f(S) \ge \frac{1}{2} \cdot \frac{1}{2} \cdot f(S^k) = \frac{1}{4}f(S^k),
     \]
     as needed.
\end{proof}

The proof of Theorem~\ref{thm:const_approx_no_small}  follows easily from the two lemmas above, as follows.

\begin{proof}
    Let $\constar$ be an equal-pay contract such that $\sum_{i \in A} \alpha^{\star}_i \le 1$ and let $S^\star$ be an equilibrium of $\constar$ such that $f(S^\star_i)\le \frac{1}{32}f(S^\star)$ for all $i\in \agents$.

    By Lemma~\ref{lem:f_approx}, there exists some $k\in \{16,\dots, n\}$ such that $f(S^k)\ge \frac{1}{620 \cdot e} f(S^\star)$, and in particular $f(S^{k^\star}) \ge \frac{1}{620 \cdot e} f(S^\star)$.
    By Lemma~\ref{lem:worst_case_equilibrium_alphak}, for every equilibrium $S$ of $\alpha^{k^{\star}}$ we have 
    \[
    \left(1-\sum_{i \in A}\alpha_i^{k^\star}\right)f(S) \ge \frac{1}{4}f(S^{k^\star}) \ge \frac{1}{4} \cdot  \left(\frac{1}{620 \cdot e}\right)
    f(S^\star) \ge  \left(\frac{1}{2480 \cdot e}\right)
    \left(1-\sum_{i \in A}\alpha_i^\star\right) f(S^\star),
    \]
    as needed.
\end{proof}

\begin{remark}
    When the reward function $f$ is gross-substitutes, the constant-factor approximation can be extended to any BEST objective (see Definition~\ref{def:goodobj_multi_multi}), including reward and social welfare.
    This result follows from the reduction between BEST objectives presented in \cite{oneactiontoomany}, which, together with our modified doubling lemma (Lemma~\ref{lemma:modified_doubling}), can be adjusted to preserve equal-pay.
\end{remark}

\section{Hardness for Equal-Pay and Unconstrained Contracts}
\label{sec:hardness}

In this section, we establish hardness results for both equal-pay contracts and unconstrained contracts. 
We begin with a simple observation that provides a general framework for proving hardness results for equal-pay and general contracts simultaneously.
\begin{observation}\label{obs:hardnesstransfer} 
Fix an approximation guarantee $K : \mathbb{N} \to [0, \infty)$.
Let $\mathcal{C}$ be a class of instances such that, for every instance $\mathcal{I} \in \mathcal{C}$, the maximum principal's utility over all unconstrained contracts is attained by an equal-pay contract.
If there exists a polynomial-time algorithm that achieves a $K(n)$-approximation to the optimal equal-pay contract or $\gamma$-equal-pay for all instances in $\mathcal{C}$, then the same algorithm achieves a $K(n)$-approximation over unconstrained contracts for all instances in $\mathcal{C}$.
\end{observation}
This observation implies that if one establishes hardness for computing optimal unconstrained contracts, and the instances used in the hardness construction admit optimal contracts that are equal-pay, then the same hardness result immediately applies to equal-pay contracts. 
Notably, three hardness results from prior work have this structure: 
\begin{enumerate}[noitemsep, topsep=0pt, label=(\textit{\roman*})]
    \item There is no constant-factor approximation algorithm for subadditive $f$ with binary actions that uses polynomially many value and demand queries \cite[Theorem 4.1]{duetting2022multi}.
    \item There is no PTAS for submodular $f$ with binary actions that uses polynomially many value and demand queries \cite[Theorem 6.1]{multimulti}.
    \item There is no constant-factor approximation algorithm for XOS $f$ with binary actions that uses polynomially many value queries \cite[Theorem 5.1]{ezra2023Inapproximability}.
\end{enumerate}
The three results above therefore imply hardness for equal-pay contracts via Observation~\ref{obs:hardnesstransfer}.

In this section, we prove two new hardness results for unconstrained contracts:
\begin{enumerate}[noitemsep, topsep=0pt,label=(\textit{\arabic*})]
\item There is no constant-factor approximation for XOS $f$ with combinatorial actions that uses polynomially many value and demand queries (Section~\ref{sec:hardness_xos}).
\item Unless \pequalsnp, there is no PTAS for gross substitutes $f$ {(or even matroid rank functions $f$)}   with combinatorial actions that uses value and demand queries (Section~\ref{sec:hardness-matroid}).
\end{enumerate}
In both cases, the constructions satisfy the conditions of Observation~\ref{obs:hardnesstransfer}, and hence each result also implies the corresponding hardness for equal-pay contracts.

\subsection{No Constant-Factor Approximation for XOS}\label{sec:hardness_xos}

In this section, we show that when $f$ is XOS, no polynomial-time constant-factor approximation is achievable.

\begin{restatable}{theorem}{XosInapprox} \label{thm:XOS_inapprox}
When $f$ is XOS, any (possibly randomized) algorithm that makes only polynomially many demand queries achieves an (expected) approximation ratio of at least $\Omega(n^{1/7})$ to the optimal profit over unconstrained contracts under combinatorial actions. 
Moreover, the same lower bound applies to equal-pay contracts.
\end{restatable}

Our result is information-theoretic: We construct an instance with a probability distribution over
XOS reward functions and establish an upper bound on the performance of any
deterministic algorithm with demand oracle access on this randomized input. By Yao's principle, this directly implies the statement of the theorem. We next describe the distribution over instances.

Fix an integer parameter $\ell \geq 5$.
We construct a family of instances $\mathcal{I}^{(G)} = \langle A, \{T_i\}_{i \in A}, f^{(G)}, c \rangle$ indexed by subsets $G \subseteq A$ with $|G| = \ell^2$. The agent set is $A = \{1, 2, \ldots, n\}$, where $n = \ell^7$. 
Each agent $i \in A$ has $m' = \ell^3$ individual actions, $T_i = \{(i-1) \cdot m' + 1, \ldots, (i-1) \cdot m' + m'\}$. 
The set of all actions is $T = T_1 \sqcup \ldots \sqcup T_n$.
Every action $j \in T_i$ incurs cost $c_{j} = 1/(2\ell^3)$. The success probability function is defined by:
\begin{align*}
    f^{(G)}(S) = \max \left\{  |A(S)|, \; (1/\ell) \cdot \big|S|_G\big|, \; \ell^3 \right\},
\end{align*}
for any non-empty action profile $S \subseteq T$, and $f(\emptyset) = 0$. When the choice of $G$ is clear from context, we suppress it and write $f$ in place of $f^{(G)}$. 
The definition of $f^{(G)}$ slightly abuses the notation since success probabilities must lie in $[0,1]$; however, by scaling both success probabilities and costs by a factor of $1/\ell^7$, the success probabilities can be made to lie in $[0,1]$.

We also define an instance $\mathcal{I}' = \langle A, \{T_i\}_{i \in A}, f', c \rangle$, which shares the same agents, actions, and costs as the family of instances above, and whose success probability is:
$f'(S) = \max\{ |A(S)|, \ell^3 \}$.    

We will show that, when $G$ is sampled uniformly at random from all $\ell^2$-size subsets of $A$, no deterministic algorithm can distinguish between $\mathcal{I}^{(G)}$ and $\mathcal{I}'$ with non-negligible probability.

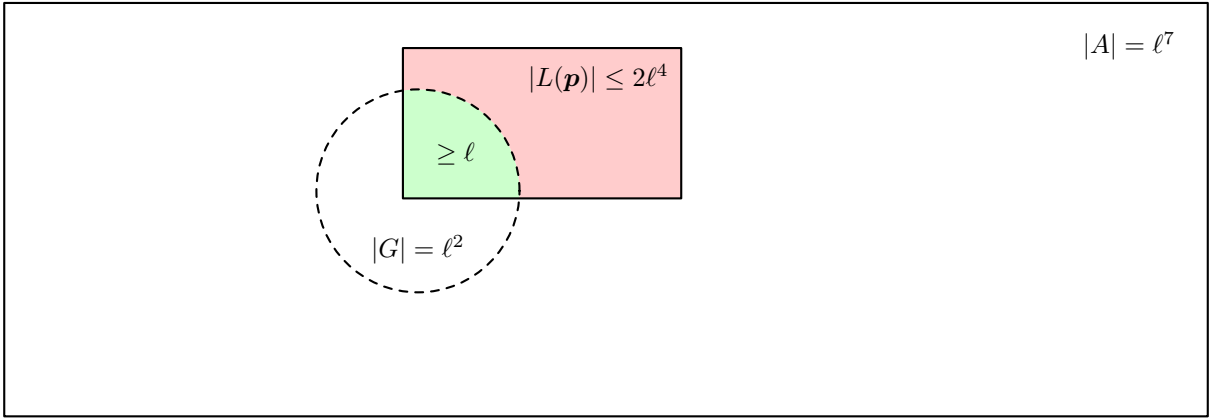
\begin{figure}
\centering
\begin{tikzpicture}[x=0.75cm,y=0.75cm,line cap=round,line join=round,thick]
  \draw (-0.5,0.5) rectangle (15.5,6.0);

  \node[anchor=north east] at (15.2,5.75) {$|A|=\ell^{7}$};

  \def\cx{5}
  \def\cy{3.5}
  \def\r{1.35}
  \def\rectanglex{8.5}

  \begin{scope}
    \clip (\cx,\cy) circle (\r);
    \fill[green!20] (4.8,5.4) rectangle (7.8,3.4);
  \end{scope}

  \begin{scope}
    \clip (4.8,5.4) rectangle (\rectanglex,3.4);
    \fill[red!20, even odd rule]
      (4.8,5.4) rectangle (\rectanglex,3.4)
      (\cx,\cy) circle (\r);
  \end{scope}

  \draw[dashed] (\cx,\cy) circle (\r);
  \node[anchor=south] at (\cx,\cy-\r+0.25) {$|G|=\ell^{2}$};

  \draw (4.8,5.4) rectangle (\rectanglex,3.4);
  \node at (\rectanglex-1.3,5) {$|L(\boldsymbol{p})| \le 2\ell^{4}$};

  \node at (5.5,4) {$\ge \ell$};
\end{tikzpicture}
\caption{Illustration of the definition of informative demand queries. 
The large rectangle represents the set of all agents, and the dashed circle represents the hidden subset of agents $G$. The smaller rectangle depicts the set $L(\boldsymbol{p})$ for a given price vector $\boldsymbol{p}$. The red region corresponds to $L(\boldsymbol{p}) \setminus G$, while the green region corresponds to $L(\boldsymbol{p}) \cap G$. A demand query $\boldsymbol{p}$ is informative if the total number of agents in the red and green regions is at most $2\ell^4$, and the green region alone contains at least $\ell$ agents.}
\label{fig:informative_demand_query}
\end{figure}

We begin by introducing a condition on demand queries that plays a central role in our analysis. An illustration of this definition is provided in Figure~\ref{fig:informative_demand_query}.

\begin{definition} [Informative Demand Query]\label{def:informative_query}
For any price vector $\boldsymbol{p} \in \mathbb{R}^T$, define:
\begin{align*}
    L(\boldsymbol{p}) = \{i \in \agents\mid p_{j} \leq 1/\ell \text{  for some  } j \in T_i \}.
\end{align*}
We call a price vector $\boldsymbol{p}$, and its associated demand query, \emph{informative} with respect to $G \subseteq A$ if it satisfies both $|L(\boldsymbol{p})| < 2\ell^4$ and $|L(\boldsymbol{p}) \cap G| > \ell$.
\end{definition}

A price vector that does not satisfy the conditions above is referred to as \emph{uninformative}.

We now state the first useful property of informative demand queries, roughly, that an uninformative price vector cannot help distinguish between an instance of the form $\I^{(G)}$ and $\I'$.

\begin{lemma}[Uninformative Queries Reveal No Information About $G$] \label{lem:demand_independent}
    For any $G \subseteq A$ with $|G| = \ell^2$ and  any  price vector $\boldsymbol{p}$ that is uninformative with respect to $G$, we have $D_{f^{(G)}}(\boldsymbol{p}) = D_{f'}(\boldsymbol{p})$.
\end{lemma}
\begin{proof}
    Let $\boldsymbol{p}$ be an uninformative price vector with respect to $G$.
   Define $\mathcal{S}_2 = \{S \subseteq T \mid f^{(G)}(S) = f'(S) \}$ and $\mathcal{S}_1 = 2^T \setminus \mathcal{S}_2$.
    Observe that:
    \begin{align*}
        f^{(G)}(S) > f'(S) \text{ for all }S \in \mathcal{S}_1 \quad \text{and} \quad f^{(G)}(S) = f'(S) \text{ for all }S \in \mathcal{S}_2.
    \end{align*}
    It is sufficient to show that $D_{f^{(G)}}(\boldsymbol{p}) \cap \mathcal{S}_1 = \emptyset$, as by the observation above, this immediately implies that $D_{f'}(\boldsymbol{p}) \cap \mathcal{S}_1 = \emptyset$ and that $D_{f^{(G)}}(\boldsymbol{p}) \cap \mathcal{S}_2 = D_{f'}(\boldsymbol{p}) \cap \mathcal{S}_2$. 
    
    Let $S \in D_{f^{(G)}}(p)$ be a demand bundle.
Suppose, towards a contradiction, that 
$S \in \mathcal{S}_1$, i.e.,
\begin{equation}\label{eq:uninfo_p}
    f^{(G)}(S) = (1/\ell) \cdot \big|S|_G\big | > \max\left\{|A(S)|, \ell^3\right\}.
\end{equation}

We claim that for every action $j \in S$, it holds that $f_S(j) \le 1/\ell$.
Indeed, for any $i \in G$ and $j \in T_i$ we have $f(S \setminus \{j\}) = (1/\ell) \cdot (\big|S|_G\big| - 1)$. For any $i \notin G$ and $j \in T_i$ we have $f(S \setminus \{j\}) = (1/\ell) \cdot \big|S|_G\big|$. In both cases, the marginal contribution of action $j$ is at most $1/\ell$, establishing the claim.

Since $S$ is a demand set, we must have $p_{j} \le f_S(j) \le 1/\ell$ for all $j \in S$. Consequently, $A(S) \subseteq L(\boldsymbol{p})$.
Additionally, Equation~\ref{eq:uninfo_p} implies that $\big|S|_G\big| > \ell^4$. Since each agent in $G$ has at most $\ell^3$ actions, it follows that $|A(S) \cap G| > \ell$.
Putting the two together we have that 
$|L(\boldsymbol{p}) \cap G| > \ell$. As $\boldsymbol{p}$ is uninformative, this implies that $|L(\boldsymbol{p})| \ge 2\ell^4$.

Now consider an action profile $S'$ defined as follows. For each $i \in L(\boldsymbol{p})$, let $S'_i = \{j\}$ for some $j \in T_i$ satisfying $p_{j} \le 1/\ell$, which exists by the definition of $L(\boldsymbol{p})$. For each $i \notin L(\boldsymbol{p})$, let $S'_i = \emptyset$.
Observe that $|S'|=|A(S')| = |L(\boldsymbol{p})| \ge 2\ell^4$. 
As $\big|S|_G\big| \le\big| \bigcup_{i \in G} T_i \big| \le \ell^5$, we have
   \begin{align*}
    \textstyle f(S')-\sum_{j\in S'} p_{j} 
    \geq
    |A(S')| \cdot (1-1/\ell)
    \geq 2\ell^4 - 2\ell^3 
    > \ell^4 
    \geq (1/\ell) \cdot \big| S|_G \big| 
    \geq f(S) - \sum_{j \in S} p_{j}.
    \end{align*}
This contradicts the assumption that $S$ is a demand set with respect to $\boldsymbol{p}$.
\end{proof}

Next, we show that a price vector is informative with exponentially small probability.

\begin{restatable}[Probability of Informativeness]{lemma}{probabilityOfInformativeness}\label{lem:exponentially_small}
Fix a price vector $\boldsymbol{p} \in \mathbb{R}^T$, and suppose that $G$ is sampled uniformly at random from $\binom{A}{\ell^2}$. Then the probability that $\boldsymbol{p}$ is informative w.r.t. $G$ is at most $\exp(-\ell)$.
\end{restatable}

\begin{proof}
If $|L(\boldsymbol{p})| \ge 2\ell^4$, then $\boldsymbol{p}$ is uninformative with respect to any $G$. We therefore assume that $|L(\boldsymbol{p})| < 2\ell^4$. Let $L_1, \ldots, L_{k}$ denote all $(\ell+1)$-element subsets of $L(\boldsymbol{p})$ where $k = \binom{|L(\boldsymbol{p})|}{\ell+1}$. Observe that
    \begin{align*}
        \prob[|L(\boldsymbol{p})\cap G| >\ell] = \prob[L_j \subseteq G \text{ for some } j] \leq \sum_{j=1}^k  \prob[L_j \subseteq G] = \binom{|L(\boldsymbol{p})|}{\ell+1} \cdot \prob[L_1 \subseteq G].
    \end{align*}
We first bound
    \begin{align*}
        \prob[L_1 \subseteq G] &= {\binom{\ell^7-(\ell+1)}{\ell^2-(\ell+1)}} \cdot {\binom{\ell^7}{\ell^2}}^{-1} = \left( \frac{(\ell^7-(\ell+1))!}{(\ell^2 - (\ell+1))! (\ell^7-\ell^2)!} \right) \cdot \left( \frac{(\ell^7)!}{(\ell^2)!(\ell^7-\ell^2)!}\right)^{-1} \\
        &= \frac{\ell^2 (\ell^2-1) \cdots (\ell^2-\ell)}{\ell^7 (\ell^7-1) \cdots (\ell^7-\ell)}         \leq \left( \frac{\ell^2}{\ell^7-\ell} \right)^{\ell+1}  \leq  \left(2 \ell^{-5} \right)^{\ell+1}.
    \end{align*}
Next, we bound
    \begin{align*}
        \binom{|L(\boldsymbol{p})|}{\ell+1} \leq \binom{2\ell^4}{\ell+1} \leq \left( \frac{2\ell^4e}{\ell+1} \right)^{\ell+1} \leq (2\ell^3 e)^{\ell+1}
    \end{align*}
where the second inequality follows from a standard binomial bound \cite{wiki:binomial_bounds}.
Combining these bounds yields:
    \begin{align*}
         \prob[|L(\boldsymbol{p})\cap G| >\ell] \leq (2\ell^3 e)^{\ell+1} (2\ell^{-5})^{\ell+1} \leq (4 \ell^{-2} e)^{\ell+1} \leq e^{-\ell}
    \end{align*}
    where the last inequality follows since $\ell \geq 5$. This completes the proof.
\end{proof}

We next establish a lower bound on the optimal value achievable under any contract.

\begin{restatable}[Lower Bound for Optimum]{lemma}{lowerBoundForOptimum}\label{lem:good_contract}
    For any parameter $G \subseteq A$ with $|G| = \ell^2$, the equal-pay contract $\contract = \eqcontract{(1/(2\ell^2))}{G}$ induces the equilibrium $T|_G$ and $(1-\sum_{i \in A} \alpha_i)f(T|_G) \ge (1/2) \cdot \ell^4$.
\end{restatable}

\begin{proof}
    Recall that $T|_G$ is the profile where each agent in $G$ performs her entire actions set and agents outside of $G$ do nothing.
    It holds that $f(T|_G) = 
    \max\left\{|A(T|_G)|, (1/\ell) \cdot \big| T|_G \big|, \ell^3\right\} = 
    \max\left\{ \ell^2, ({1}/{\ell}) \cdot \ell^2 \cdot \ell^3, \ell^3 \right\}= \ell^4$.
    
    We first argue that $T|_G$ is an equilibrium with respect to contract $\contract$. 
    Let $i \in \agents$. If $i \notin G$, then $\alpha_i = 0$, and the agent has no incentive to deviate to any non-empty set of actions.
    Now suppose $i \in G$. In this case,
    \begin{align*}
        u_i(\contract, T|_G) = \alpha_i \cdot f(T|_G) - c(T_i) = (1/(2\ell^2)) \cdot \ell^4 - \ell^3 \cdot (1/(2\ell^3)) = (1/2) \cdot \ell^2 - 1/2.
    \end{align*}
    Moreover, for any alternative action set $S_i' \subseteq T_i$, we have
    \begin{align*}
         u_i(\contract, S_i' \sqcup T|_{G \setminus \{i\}}) &= \alpha_i \cdot f( S_i' \sqcup T|_{G \setminus \{i\}}) - c(S'_i) \\
         &= (1/(2\ell^2)) \cdot \left(\ell^4 - (1/\ell) \cdot (\ell^3 -|S_i'|)\right)  - |S_i'| \cdot (1/(2\ell^3)) 
         = (1/2) \cdot \ell^2 - 1/2.
    \end{align*}
Thus, no agent can improve her utility by deviating, and $T|_G$ is indeed an equilibrium.    

    The principal's utility from the contract $\contract$ at the equilibrium $T|_G$ satisfies:
    \[
    \textstyle
    u_P(\contract, T|_G) = \left(1-\sum_{i=1}^n\alpha_i\right) \cdot f(T|_G)= (1 - \ell^2 \cdot (1/(2\ell^2))) \cdot f(T|_G) =  (1/2) \cdot \ell^4.
    \]
    This completes the proof.
\end{proof}

We now define the notion of aligned contracts.

\begin{definition}[Aligned Contract]\label{def:alignment}
We say that a contract $\contract$ is aligned with respect to $G \subseteq A$ if:
\begin{align*}
    \textstyle
     |\{ i\in G \mid \alpha_i \ge {1}/(2\ell^2) \}| > \ell \quad\text{ and }\quad \sum_{i \in A} \alpha_i \leq 1.
\end{align*}
\end{definition}

Any contract that fails to satisfy the condition above is called unaligned.

Next, we derive an upper bound on the value of any unaligned contract. Combined with the lower bound on the optimum from Lemma~\ref{lem:good_contract}, the following lemma implies that achieving a constant-factor approximation requires returning an aligned contract.

\begin{restatable}[Upper Bound for Unaligned Contracts]{lemma}{badContract}\label{lem:bad_contract}
    Let $G \subseteq A$ satisfy $|G| = \ell^2$. For any contract $\contract$ that is unaligned with respect to $G$ and for any equilibrium $S$, it holds that $f(S) \le 2\ell^3$.
\end{restatable}

\begin{proof}
    Let $\contract$ be a contract such that $|\{ i\in G \mid \alpha_i \ge {1}/(2\ell^2) \}| \le \ell$, and let $S$ be an equilibrium.
    Observe that for any action $j\in S$ with $j \in T_i$, it 
    must be that:
    \begin{align*}
        &u_i(\contract, S) \geq u_i(\contract, S \setminus \{j\}) \\
        \Longrightarrow \quad & \alpha_i \cdot (f(S) - f(S \setminus \{j\})) \geq c_{j} = 1/(2\ell^3) \\
        \Longrightarrow \quad & \alpha_i \geq 1/(2\ell^3)
    \end{align*}
    where the last implication follows from the fact that $f(S) - f(S \setminus \{j\}) \leq 1$ for any $j \in S$.
    Hence, by the assumption that $\sum_{i \in A} \alpha_i \leq 1$, we obtain that $|A(S)| \leq 2\ell^3$.
    
    Suppose that $f(S) > 2\ell^3$.
    Since $f(S) = \max \{ |A(S)|, (1/\ell) \cdot \big| S|_G \big|, \ell^3 \}$, and the first and third terms are at most $2\ell^3$, 
    it follows that $f(S) = (1/\ell) \cdot \big| S|_G \big|$. Moreover, for any action $j \in S$ with $j \in T_i$ and $i \in G$, we have $f(S \setminus \{j\}) = (1/\ell) \cdot (\big| S|_G \big| - 1)$, and thus:
        \begin{align*}
        &u_i(\contract, S) \geq u_i(\contract, S \setminus \{j\}) \\
        \Longrightarrow \quad & \alpha_i \cdot (f(S) - f(S \setminus \{j\})) \geq c_{j} = 1/(2\ell^3) \\
        \Longrightarrow \quad & \alpha_i \cdot (1/\ell) \geq 1/(2\ell^3)  \\
        \Longrightarrow \quad & \alpha_i \geq 1/(2\ell^2).
    \end{align*}
   Therefore, any agent $i \in G$ with $S_i \neq \emptyset$ satisfies $\alpha_i \ge 1/(2\ell^2)$.
By assumption, this implies that there are at most $\ell$ agents in $G$ with non-empty action sets in $S$, and hence:
    \[
    f(S) = (1/\ell) \cdot |S_G|  \leq (1/\ell) \cdot \ell \cdot \ell^3 = \ell^3
    \]
which yields a contradiction and completes the proof.
\end{proof}

We use the bound on the probability of making an informative demand query to derive a corresponding bound on the probability that a contract is aligned.

\begin{restatable}[Probability of Alignment]{lemma}{probabilityOfAlignment}\label{lem:alignment_prob}
    Fix a contract $\contract$, and suppose that $G$ is sampled uniformly at random from $\binom{A}{\ell^2}$. Then the probability that $\contract$ is aligned with respect to $G$ is $\exp(-\ell)$.
\end{restatable}

\begin{proof}
If $\sum_{i \in A} \alpha_i > 1$, then by the second condition of Definition~\ref{def:alignment} the contract $\contract$ is aligned with probability zero. We therefore assume that $\sum_{i \in A} \alpha_i \leq 1$.
Define a price vector $\boldsymbol{q} \in \mathbb{R}^T$ by setting
$q_j = (1/\ell)\cdot \indicator[\alpha_i \geq 1/(2\ell^2)] + 2 \cdot \indicator[\alpha_i < 1/(2\ell^2)]$
for every agent $i \in A$ and action $j \in T_i$.

We claim that for any $G \subseteq A$, if $\contract$ is aligned with respect to $G$, then $\boldsymbol{q}$ is informative with respect to $G$. Once this claim is established, it follows that the probability that $\contract$ is aligned is at most the probability that $\boldsymbol{q}$ is informative, which by Lemma~\ref{lem:exponentially_small} is at most $\exp(-\ell)$.

Fix an arbitrary $G \subseteq A$. By construction of $\boldsymbol{q}$,
    \begin{align*}
    L(\boldsymbol{q}) = \{i \in \agents\mid q_{j} \leq 1/\ell \text{  for some  } j \in T_i \} =  \{ i\in \agents\mid \alpha_i \ge {1}/(2\ell^2) \}.    
    \end{align*}
Since $\sum_{i \in A} \alpha_i \le 1$, we have $|L(\boldsymbol{q})| \le 2\ell^2 \le 2\ell^4$, and thus $\boldsymbol{q}$ satisfies the first condition of Definition~\ref{def:informative_query}. Moreover, the second condition of Definition~\ref{def:informative_query} holds because
    \begin{align*}
        |L(\boldsymbol{q}) \cap G| = |\{ i\in G \mid \alpha_i \ge {1}/(2\ell^2) \}| > \ell
    \end{align*}
where the inequality follows from the assumption that $\contract$ is aligned. This completes the proof.
\end{proof}

The lemmas above can be combined to yield the following lemma.

\begin{lemma}\label{lem:small_probability}
    Let $\alg$ be a deterministic algorithm which makes at most $M$ demand queries and no value queries. 
    When $G$ is sampled uniformly at random among subsets of $\agents$ of size $\ell^2$, the probability of $\alg$ achieving an $(\ell/4)$-approximation to to the optimal profit on instance $\mathcal{I}^{(G)}$ is at most $(M+1) \cdot \exp(-\ell)$.
\end{lemma}
\begin{proof}    
Let $\alg$ be a deterministic algorithm that makes at most $M$ demand queries, and let $\contract^{(G)}$ denote the contract returned by $\alg$ on instance $\mathcal{I}^{(G)}$.
    
By Lemma~\ref{lem:bad_contract,lem:good_contract}, the algorithm $\alg$ achieves an $(\ell/4)$-approximation on instance $\mathcal{I}^{(G)}$ only if the returned contract $\contract^{(G)}$ is aligned with respect to $G$.

Let $\contract'$ denote the contract returned by $\alg$ on instance $\mathcal{I}'$, and let $\boldsymbol{p}'_1,\ldots,\boldsymbol{p}'_k$ be the sequence of demand queries issued by $\alg$ on $\mathcal{I}'$. Since $\alg$ makes at most $M$ demand queries on any instance, we have $k \leq M$.

Fix a subset $G \subseteq A$. If all demand queries in $\boldsymbol{p}'_1,\ldots,\boldsymbol{p}'_k$ are uninformative with respect to $G$, then $\alg$ issues the same sequence of demand queries on instance $\mathcal{I}^{(G)}$, and hence returns  contract $\contract'$.
This follows directly from Lemma~\ref{lem:demand_independent}, as the responses to all uninformative demand queries are identical in $\mathcal{I}^{(G)}$ and $\mathcal{I}'$.

By the union bound and Lemma~\ref{lem:exponentially_small}, the probability that at least one of the demand queries in $\boldsymbol{p}'_1,\ldots,\boldsymbol{p}'_k$ is informative is at most $k \cdot \exp(-\ell)$. Moreover, by Lemma~\ref{lem:alignment_prob}, the probability that $\contract'$ is aligned with respect to $G$ is at most $\exp(-\ell)$. Applying the union bound once more, the probability that $\alg$ returns an aligned contract is at most $(k+1)\cdot \exp(-\ell)$, which completes the proof.
\end{proof}

We are now ready to prove Theorem~\ref{thm:XOS_inapprox}.

\begin{proof}[Proof of Theorem~\ref{thm:XOS_inapprox}]
First, note that $f$ assumes only integer values within $[\ell^7]$. Consequently, by a standard result of \cite[Lemma 11.22]{nisan2007algorithmic}, any value query can be simulated with a polynomial number of demand queries. Hence, without loss of generality, we may suppose that the algorithm uses only demand queries.
Let $M$ denote the number of such queries. Fix $n$ sufficiently large and set $\ell = n^{1/7}$. Consider an arbitrary deterministic algorithm and a random instance $\mathcal{I}^{(G)}$, where $G$ is sampled uniformly at random from $\binom{A}{\ell^2}$. By Lemma~\ref{lem:small_probability}, this deterministic algorithm achieves an $(\ell/4)$-approximation with probability at most
\begin{align*}
(M+1) \cdot \exp(-\ell) = (M+1) \cdot \exp(-n^{1/7}) < 1/(8\ell)    
\end{align*}
for sufficiently large $n$, since $M$ and $\ell$ are polynomial in $n$. 
By symmetry, all instances $\mathcal{I}^{(G)}$ share the same optimal value; denote this value by $\textsc{Opt}$.
It follows that:
\begin{align*}
     \mathbb{E}_{G \sim \textsc{Uniform} \binom{A}{\ell^2}}[\alg(\mathcal{I}^{(G)})] \leq \textsc{Opt} \cdot (1/(8\ell)) + (4/\ell) \cdot \textsc{Opt} \cdot (1-1/(8\ell)) \leq (8/\ell) \cdot \textsc{Opt}.
\end{align*}
By Yao's principle, for any randomized algorithm there exists an instance on which the expected approximation ratio is at least $\ell/8 = \Omega(n^{1/7})$.
\end{proof}

\subsection{No PTAS for Matroid Rank Functions}
\label{sec:hardness-matroid}

In this section, we establish that maximizing profit admits no PTAS unless \pequalsnp, even when the success probability is a matroid rank function, which is a special case of gross substitutes.

\begin{restatable}{theorem}{matroidHardness}\label{thm:matroid_hardness}
    When the success probability function $f$ is a normalized unweighted matroid rank function, it is NP-hard to find a $0.7$-approximation to the optimal profit over unconstrained contracts under combinatorial actions. Moreover, the same hardness result applies to equal-pay contracts.
\end{restatable}

The argument proceeds via a reduction from a promise variant of the maximum coverage problem, introduced
in \cite{ezra2023Inapproximability}, which generalizes the earlier hardness result of \cite{Feige98}.

\begin{proposition}[\cite{ezra2023Inapproximability}]\label{prop:coverage_hardness}
Let $k\in\mathbb{N}_{+}$, and let $(U,A,h)$ be a set system with universe $U$ and family of sets $h : A \to 2^U$ such that $|h(i)| = |U|/k$ for all $i \in A$.
Define the normalized unweighted coverage function $\tilde{f}:2^{A}\to[0,1]$ by $\tilde{f}(B)={\left|\bigcup_{i\in B} h(i)\right|}/{|U|}$.
Consider the following promise problem on input $(U,A,h,k)$, where exactly one of the two conditions below holds.
\begin{enumerate}
    \item \emph{The good case}: There exists a set $B\subseteq A$ with $|B|=k$ such that $\tilde{f}(B)=1$.
    \item \emph{The bad case}: For every set $B\subseteq A$ with $|B|\le 2k$ it holds that $\tilde{f}(B) \le 1 - e^{-|B|/k} + 0.01$.
\end{enumerate}
It is NP-hard to distinguish the good case from the bad case.
\end{proposition}

We now prove the main theorem of this section.

\begin{proof}[Proof of Theorem~\ref{thm:matroid_hardness}]
Let $(U, A, h, k)$ be an instance from Proposition~\ref{prop:coverage_hardness}.
We define a multi-agent combinatorial-actions instance $\multiInstance$ by setting $T_i = \{ (i, u) \mid u \in h(i) \}$.
The success probability for all $S \subseteq T = \bigcup_{i \in A} T_i$ is given by
$f(S) = |\bigcup_{(i,u) \in S} \{u\}| / |U|$, and each action $(i,u) \in T$ has cost $c_{(i,u)} = 1/(2k|U|)$.

Note that $f$ is a normalized unweighted matroid rank function, since $|U|\cdot f$ is the rank function of the partition matroid on ground set $T$ with parts 
$T(u)=\{(i,v) \in T \mid v=u\}$,
where independent sets contain at most one element from each part.

We analyze the optimal profit in the two cases from Proposition~\ref{prop:coverage_hardness}.

In the good case, there exists a set $B \subseteq A$ of size $k$ such that $\tilde{f}(B) = 1$.
By the definition of $\tilde{f}$, this implies that $\bigcup_{i \in B} h(i) = U$, and since $|h(i)| = |U|/k$ for all $i \in A$, it must be the case that $\{h(i)\}_{i \in B}$ are disjoint. 
Consider the contract $\alpha_i = 1/(2k)$
for each $i \in B$ and $\alpha_i = 0$ for each $i \notin B$.
Since agents outside of $B$ receive zero payment and all action costs are strictly positive, no agent outside of $B$ takes any action in any Nash equilibrium.
Fix any agent $i \in B$ and any action profile $Q = \bigsqcup_{j \in B} Q_j$.
By the fact that the sets $\{h(i)\}_{i \in B}$ are disjoint, we have that $f(Q) = | \bigcup_{(i,u) \in Q} \{u\}| / |U|  = |Q| / |U|$. Therefore:
\begin{align*}
\textstyle
 u_i(\contract, Q) =  \alpha_i \cdot f(Q) - \sum_{j \in Q_i} c_j
  &= (1/(2k)) \cdot (|Q| / |U|) - |Q_i| \cdot (1/(2k|U|)) \\
  &= (1/(2k|U|)) \cdot (|Q_i| + |Q_{-i}|) - |Q_i| \cdot (1/(2k|U|)) \\
  &= (1/(2k)) \cdot (|Q_{-i}| / |U|).
\end{align*}
Hence, agent $i$'s utility is independent of its chosen set of actions $T_i$. 
Thus, the action profile $T|_B$, where each agent $i \in B$ takes all actions in $T_i$, is an equilibrium that satisfies:
\[
    u_P(\contract, T|_B) = (1 - k \cdot (1/(2k))) \cdot f(T|_B) = (1 - k \cdot (1/(2k))) \cdot \tilde{f} (B) = 0.5,
\]
where $\tilde{f}$ is defined as in Proposition~\ref{prop:coverage_hardness}.

In the bad case, observe that incentivizing any agent to take a non-empty set of actions requires a payment of at least $1/(2k)$. Indeed, for any $\alpha_i < 1/(2k)$ and any action profile $Q = \bigsqcup_{j \in A} Q_j$ with $Q_i \neq \emptyset$, we have
\begin{align*}
\textstyle
    u_i(\contract, Q) = \alpha_i \cdot f(Q) - \sum_{j \in Q_i} c_j &= \alpha_i \cdot \left( f(Q) - f(Q_{-i}) \right) + \alpha_i \cdot f(Q_{-i}) - |Q_i| \cdot (1/(2k|U|))  \\
    &\leq \alpha_i \cdot (|Q_i| / |U|) + \alpha_i \cdot f(Q_{-i}) - |Q_i| \cdot (1/(2k|U|))  \\
    &< \alpha_i \cdot f(Q_{-i}).
\end{align*}
Therefore, agent $i$ strictly prefers deviating to the empty action set, which proves the claim.
Thus, any action profile $Q$ with $|A(Q)| \geq 2k$ yields non-positive principal's utility, since the total payment required to incentivize agents in $A(Q)$ is $|A(Q)| \cdot (1/(2k)) \geq 1$. Moreover, for any action profile $Q$ with $|A(Q)| < 2k$ induced by contract $\contract$, the assumption in case (2) of Proposition~\ref{prop:coverage_hardness} implies
\begin{align*}
    u_P(\contract, Q) \leq (1 - |A(Q)| \cdot (1/(2k))) \cdot f(Q) &\leq (1 - |A(Q)| \cdot (1/(2k))) \cdot \tilde{f}(A(Q)) \\
    &< (1-|A(Q)| \cdot (1/(2k))) \cdot (1-e^{-|A(Q)|/k} + 0.01) \\
    &< 0.35,
\end{align*}
where the third inequality is since $(1 - x/2)(1 - e^{-x} + 0.01) < 0.35$ for all $x$.

Since $0.35/0.5 = 0.7$, the existence of a $0.7$-approximation algorithm for maximizing profit under matroid rank success probabilities would allow one to distinguish between the good and bad cases. Therefore, finding a $0.7$-approximation is NP-hard by Proposition~\ref{prop:coverage_hardness}.
\end{proof}

\section{\Price }\label{sec:pof_xos}

In this section we quantify the worst-case loss to the principal's utility by restricting to an equal-pay contract.
In Section~\ref{subsec:pof_lower} we present a $\Omega\left({\log n} / {\log \log n}\right)$ lower bound on the price of equality for additive $f$, where each agent can either work or shirk (binary actions). 
This bound is tight even for instances with XOS $f$ and combinatorial actions, as we show in Section~\ref{subsec:pof_xos}.
In Section~\ref{subsec:pof_sa} we show a separation between instances with XOS and subadditive $f$ and give a lower bound of $\Omega(\sqrt{n})$ on the \price\ for the latter, which holds even in the binary-actions case. 
We begin by formally defining \price.

\begin{definition}[\Price ]\label{def:pof}
    Given an instance $\I = \multiInstance$, the \price\ is the ratio between the maximal principal's utility from an unconstrained contract, and the maximal principal's utility under an equal-pay contract. The measure is defined with respect to the best equilibrium the contracts induce.
    When considering a family of instances $\F$ (e.g., all instances with additive $f$), the \price\ is taken with respect to the worst-case instance in the family, i.e.,  
    {
    \[
    \POE_\I = \frac{\max_{\contract, S \in \nash(\contract)} (1-\sum_{i \in \agents}\alpha_i)f(S)}
    {\max_{\text{equal-pay }\contract, S \in \nash(\contract)} (1-\sum_{i \in \agents}\alpha_i)f(S)}
    \ \ \ \  \mbox{and}
 \ \ \ \ 
    \POE_\F = \max_{\I \in \F} \POE_\I.
    \]
    }
\end{definition}

\subsection{Lower Bound of $\log (n) / \log \log (n)$ on Price of Equality for Additive $f$}\label{subsec:pof_lower}

We provide the following lower bound on the price of equality for additive rewards.

\begin{restatable}{proposition}{pofAdd}\label{prop:pof_add}
There exists a binary-action instance with additive $f$ such that the \price\ is $\Omega\left({\log n}/{\log \log n}\right)$.
\end{restatable}

\begin{proof}
    Consider an instance where each agent $i \in \agents$ has a single non-shirking actions $S_i = \{i\}$.
    Let $f_i = \frac{1}{i}$ and $c_i =\frac{1}{2i^2H_n}$, where $H_n$ denotes the $n$th harmonic number. 

    We first show that the optimal unconstrained contract has an equilibrium which yield the principal a utility of at least $\frac12 \cdot H_n$. 
    In particular, consider the contract that pays $\alpha_i = 1/(2iH_n)$ to agent $i$, for a total payment of $\sum_{i=1}^n 1/(2iH_n)=H_n/(2H_n)=\frac{1}{2}$.
    It is easy to observe that this contract induces the Nash equilibrium $T = \{1,\dots,n\}$.
    Let $(\contract^\star,S^\star)$ be the optimal contract and equilibrium. It holds that
    \[
        u_P(\constar,\eqstar)
        \ \ge \ u_P(\contract,\{1,\dots,n\})  
        \ = \ \left(1- \sum_{i \in \agents} \frac{1}{2iH_n} \right)f(\{1,\dots,n\}) 
        \ = \ \frac12 \cdot H_n,
    \]
    where the first inequality follows by the optimality of $(\constar,\eqstar)$.

    We next show that the optimal equal-pay contract yields an $f$ value of at most  $\ln (3 \ln n) + o(1)$.
    Observe that agent $i$ is incentivized to take action $i$ if and only if $\alpha_i \ge \frac{c_i}{f_i} = \frac{1}{2iH_n}$.

    Consider an equal-pay contract $\contract$, and let $S$ be an equilibrium induced by $\contract$, and note that 
    it is without loss of generality to assume that for any $i \in \agents$ such that $i \notin S$, we have $\alpha_i = 0$.
    
    If $\contract$ is the zero contract, then $S=\emptyset$ is the unique equilibrium induced by $\contract$, and we have $f(\emptyset)=0$. 

    Otherwise, let $j \in \agents$ 
    be the agent with the maximal non-zero $\alpha_j$, breaking ties in favor of the smallest index. 
    It is without loss of generality to assume that $\alpha_j = \frac{1}{2jH_n}$ and that all agents $i<j$ have $\alpha_{i}=0$.
    As $\contract$ is an equal-pay contract, any other agent $i \ge j$ such that $i \in S$ must have $\alpha_i = \alpha_j = \frac{1}{2jH_n}$.
    Since the total payment cannot exceed $1$, the number of agents who choose the non-shirking action is at most $2jH_n$.
    Thus, for the equilibrium $S \in \nash (\contract)$ it holds that
        \begin{align*}
        f(S)
        &=
        \sum_{i \in S} \frac{1}{i} 
        \le
        \sum_{i=j}^{j+2jH_n} \frac{1}{i} 
        =
        H_{j+2jH_n} - H_j 
        =
        \ln({j+2jH_n}) - \ln(j) + o(1) \\
        &=
        \ln({2H_n+1}) + o(1) 
        \le
        \ln (3 \ln n) + o(1),
    \end{align*}

    We conclude that 
    \[
    \POE_{\text{additive }f} = \frac{u_P(\contract^\star,S^\star)}{u_P(\contract,S) } \ge \frac{H_n}{2\ln (3 \ln n) + o(1)} = \Omega\left(\frac{\ln n}{\ln ( \ln n)}\right)
    \]
    which completes the proof.
\end{proof}

\subsection{Upper Bound of $\log (n) / \log \log (n)$ on Price of Equality for XOS $f$}\label{subsec:pof_xos}

\newcommand{\eqprice}[1]{p(#1)}

We provide the following upper bound on the price of equality with XOS rewards.

\begin{restatable} {theorem}{pofxos}\label{thm:pof_xos}
    For any combinatorial-actions instance with XOS $f$, 
    any contract $\contract^\star$, and any equilibrium $S^\star \in \nash(\contract^\star)$, 
    there exists an equal-pay contract $\contract$ and $S \in \nash(\contract)$ such that 
    \begin{align*}
        \textstyle
       \left(1-\sum_{i \in \agents}\alpha_i\right)f(S) \ge 
    \Omega\left(\frac{\log \log n}{\log n}\right)
    \left(1-\sum_{i \in \agents} \alpstar_i\right)f(\eqstar). 
    \end{align*}
\end{restatable}

To prove Theorem~\ref{thm:pof_xos}, we propose a two-stage bucketing procedure described in Algorithm~\ref{alg:pof_xos}.
First, we examine whether a single large agent (whose payment under $\constar$ is greater that $1/2$) can achieve a good approximation by itself. 
Otherwise, we find a good subset of the other (small) agents via a bucketing procedure and the scaling-for-existence lemma (Lemma~\ref{lem:scaling_for_exist}).
In Section~\ref{subsec:pof_xos_large_agent} we analyze the large agent case, which uses tools from \cite{multimulti}. 
In Section~\ref{subsec:pof_xos_small_agents} we analyze our bucketing procedure and conclude with the correctness of Theorem~\ref{thm:pof_xos}.

\begin{algorithm}[h]
\caption{An $O(\log n/ \log \log n)$-approximation to profit via an equal-pay contract for XOS $f$}\label{alg:pof_xos}
\KwIn{A contract $\constar$ with $\sum_i \alpstar_i <1$, a profile $\eqstar \in \nash(\constar)$, costs $c_1,\dots,c_m \ge 0$, value oracle access to an XOS set function $f:2^{[m]}\rightarrow \reals_{\ge 0}$}
\KwOut{An equal-pay contract $\contract$ for which there exists a profile $S \in \nash(\contract)$ 
such that $(1-\sum_i \alpha_i)f(S) \ge 
\Omega({\log \log n}/{\log n})
(1-\sum_i \alpstar_i)f(\eqstar)$}
let $N \gets A(\eqstar)$ 
\tcp*[l]{\textbf{Ignore all agents that shirk in $\eqstar$}} 

let $Z \gets \{i \in N \mid \alpstar_i > 1/2\}$ \tcp*[l]{\textbf{Large agent ($Z$ can be of size 0 or 1)}} \label{line:pof_large_def}
\If{$\exists z \in Z \text{ s.t. }(1-\alpstar_z)f(\eqstar_z) \ge 4 f(\eqstar_{-z})$}{\label{line:pof_if_large}
$\alpha_z \gets ({\alpstar_z+1})/{2}$\;
For any $i \ne z$, $\alpha_i \gets 0$\;
\Return $\contract$\;\label{line:pof_return_large}
}

$N \gets N \setminus Z$, rename remaining agents so that $\alpstar_1 \ge \alpstar_2 \ge \dots$ \tcp*[l]{\textbf{Small agents}}
$X \gets \{i \in N \mid \alpstar_i < 1/(2n)\}$, \,\, $k \gets |N \setminus X|$\;
$h \gets 0$, \,\, $b_0 \gets 0$\;
\While{$b_h + 1 \leq k$}{
     $h \gets h+1$\;
    $a_h \gets b_{h-1} +1$, \,\, $b_h \gets \min\{a_h + \floor{{1}/({2\alpstar_{a_h}})},k+1\}$\;
     $B_h \gets \{a_h, \dots, b_h-1\}$,  \,\, $B'_h \gets \{b_h\}$ \tcp*[l]{if $b_h=k+1$, we set $B'_h \gets \emptyset$}}
$\B \gets \{X,B_1,\dots,B_h,B'_1,\dots B'_h\}$\; \label{line:pof_all_buckets}
$B \gets \argmax_{Q \in \B} f(\eqstarset{Q})$ \tcp*[l]{\textbf{$\eqstarset{Q}$ are the actions in $\eqstar$ performed by agent in $Q$}}\label{line:best_bucket}
\label{line:pof_bucket}
$p_B \gets \max\{\alpstar_i \mid i \in B\}$\; \label{line:pof_bucket_price}

\Return $\eqcontract{(3/2) \cdot p_B}{B} $\; \label{line:pof_return_bucket}
\end{algorithm}

\subsubsection{Analyzing the Large Agent Case}\label{subsec:pof_xos_large_agent}
The analysis of the large agent case resembles the analysis in \cite{multimulti}.
Specifically, we use the following lemma:
\begin{lemma}[Case 1 in Lemma 5.2 in \cite{multimulti}]\label{lem:sa_big_agent}
    Consider a subadditive $f$ and a contract $\constar$ with $\sum_{i} \alpstar_i <1$. 
    Let $S^\star\in \nash(\constar)$ be an equilibrium such that exists $i \in \agents$ with $(1-\alpstar_i)f(\eqstar_i)\ge 4f(\eqstar_{-i})$. 
    The contract $\alpha_i = \frac{1+\alpha_i^*}{2}$ and $\alpha_j=0$ for any other $j$ has an equilibrium $S$ such that $S_j=\emptyset$ for any $j \ne i$ and $f(S_i) \ge \frac12 f(\eqstar_i)$.
\end{lemma}

Through a similar analysis to Theorem 5.2 in \cite{multimulti}, we prove the following guarantees for the large agent case.

\begin{restatable}{lemma}{pofLargeAgentApprox}\label{lem:pof_large_agent_approx}
    Consider any subadditive $f$, any $\constar$ such that $\sum_i \alpstar_i<1$ and any $\eqstar \in \nash(\constar)$.
    If Algorithm~\ref{alg:pof_xos} returns a contract $\contract$ in line \ref{line:pof_return_large}, it is equal-pay, and there exists an equilibrium $S \in \nash(\contract)$ such that  $(1-\sum_{i=1}^n\alpha_i)f(S)\ge(2/9)(1-\sum_{i=1}^n\alpstar_i)f(\eqstar)$.
\end{restatable}
\begin{proof}
    Let $z \in \agents$ be the agent for which the condition in line \ref{line:pof_if_large} of Algorithm~\ref{alg:pof_xos} is met.
    In line \ref{line:pof_return_large}, the algorithm returns the contract $\alpha_z=\frac{1+\alpstar_z}{2}$ and $\alpha_i=0$ for any other $i \ne z$.
    Observe that for any $i \ne z$, $S_i = \emptyset$ is a best-response regardless of others' actions.
    By Lemma~\ref{lem:sa_big_agent}, $f(S_z) \ge f(\eqstar_z)/2$, which implies that
    \begin{equation}\label{eq:sa_big_agent}
    (1-\alpha_z)f(S_z)\ =\ \frac12(1-\alpstar_z)f(S_z)\ \ge\ \frac14(1-\alpstar_z)f(\eqstar_z)     
    \end{equation}

    Because agent $z$ meets the conditions in line \ref{line:pof_if_large}, it holds that $\alpstar_z>1/2$, and $(1-\alpstar_z)f(\eqstar_z)\ge4f(\eqstar_{-z})$. We get,
    \begin{align*}
        \frac12 f(\eqstar_z) \ >\
        (1-\alpstar_z)f(\eqstar_z) \ \ge\ 4 f(\eqstar_{-z})\ \ge\ 4 (f(\eqstar)-f(\eqstar_z)),
    \end{align*}
    where the last inequality is by subadditivity. Rearranging, we get
    that $f(\eqstar_z) > \frac{8}{9}f(\eqstar)$.
    Together with Equation~\ref{eq:sa_big_agent}, we get:
    \[
    (1-\alpha_z)f(S_z)\ \ge\ \frac14(1-\alpstar_z)f(\eqstar_z)\ \ge\ \frac14\left(1-\sum_{i \in \agents} \alpstar_i\right)f(\eqstar_z)
    \ \ge\ \frac{2}{9}\left(1-\sum_{i \in \agents}\alpstar_i\right)f(\eqstar),
    \]
   as needed.
\end{proof}

\subsubsection{Analyzing the No Large Agent Case}\label{subsec:pof_xos_small_agents}

We begin the analysis of the no large agent case by a bound on the total payment returned in this case.

\begin{lemma}\label{lem:pof_small_agents_payment}
    The contract $\contract$ returned in line \ref{line:pof_return_bucket} of Algorithm~\ref{alg:pof_xos} is equal-pay, and $\sum_{i \in \agents} \alpha_i \leq 3/4$.
\end{lemma}
\begin{proof}
    Let $\contract$ be the equal-pay contract returned in line \ref{line:pof_return_bucket}. 
    Consider a possible set of agents (bucket) returned in line \ref{line:best_bucket}, i.e., some $B \in \B$ defined in line \ref{line:pof_all_buckets}.
    Let $p_B = \max_{i \in B}\{\alpstar_i\}$, as in line \ref{line:pof_bucket_price}.
    Since $\contract$ pays $(3/2)p_B$ to every agent in $B$ and 0 to all others, it suffices to show that $|B|p_B \le 1/2$.

    First, if $B=X$, the statement clearly holds, as by definition $|X|p_X < n \cdot \frac{1}{2n} = 1/2$.
    Second, observe that in this point of the algorithm we only consider small agents, those with $\alpstar_i \le 1/2$.
    Thus, if $|B|=1$, the claim holds as well.
    Lastly, a set $B$ with more than one agent is of the form $B = \{a_j,\dots,b_j-1\}$, for some $j \in [h]$. Since agents are ordered we have $p_B = \alpstar_{a_j}$. Observe that by definition of $b_{j}$ it holds that
    $p_B\cdot |B| 
    = p_B(b_j-a_j) 
    \le 
    p_B(\floor{1/(2p_B)} + a_j -a_j) 
    \le 1/2$.
\end{proof}

Next, we bound the maximum possible number of buckets.

\begin{lemma}\label{lem:pof_num_buckets}
    The number of buckets in line \ref{line:pof_all_buckets} is $2h+1 \le 33 \cdot {\ln n}/{\ln \ln n}$.
\end{lemma}
\begin{proof}
    Let $h$ be the total number of buckets in line \ref{line:pof_all_buckets}, denote $h'=h-1$.
    For any $j \in [h']$ denote $C_j = B_j \cup B'_j = \{a_j,\dots,b_j\}$. 
    Observe that by construction for any $j \in [h']$, 
    it holds that $b_j = a_j + \lfloor 1/(2\alpha^\star_{a_j}) \rfloor$ and so
    $|C_j| \cdot \alpstar_{a_j} = (b_j - a_j + 1) \cdot \alpstar_{a_j} = (\lfloor 1/(2\alpha^\star_{a_j}) \rfloor + 1) \cdot \alpstar_{a_j}  \ge 1/2$, and thus 
    \begin{align*}
        1/2 
        \le \alpstar_{a_j} \cdot |C_j|
        \le 
         \alpstar_{a_j} \cdot \sum_{i \in C_j} 
        \frac{\alpstar_i}{\alpstar_{b_j}}
        \le \frac{\alpstar_{a_j}}{\alpstar_{a_{j+1}}} \cdot \sum_{i\in C_j} \alpstar_i,
    \end{align*}
    where the second inequality follows from the ordering of the agents by payment.
    Rearranging,
    \begin{equation}\label{eq:buckets_payment_ratio}
     \textstyle \frac{\alpstar_{a_{j+1}}}{\alpstar_{a_j}} \leq 2 \cdot \sum_{i\in C_j} \alpstar_i   
    \end{equation}
    We consider the (telescopic) product of the left-hand side of Equation~\ref{eq:buckets_payment_ratio}, for any $j \in [h']$. Using the fact that small agents receive payment at most $1/2$, we get
     \begin{align*}
        \textstyle \prod_{j=1}^{h'} \frac{\alpstar_{a_{j+1}}}{\alpstar_{a_j}} = \frac{\alpstar_{a_{h'+1}}}{\alpstar_{a_1}} \ge \frac{1/2n}{1/2} = \frac{1}{n}.
    \end{align*}
    The product of the right-hand side of Equation~\ref{eq:buckets_payment_ratio} can be upper bounded using the AM-GM inequality, and the assumption that $\sum_{i \in \agents} \alpstar_i \leq 1$, we get:
    \begin{align*}
        \textstyle \prod_{j=1}^{h'} \left(  2 \cdot \sum_{i\in C_j} \alpstar_i \right) \leq \left( \frac{1}{h'} \cdot \sum_{j=1}^{h'} \left(2 \cdot \sum_{i \in C_j} \alpstar_i \right) \right)^{h'}  \leq \left( \frac{2}{h'} \cdot \sum_{i \in \agents} \alpstar_i \right)^{h'} \leq \left( \frac{2}{h'} \right)^{h'}.
    \end{align*}
    Combining these inequalities, we obtain $1/n \leq (2/h')^{h'}$, which implies $n \geq (h'/2)^{h'}$, and therefore $\ln n \geq h' \ln (h'/2)$.
    Since the function $x / \ln x$ is increasing for $x \in [e, \infty)$, it follows that:
    \begin{align*}
        \frac{\ln n}{\ln \ln n} &\ge \frac{h' \ln (h'/2)}{\ln(h'\ln (h'/2))} = \frac{h' \ln (h'/2)}{\ln h' + \ln \ln (h'/2)} \geq \frac{h' \ln (h'/2)}{3\ln (h'/2)} = \frac{h'}{3},
    \end{align*}
    where the last inequality holds for $h' \ge 2e^2 \approx 14.7$ 
We conclude that $h = h'+1\leq \max\{ {3 \ln n}/{\ln \ln n}+1, 16\} \le {16 \ln n}/{\ln \ln n}$.
\end{proof}

We combine the lemmas above to obtain the following  guarantee in the no large agent case.

\begin{lemma}\label{lem:pof_small_agents_approx}
Let $\contract$ be the contract returned in line \ref{line:pof_return_bucket}, there exists an equilibrium $S \in \nash(\contract)$ such that $\left(1-\sum_{j \in \agents} \alpha_j \right)f(S) \ge ({\ln \ln n}/({1980\ln n}))(1-\sum_{j \in \agents} \alpstar_j )f(\eqstar)$.
\end{lemma}

\begin{proof}
    Let $B \subseteq \agents$ be the subset of agents picked in line \ref{line:pof_bucket}, and let $p_B$ be the price of the most expensive agent in $B$, as in line \ref{line:pof_bucket_price}.
Denote by $\eqstarset{B}$ the actions taken by agents of $B$ in the profile $\eqstar$, namely $\eqstarset{B} = \eqstar\cap (\bigcup_{i \in B}T_i)$.

Since $\eqstar$, the input to Algorithm~\ref{alg:pof_xos}, is a Nash equilibrium with respect to $\constar$, it is also dropout stable.
Let $\contract^\dagger$ denote a contract defined by $\contract^\dagger_j = \contract^\star_j$ for every $j \notin B$ and $\contract^\dagger_j = p_B = \max_{i \in B} \contract^\star_i$ for each $j \in B$. Because $S^\star$ is dropout stable with respect to $\contract^\star$, it remains dropout stable under $\contract^\dagger$.
Thus, applying Lemma~\ref{lem:scaling_for_exist} to $(\contract^\dagger,\eqstar)$ with $B \subseteq \agents$ and $\gamma=3/2$, yields exactly the contract $\contract$ in line \ref{line:pof_return_bucket}. 
It is guaranteed that there exists an equilibrium $S \in \nash(\contract)$ such that $f(S) \ge \frac13f(\eqstarset{B})$.

Since $B$ is the bucket with the maximal value of $f$, by subadditivity of $f$, $f(\eqstarset{B}) \ge \frac{1}{2h+1}f(\eqstar\setminus \eqstarset{Z}) \ge \frac{\ln \ln n}{33\ln n}f(\eqstar\setminus \eqstarset{Z})$ where $\eqstarset{Z}$ is the action taken by the agents in $Z$ in the profile $\eqstar$.

Lastly, observe that $Z$ is either a singleton or an empty set. 
If $Z = \{z\}$, then, because the condition in line \ref{line:pof_if_large} is not met, we have $(1-\alpstar_z)f(\eqstar_z) < 4f(\eqstar_{-z})$, and so
\begin{align*}
\textstyle
    \left(1-\sum_{i \in \agents} \alpstar_i \right)f(\eqstar)
    &\le
    \textstyle
    \left(1-\sum_{i \in \agents} \alpstar_i \right)f(\eqstar_z) + \left(1-\sum_{i \in \agents} \alpstar_i \right)f(\eqstar_{-z}) && (\text{subadditivity of } f)\\
    &\le
    \left(1- \alpstar_z \right)f(\eqstar_z) + f(\eqstar_{-z}) && (\text{non-negativity of } \contract)\\
    &<
    5f(\eqstar_{-z}) = 5f(\eqstar \setminus \eqstarset{Z}).
\end{align*}
If $Z = \emptyset$, then clearly $\left(1-\sum_{i \in \agents} \alpstar_i \right)f(\eqstar) \le f(S^\star) \le 5f(\eqstar \setminus \eqstarset{Z})$.
Overall we get,
\[
\textstyle
f(S) \ge (1/3) f(\eqstarset{B}) 
\ge ({\ln \ln n}/({99\ln n}))f(\eqstar \setminus \eqstarset{Z}) 
\ge ({\ln \ln n}/({495\ln n}))(1-\sum_{j \in \agents} \alpstar_j )f(\eqstar).
\]
By Lemma~\ref{lem:pof_small_agents_payment}, $\sum_i \alpha_i \le \frac{3}{4}$, and $\left(1-\sum_{j \in \agents} \alpha_j \right)f(S) \ge \frac14 f(S) \ge \frac{\ln \ln n}{1980\ln n}(1-\sum_{j \in \agents} \alpstar_j )f(\eqstar)$.
\end{proof}

We are now ready to prove Theorem~\ref{thm:pof_xos}.

\begin{proof}[Proof of Theorem~\ref{thm:pof_xos}]
    Let $\constar$ be any contract. If $\sum_i \alpstar_i \ge 1$, the principal's utility is zero, and the all-zero contract satisfies the claim. Otherwise, we can run Algorithm~\ref{alg:pof_xos}. Let $\contract$ be the contract returned in line \ref{line:pof_return_bucket}. 
    
    If Algorithm~\ref{alg:pof_xos} returns a contract in line \ref{line:pof_return_bucket}, then by Lemma~\ref{lem:pof_small_agents_approx}, there exists $S \in \nash(\contract)$ such that
    $$\textstyle \left(1-\sum_{j \in \agents} \alpha_j \right)f(S) \ge \left(({\ln \ln n})/({1980\ln n})\right) \cdot \left(1-\sum_{j \in \agents} \alpstar_j \right)f(\eqstar).$$
    Otherwise, Algorithm~\ref{alg:pof_xos} returns a contract in line \ref{line:pof_return_large}, then by Lemma~\ref{lem:pof_large_agent_approx}, we have $$\textstyle \left(1-\sum_{j \in \agents} \alpha_j \right)f(S) \ge ({2}/9) \left(1-\sum_{j \in \agents} \alpstar_j \right)f(\eqstar).$$
    This concludes the proof.
\end{proof}

\subsection{Lower bound of $\Omega(\sqrt{n})$ on \Price\ for Subadditive $f$}\label{subsec:pof_sa}

We provide the following lower bound on the \price\ with subadditive rewards.

\begin{proposition}[Lower Bound on Price of Equality for Subadditive $f$]\label{prop:pof_subadditive}
There exists an instance with subadditive $f$, for which the \price\ is $\Omega(\sqrt{n})$
\end{proposition}

\begin{proof}
Assume without loss of generality that $n \geq 16$.
We will show that there exists an (unconstrained) contract $\constar$ and equilibrium $S^\star$ which satisfy
$u_P(\contract^\star, S^\star) \ge \frac{1}{8}$,
and that for any equal-pay contract $\contract$ with $\sum_{i=1}^n\alpha_i \le 1$ and equilibrium $S$ it holds that 
$u_P(\contract, S) \leq f(S) \leq 2/\sqrt{n}$.

   Consider an instance with $n$ agents, each has one non-shirking action. 
   The cost of the agents $[n-1]=\{1,\dots,n-1\}$ is $c_i = 1 / (2\cdot n\cdot \sqrt{n})$, and the cost of the last agent is $c_i=1 / (4 \sqrt{n})$. The reward function is defined as: 
    \begin{align*}
        f(S) = \begin{cases}
             1/\sqrt{n} + |S|/n & \text{if }  |S| \leq n-1\\
             2/\sqrt{n} + \frac{n-1}{n} & \text{if } |S| = n
        \end{cases}
    \end{align*}
    
    Let us first argue that \( f \) is subadditive. We need to verify that for \( S_1, S_2 \subseteq \agents \), we have \( f(S_1 \cup S_2) \leq f(S_1) + f(S_2) \). Note that \( f(S) \leq 2/\sqrt{n} + |S|/n \) for every \( S \subseteq \agents \).  
    If \( |S_1| = n \) or \( |S_2| = n \), then we have \( f(S_1 \cup S_2) = f(S_1) \) or \(  f(S_1 \cup S_2) = f(S_2)\), respectively.
    If \( |S_1| \leq n-1 \) and \( |S_2| \leq n-1 \), then
    \begin{align*}
        f(S_1 \cup S_2) \leq 2/\sqrt{n} + |S_1 \cup S_2|/n \leq (1/\sqrt{n} + |S_1|/n) + (1/\sqrt{n} + |S_2|/n) = f(S_1) + f(S_2).
    \end{align*}
 Therefore, $f$ is subadditive.

 Consider the (unconstrained) contract $\constar$ defined by $\alpha^\star_i = 1/(2n)$ for any $i \in [n-1]$, and $\alpha^\star_n = 1/4$.
 Observe that $S^\star = [n]$ is an equilibrium of $\constar$, since for every $i\in S^\star$ we have $\alpha_i^\star f(i\mid S^\star\setminus \{i\}) = c_i$. Additionally, 
 \[
 u_P(\constar,\eqstar)
 = \left( 1 - \sum_{i \in [n]} \alpha^\star_i \right) f(S^\star) =  \left(1-\frac{1}{2n}\cdot (n-1) - 1/4\right) \left( \frac{2}{\sqrt{n}} + \frac{n-1}{n} \right) \ge \frac{1}{8}.
 \]
 
On the other hand, let $\contract$ be an equal-pay contract such that $\sum_{i=1}^n \alpha_i \le 1$ and let $S \in \nash(\contract)$.

If $S \subseteq [n-1]$, then for every $i \in S$ we have $f_S(i) = 1/n$, and $c_i/f_S(i) = 1/\sqrt{n}$, which is the minimal payment required to incentivize $i$.
Thus, for the principal's utility to be non-negative it must be that $|S|\le \sqrt{n}$ and the reward from this equilibrium is $f(S) \le 2/\sqrt{n}$.

Otherwise, agent $n$ exerts effort, and his payment is at least
$\frac{c_n}{f_S(n)} \ge \frac{1}{4\sqrt{n}}\cdot \sqrt{n} = 1/4$.
which implies that there are at most $4$ agents, as each of them must be paid at least $1/4$. 
Thus, the total reward generated by $S$ is $f(S) \le 1/\sqrt{n} + 4/n \le 2/\sqrt{n}$, where the last inequality holds since $n \ge 16$.
We conclude that 
$$\POE \ge \frac{u_P(\constar,\eqstar)}{u_P(\contract,S)} \ge \frac{u_P(\constar,\eqstar)}{f(S)} \ge \frac{\sqrt{n}}{16}$$
which completes the proof.
\end{proof}

\bibliographystyle{alpha}
\bibliography{refs}

\appendix
\section{Reduction from $\gamma$-Equal-Pay to Equal-Pay}

\begin{definition}[$\gamma$-Equal-Pay Contract]
    Let $\gamma \ge 1$. A contract $\noindexcontract$ is $\gamma$-equal-pay contract if for any two non-zero payments $\alpha_i \ge \alpha_j$, it holds that $\alpha_i/\alpha_j \le \gamma$.
\end{definition}

\begin{theorem}\label{thm:gamma_reduction}
    When $f$ is XOS, for any $\gamma$-equal-pay contract $(\alpha^\star, S^\star)$, Algorithm~\ref{alg:gamma_xos} runs in polynomial time with value oracle access to $f$ and outputs an equal-pay contract $\alpha$ such that there exists an equilibrium $S$ which satisfies
    \[
    \left(1-\sum_{i=1}^n \alpha_i\right) f(S) \ge \Omega\left(\frac{1}{\gamma}\right) \left(1-\sum_{i=1}^n \alpha^\star_i\right) f(S^\star).
    \]
\end{theorem}

\begin{algorithm}[h]
\caption{An $O(\gamma)$-approximation to $\gamma$-equal-pay via an equal-pay contract for XOS $f$}\label{alg:gamma_xos}
\KwIn{A $\gamma$-equal-pay contract $\constar$ with $\sum_i \alpstar_i <1$, a profile $\eqstar \in \nash(\constar)$, costs $c_1,\dots,c_m \ge 0$, value  oracle access to an XOS set function $f:2^{[m]}\rightarrow \reals_{\ge 0}$}
\KwOut{An equal-pay contract $\contract$ for which there exists a profile $S \in \nash(\contract)$ 
such that $(1-\sum_i \alpha_i)f(S) \ge \Omega\left(\frac{1}{\gamma}\right)
(1-\sum_i \alpstar_i)f(\eqstar)$}
let $N \gets A(\eqstar)$ \tcp*[l]{ignore all other agents, i.e., those that shirk in $\eqstar$} 
let $Z \gets \{i \in N \mid \alpstar_i > 1/2\}$ \tcp*[l]{\textbf{Large agent (if any)}}
\If{$\exists i \in Z \text{ s.t. }(1-\alpstar_i)f(\eqstar_i) \ge 4 f(\eqstar_{-i})$}{
$\alpha_i \gets \frac{\alpstar_i+1}{2}$\;
For any $j \ne i$, $\alpha_j \gets 0$\;
\Return $\contract$\;\label{line:gamma_huge_agent} 
}
$C\gets \{i\in \agents\mid \alpha^\star_i = 0\}$\;
\If{$f(S^\star|_C) \ge \frac{1}{2}f(S^\star|_{\agents\setminus Z})$\label{line:gamma_free_if}} {
For any $i\in \agents$, $\alpha_i \gets 0$\;
\Return $\alpha$\; \label{line:gamma_free}
}
$N \gets N \setminus (Z\cup C)$\;\label{line:gamma_N}
\If{$\exists i \in N \text{ s.t. }f(\eqstar_i) \ge \frac{1}{4 \gamma} f(\eqstar_{N})$\label{line:gamma_if_big}}{
$\alpha_i \gets \frac{3}{4}$\;
For any $j \ne i$, $\alpha_j \gets 0$\;
\Return $\contract$\; \label{line:gamma_big_agent}
}
Divide the agents of $N$ into $\lceil 4\gamma \rceil$ groups $A_1,\dots, A_{\lceil4 \gamma \rceil}$, such that $|A_j| \le \frac{1}{2\cdot \gamma}|N|$ for all  $j\in [\lceil 4 \gamma \rceil]$\;
Let $j=\argmax_{j\in [\lceil4 \gamma \rceil]} f(S^\star_{A_j})$\;
$\alpha_i \gets \frac{3}{2}\max_{i'\in \agents} \alpha^\star_{i'}$ for all $i\in A_j$\;
$\alpha_i \gets 0$ for all $i\notin A_j$\;
\Return $\contract$\; \label{line:gamma_last} 
\end{algorithm}

\begin{proof}
If Algorithm~\ref{alg:gamma_xos} returns on Line~\ref{line:gamma_huge_agent}, since until that point its execution is identical to that of Algorithm~\ref{alg:pof_xos}, we are done by Lemma~\ref{lem:pof_large_agent_approx}.

Otherwise, we start by claiming that $f(S^\star|_{\agents\setminus Z}) \ge \frac{1}{5} (1-\sum_{i=1}^n \alpha^\star_i) f(S^\star)$. Indeed, if $Z=\emptyset$, this is holds trivially. Otherwise, $Z=\{i_0\}$, and since the if condition in Line~3 wasn't met $4f(S^\star_{-{i_0}}) >  (1-\alpha^\star_{i_0}) f(S^\star_{i_0})$, and thus 
\[
\left(1-\sum_{i=1}^n \alpha^\star_i\right) f(S^\star) \le \left(1-\sum_{i=1}^n \alpha^\star_i\right) (f(S^\star_{i_0})+f(S^\star_{-i_0}) \le (1-\alpha^\star_i) f(S^\star_{i_0}) + f(S^\star_{-i_0}) < 5f(S^\star_{-i_0})=5f(S^\star|_{\agents\setminus Z}),
\]
as desired.

If Algorithm~\ref{alg:gamma_xos} returns on Line~\ref{line:gamma_free}, observe that all the agents $C$ are working for free, and thus $S^\star|_C$ is also an equilibrium under the contract $\alpha$ which pays $0$ to everyone. The utility from this is 
\[
\left(1-\sum_{i=1}^n \alpha_i\right) f(S^\star|_C) = f(S^\star|_C) \ge \frac{1}{2} f(S^\star|_{\agents\setminus Z}) \ge \frac{1}{10} f(S^\star) \ge \frac{1}{10}\left(1-\sum_{i=1}^n \alpha^\star_i\right)f(S^\star),
\]
as needed.

If Algorithm~\ref{alg:gamma_xos} reaches Line~\ref{line:gamma_N}, then observe that $f(S^\star|_N) \ge \frac{1}{2} f(S^\star|_{\agents\setminus Z}) \ge \frac{1}{10} (1-\sum_{i=1}^n \alpha^\star_i)f(S^\star)$.

Now, if Algorithm~\ref{alg:gamma_xos} returns on Line~\ref{line:gamma_big_agent}, then $\alpha_i =\frac{3}{4}$ for some $i$ such that $f(S^\star_i) \ge \frac1{4\gamma} f(S^\star|_N)$ and $\alpha^\star_i \le \frac{1}{2}$. Let $S$ be an equilibrium of $\alpha$ where only agent $i$ acts, we have that 
\[
f(S_i) \ge \alpha_i f(S_i) -c(S_i) \ge \alpha_i f(S^\star_{i}) - c(S^\star_i) \ge (\alpha_i - \alpha^\star_i) f(S^\star_i) + \alpha^\star_i f(S^\star_i) - c(S^\star_i) \ge (\alpha_i - \alpha^\star_i) f(S^\star_i) \ge \frac{1}{4}f(S^\star_i),
\]
and thus, 
\[
(1-\alpha_i) f(S) \ge \frac{1}{4} \cdot \frac{1}{4}f(S^\star_i) \ge \frac{1}{16} \frac{1}{4\gamma} f(S^\star|_N) \ge \frac{1}{640 \gamma} \left(1-\sum_{i=1}^n \alpha^\star_i\right) f(S^\star),
\]
as needed.

Lastly, it remains to consider the case where Algorithm~\ref{alg:gamma_xos} returns on Line~\ref{line:gamma_last}. Observe that, because the if condition of Line~\ref{line:gamma_if_big} was never met, $|N| \ge \frac{1}{4\gamma}$, and thus a partition $A_1,\dots,A_{\lceil 4\gamma\rceil}$ of $N$ such that $|A_j|\le \frac{1}{2 \cdot \gamma}|N|$ for all $j\in [\lceil 4\gamma \rceil]$ is possible.
Let $j=\argmax_{j\in [\lceil 4 \gamma \rceil]} f(S^\star_{A_j})$. Observe that, by subadditivity, $f(S^\star_{A_j}) \ge \frac{1}{\lceil 4 \gamma \rceil} f(S^\star|_N) \ge \frac{1}{8\gamma }f(S^\star|_N) \ge \frac{1}{80 \gamma }(1-\sum_{i=1}^n \alpha^\star_i) f(S^\star)$. Additionally, since $\alpha^\star$ is $\gamma$-equal-pay, it holds that $\alpha^\star_i \ge \frac{1}{\gamma} \max_{i'\in \agents}\alpha^\star_{i'}$ for all $i\in N$, and in particular, $|N|\cdot \max_{i'\in \agents} \alpha^\star_{i'} \le \gamma$. Thus,
\[
\sum_{i=1}^n \alpha_i = \frac{3}{2} \cdot |A_j| \cdot \max_{i'\in \agents}\alpha^\star_{i'}  \le \frac{3}{2}\frac{|N|}{2 \gamma}\max_{i'\in \agents}\alpha^\star_{i'}\le \frac{3}{4}.
\]
By Lemma~\ref{lem:scaling_for_exist}, there exists an equilibrium $S$ of $\alpha$ such that $f(S) \ge \frac{2}{3} f(S^\star_{A_j})$, and thus the utility is
\[
\left(1-\sum_{i=1}^n \alpha_j\right) f(S) 
\ge \left(1-\frac{3}{4}\right) \frac{2}{3}f(S^\star_{A_j}) 
\ge 
\frac16
\cdot \frac{1}{80\gamma}\left(1-\sum_{i=1}^n \alpha^\star_i\right) f(S^\star) 
\ge 
\frac{1}{640 \gamma}
\left(1-\sum_{i=1}^n \alpha^\star_i\right) f(S^\star),
\]
as needed.
\end{proof}
\section{Extending results to BEST Objectives}\label{app:best}

In this section we consider a wider class of objectives, beyond the principal's utility (profit).
Our focus is ont he class of BEST objectives, introduced by \cite{feldman2025budget} for the multi-agent binary-actions model and generalized by \cite{oneactiontoomany} to the combinatorial actions setting.

In Appendix~\ref{subsec:best_prelims} we present the necessary preliminaries for this section, including the definition of BEST objectives (Definition~\ref{def:goodobj_multi_multi}).
In Appendix~\ref{subsec:approx_equal_pay_gs_best} we give a poly-time algorithm for computing a constant-factor approximation to the optimal equal-pay contract with respect to any BEST objective when $f$ is gross-substitutes.
In Appendix~\ref{subsec:hardness_xos_best} we extend the inapproximability results for XOS $f$ from Appendix~\ref{sec:hardness_xos} to any BEST objective.

In Appendix~\ref{subsec:poe_lower_bounds_best} we extend the lower bounds on the \price\ for any BEST objectives, generalizing the results for additive and subadditive $f$ in Appendix~\ref{sec:pof_xos}.
In Appendix~\ref{subsec:poe_gs_best} we give a an upper bound of $O\big(\frac{\log n}{\log \log n}\big)$ on the \price for any BEST objective when $f$ is gross-substitutes, and show its tightness in Appendix~\ref{subsec:poe_submod_unbounded}, by showing that when $f$ is submodular the \POE\ is unbounded. 

We remark that in the case of BEST objectives we implicitly assume the restriction of all contracts to those which are individually rational for the principal, i.e., $\sum_{i=1}^n \alpha_i \le 1$. 

\subsection{Preliminaries for \goodobj\ Objectives}\label{subsec:best_prelims}

Below we repeat the definition of an objective and BEST objective as introduced by \cite{oneactiontoomany}.
One may observe that it captures the principal's utility (profit), social welfare, reward and others.

\begin{definition} [Objectives in the Multi-Agent Combinatorial-Actions Model]\label{def:obj_multi_multi}
    An \emph{objective} $\varphi$ is defined by a poly-time algorithm that, given a problem instance $\multiInstance$, a contract $\noindexcontract$, and a subset of {actions} $S\subseteq T$, outputs a non-negative real number, denoted $\varphi_{\multiInstance}(\noindexcontract,S)$. This algorithm has value oracle access to $f$. 
    We omit the subscript if the instance is clear from context.
\end{definition}

\begin{definition}[{BEST Objectives}]  \label{def:goodobj_multi_multi}
    An objective $\varphi$ belongs to the class of  beyond standard (BEST) objectives if, for any instance $\multiInstance$, it is:
    \begin{enumerate}[label=(\roman*)]
        \item \emph{Sandwiched between profit and reward:} For any $\noindexcontract$ and $S \subseteq \actions$,
        $u_P(\noindexcontract,S)\le \varphi(\noindexcontract, S) \le f(S)$. \label{def:sandwichproperty}
        \item \emph{Decomposable:} For any $\noindexcontract$, any $S \subseteq \actions$, and any $i\in \agents$,  $\varphi(\noindexcontract, S) \le f(S_{-i})+\varphi(\noindexsubcontract{i}, S_i)$.\label{def:decompositionproperty} 
        \item \emph{Weakly increasing in $S$:} For any $\noindexcontract$ and any $S\subseteq S'\subseteq \actions$, $\varphi(\noindexcontract, S) \le \varphi(\noindexcontract, S')$. \label{def:weakly_increasing_property}
        \item \emph{Weakly decreasing in $\noindexcontract$:} For any  $\noindexcontract \le \noindexcontract'$ (coordinate-wise) and $S\subseteq \actions$,  $\varphi(\noindexcontract, S) \ge \varphi(\noindexcontract', S)$. \label{def:weakly_decreasing_property}
    \end{enumerate}
\end{definition}

Note that the definition of the \price\ (Definition~\ref{def:pof}), genenralizes to any BEST obejective $\varphi$ be replacing $u_P(\noindexcontract,S)$ with $\varphi(\noindexcontract,S)$.

In this section we utilize the best-response monotonicity of contracts when $f$ is GS, due to \cite{oneactiontoomany}.
\begin{lemma}[\cite{oneactiontoomany}]
\label{lem:bestresponsemonotonicity}
    Consider an instance with a gross substitutes $f$.
    For any contract $\noindexcontract$, equilibrium $S \in \nash(\noindexcontract)$, and agent $i \in \agents$,
    there exists an equilibrium $S' \in \nash(\noindexsubcontract{i})$ such that $S_i \subseteq S_i'$ and $S'_j = \emptyset$ for any $j \ne i$.
\end{lemma}

\subsection{Constant Approximation of the Optimal Equal-Pay Contract for GS $f$}\label{subsec:approx_equal_pay_gs_best}
In this section we show an efficient constant-factor approximation algorithm to the optimal equal-pay contract for any BEST objective.
\begin{theorem}\label{thm:o(1)-approx-best}
    Let $\varphi$ be a BEST objective. When $f$ is gross substitutes, there exists a poly-time algorithm which computes an $O(1)$-approximation to the optimal equal-pay contract with respect to $\varphi$, using value oracle access to $f$ and $\varphi$. 
\end{theorem}

To show Theorem~\ref{thm:o(1)-approx-best}, we present modified versions of Theorem~\ref{thm:modified_defk_cases}, Theorem~\ref{thm:single-fptas}, and Theorem~\ref{thm:const_approx_no_small} for gross substitutes $f$ and with respect to BEST objectives. 

We start with the following lemma, which corresponds to Theorem~\ref{thm:modified_defk_cases}:
\begin{lemma}\label{lem:modified_cases_gs}
    Let $f$ be a gross substitutes success probability function. Let $\contract^\star$ be any equal-pay contract such that $\sum_{i\in A} \alpha_i\le 1$ and $S^\star$ be an equilibrium of $\contract^\star$. For any $0\le \phi \le 1$, there exists an equal-pay contract $\contract'$ and an equilibrium $S'$ of $\contract'$ fulfilling 
    $\varphi(\contract', S') \ge \frac{\phi}{36} \cdot \varphi(\contract^\star, S^\star)$ and
    \begin{enumerate}
        \item \textbf{(no large agent)} $f(S_i') \le \phi f(S')$ for all $i\in A$, or
        \item \textbf{(single agent)} $\alpha'_i =0$ and $S_i=\emptyset$ for all but one $i\in A$.
    \end{enumerate}
\end{lemma}
\begin{proof}
    We consider the following two cases:
    
    \paragraph{Case 1: For all $i\in A$ it holds that $\alpha_i \le \frac{2}{3}$. } 
    If for all $i\in A$ it also holds that $f(S^\star_i) \le \phi f(S^\star)$, we are done. Otherwise, let $i\in A$ be such that $f(S^\star_i) > \phi f(S^\star)$. 
    By Lemma~\ref{lem:bestresponsemonotonicity}, there exists an equilibrium $S'\subseteq T_i$ of $\noindexsubcontractstar{i}$ such that $S^\star_i\subseteq S'$, and thus, by Definition~\ref{def:goodobj_multi_multi},
    \[ \varphi(\noindexsubcontractstar{i},S') ~\ge~ \varphi(\noindexsubcontractstar{i}, S^\star_i)~\ge~(1-\alpha_i)f(S^\star_i) ~\ge~\frac{\phi}{3} f(S^\star) ~\ge~ \frac{\phi}{3} \varphi(\contract^\star, S^\star),
    \]
    and we are done.
    \paragraph{Case 2: There exists $i\in A$ such that $\alpha_i > \frac{2}{3}$.}
    By Definition~\ref{def:goodobj_multi_multi}\ref{def:decompositionproperty} it holds that
    \[
    \varphi(\contract^\star, S^\star) \leq f(S_{-i}^\star) +\varphi(\noindexsubcontractstar{i}, S_i^\star).
    \]
    If $\varphi(\noindexsubcontractstar{i}, S^\star_i) \ge f(S^\star_{-i})$, then by Lemma~\ref{lem:bestresponsemonotonicity}, there exists an equilibrium $S'\subseteq T_i$ of $\noindexsubcontractstar{i}$ such that $S^\star_i\subseteq S'$, and thus, by Definition~\ref{def:goodobj_multi_multi}\ref{def:weakly_increasing_property} \[
    \varphi(\noindexsubcontractstar{i},S') ~\ge~ \varphi(\noindexsubcontractstar{i}, S^\star_i)~\ge~\frac{1}{2} \varphi(\contract^\star, S^\star),
    \]
    and we are done. Otherwise, if $f(S^\star_{-i}) > \varphi(\noindexsubcontractstar{i}, S^\star_i)$, we apply Lemma~\ref{lem:scaling_for_exist} on $\contract^\star$ and $A\setminus \{i\}$ with $\gamma=2$, and get an equal-pay contract $\contract$ and equilibrium $S$ such that $\sum_{j\in A} \alpha_j =2\sum_{j\in A\setminus \{i\}} \alpha^\star_j \le \frac{2}{3}$ and $f(S) \ge \frac{1}{2}f(S^\star_{-i})$. Thus, 
    \[
    \varphi(\contract, S) ~\ge~ \left(1-\sum_{i\in A} \alpha_i\right) f(S) ~\ge~ \frac{1}{3}\cdot \frac{1}{2} f(S^\star_{-i}) ~\ge~ \frac{1}{12} \varphi(\contract^\star, S^\star).
    \]
    By applying the case 1 analysis on $\alpha, S$ we get an equal-pay contract $\contract'$ and equilibrium $S'$ such that one of the two conditions of Lemma~\ref{lem:modified_cases_gs} holds that
    \[
    \varphi(\contract', S') ~\ge~ \frac{\phi}{3} \varphi(\contract, S) ~\ge \frac{\phi}{36}\varphi(\contract^\star, S^\star),
    \]
    as needed.
\end{proof}

The next observation from \cite{oneactiontoomany} corresponds to Theorem~\ref{thm:single-fptas}.
\begin{observation}[\cite{oneactiontoomany}]\label{obs:single-agent}
    When $f$ is gross substitutes, the optimal single-agent contract with respect to any BEST objective $\varphi$ can be computed in polynomial time, with value oracle access to $f$ and $\varphi$.
\end{observation}

Finally, the proof of Theorem~\ref{thm:const_approx_no_small} in Section~\ref{sec:const_approx} actually shows the following stronger claim with respect to BEST objectives.
\begin{theorem}\label{thm:const_approx_no_small_gs}
    When $f$ is submodular, Algorithm~\ref{alg:approx_equal_pay} runs in polynomial time with value oracle access to $f$, and returns an equal-pay contract $\contract$ such that $\sum_{i\in [n]}\alpha_i \le \frac{1}{2}$ and in any equilibrium $S$, $f(S)\ge \Lambda$ with the following guarantee.
    For any equal-pay contract $\constar$ and any equilibrium $S^\star$ of $\constar$ such that $f(S^\star_i) \le \frac{1}{16}f(S^\star)$ for all $i\in [n]$, it holds that 
    \[
    \Lambda \ge \Omega(1) f(S^\star).
    \]
\end{theorem}
In particular, for the contract-equilibrium pairs $(\contract, S)$ and $(\constar, S^\star)$ per Theorem~\ref{thm:const_approx_no_small_gs} we have, for any BEST objective
\[
\varphi(\contract, S) ~\ge~ \left(1-\sum_{i=1}^n \alpha_i\right) f(S) ~\ge~ \frac{1}{2}f(S) ~\ge~ \frac{1}{2}\Lambda = \Omega(1)f(S^\star)~\ge~ \Omega(1) \varphi(\constar, S^\star).
\]
Together with Lemma~\ref{lem:modified_cases_gs} and Observation~\ref{obs:single-agent} this proves Theorem~\ref{thm:o(1)-approx-best}.

\subsection{No Constant-Factor Approximation for XOS $f$}\label{subsec:hardness_xos_best}

In this section we extend the results of Theorem~\ref{thm:XOS_inapprox} to any BEST objective. Namely,
\begin{theorem}
Fix any BEST objective $\varphi$. When $f$ is XOS, any (possibly randomized) algorithm that makes only polynomially many demand queries achieves an (expected) approximation ratio of at least $\Omega(m^{1/10})$ to $\varphi$ over unconstrained contracts, where $m$ denotes the total number of actions. Moreover, the same lower bound applies to equal-pay contracts.
\end{theorem}

\begin{proof}
First, Lemma~\ref{lem:good_contract} implies that for any BEST objective $\varphi$, the equal-pay contract 
$\eqcontract{(1/(2\ell^2))}{G}$
induces the equilibrium $T|_G$ and satisfies
$\varphi(\contract, T|_G) \ge (1-\sum_{i \in A} \alpha_i)f(T\mid_G) \ge (1/2) \cdot \ell^4.$

Similarly, Lemma~\ref{lem:bad_contract} implies that, for any contract $\contract$ that is unaligned with respect to $G$ and for any equilibrium $S$, it holds that $\varphi(\contract, S) \le 2\ell^3$.

Thus, using the same argument as in Lemma~\ref{lem:small_probability}, and algorithm has an exponentially small probability to get an $(\ell/4)$-approximation to any BEST objective, when $G$ is sampled uniformly at random.

The rest of the proof is identical to that of Theorem~\ref{thm:XOS_inapprox}.
\end{proof}

\subsection{\Price\ Lower Bounds}\label{subsec:poe_lower_bounds_best}

Below we adjust the proof of Proposition~\ref{prop:pof_add} so that the lower bound for additive $f$ is valid all BEST objectives.
\begin{proposition}\label{prop:poe_add_best}
There exists an instance with additive $f$ such that the price of equality for any \goodobj\  objective $\varphi$ is $\Omega\left(\frac{\log n}{\log \log n}\right)$.
\end{proposition}
\begin{proof}
    Consider the same instance as in the proof of Proposition~\ref{prop:pof_add} and let $(\contract^\star,S^\star)$ be the optimal contract and equilibrium with respect to $\varphi$.
    \[
        \varphi(\contract^\star,S^\star) 
        \ \ge \ 
        \varphi(\contract,\{1,\dots,n\})  
        \ \ge \ u_P(\contract,\{1,\dots,n\})  
        \ = \ \left(1- \sum_{i \in \agents} \frac{1}{i2H_n} \right)f(\{1,\dots,n\}) 
        \ = \ \frac12 \cdot H_n,
    \]
    where $\contract$ is the contract for which $\alpha_i = \frac{1}{i2H_n}$.

    Additionally, it is shown in the proof of Proposition~\ref{prop:pof_add} that for any equal-pay $\contract$, and for any $S \in \nash(\contract)$, it holds that $f(S) \le \ln(3\ln n)+o(1)$.
    Thus, we get for any such pair $(\contract,S)$, $\varphi(\contract,S) \le f(S) \le \ln(3\ln n)+o(1)$.

    We conclude that 
    \[
    \POE_{\text{additive }f} = \frac{\varphi(\contract^\star,S^\star)}{\varphi(\contract,S) } \ge \frac{H_n}{2\ln (3\gamma \ln n) + o(1)} = \Omega\left(\frac{\ln n}{\ln (\gamma \ln n)}\right) =\Omega\left(\frac{\log n}{\log (\gamma \log n)}\right).
    \]
\end{proof}

We also modify Proposition~\ref{prop:pof_subadditive} so that all BEST objectives are included.
\begin{proposition}
There exists an instance with subadditive $f$ such that for any BEST objective $\varphi$, there exists an (unconstrained) contract $\alpha^\star$ and equilibrium $S^\star$ which satisfy
\[
\varphi(\alpha^\star, S^\star) \ge \frac{1}{8},
\]
but such that for any equal-pay contract $\alpha$ with $\sum_{i=1}^n\alpha_i \le 1$ and equilibrium $S$
\[
\varphi(\alpha, S) \le \frac{2}{\sqrt{n}}.
\]
\end{proposition}
\begin{proof}
The proof of Proposition~\ref{prop:pof_subadditive} suggests that there exists an unconstrained contract and an equilibrium $(\constar,\eqstar)$ such that $u_P(\constar,\eqstar) \ge 1/8$, this immplies that $\varphi(\constar,\eqstar) \ge u_P(\constar,\eqstar) \ge 1/8$.
Additionally, for any equal-pay contract $\contract$ and any $S \in \nash(\contract)$ it holds that $\varphi(\contract,S) \le f(S) \le 2/\sqrt{n}$, where the last inequality follows from the proof of Proposition~\ref{prop:pof_subadditive}.
\end{proof}

\subsection{Upper Bound of $O(\log n/ \log \log n)$ on \Price\ for GS $f$ }\label{subsec:poe_gs_best}

Below we show how to achieve an upper bound of $O(\log n/ \log \log n)$ on \price\ for ant BEST objective under gross substitutes $f$, matching the lower bound for the additive case established in Proposition~\ref{prop:poe_add_best}.
This result is tight in the sense that for submodular $f$ the \price is unbounded, see Theorem~\ref{thm:unbouded_poe_sm}.

\begin{algorithm}[h]
\caption{An $O(\log n/\log \log n)$-approximation to any \goodobj\ objective via an equal-pay contract for GS $f$}\label{alg:pof_gs}
\KwIn{A contract $\constar$ with $\sum_i \alpstar_i <1$, a profile $\eqstar \in \nash(\constar)$, costs $c_1,\dots,c_m \ge 0$, value oracle access to a GS set function $f:2^{[m]}\rightarrow \reals_{\ge 0}$ and the objective  function $\varphi$}
\KwOut{An equal-pay contract $\contract$ 
and a profile $S \in \nash(\contract)$ 
such that $O(\log n/\log \log n)(1-\sum_i \alpha_i)f(S) \ge 
(1-\sum_i \alpstar_i)f(\eqstar)$}
let $N \gets A(\eqstar)$ and ignore all other agents, i.e., those that shirk in $\eqstar$\; 

\If{$\exists i \in [n]$ such that $\alpstar_i > 1/2$}{
$\contract^\dagger \gets \noindexsubcontractstar{i}$\;
Let $S^\dagger \gets \argmax_{S^\dagger_i \subseteq T_i}f(S'_i) -  \sum_{j \in S'_i} \frac{c_j}{\alpha_i}$ \tcp*[l]{where $\frac{0}{0}=0$ and $\frac{c}{0}=\infty$ for $c >0$}
$N \gets N \setminus \{i\}$
}
Rename remaining agents so that $\alpstar_1 \ge \alpstar_2 \ge \dots$
\tcp*[l]{\textbf{Only small agents}}
$X \gets \{i \in N \mid \alpstar_i < 1/(2n)\}$, $k \gets |N \setminus X|$\;
$h \gets 1$, $a_h \gets 1$\;
\While{$a_h < k$}{
     $b_h \gets \min\{a_h + \floor{\frac{1}{2\alpstar_{a_h}}},k+1\}$\;
     $B_h \gets \{a_h, \dots, b_h-1\}$\;
     $B'_h \gets \{b_h\}$ \tcp*[l]{if $b_h=k+1$, we set $B'_h \gets \emptyset$}
     $h \gets h+1$, $a_h \gets b_h +1$\;
}
$\B \gets \{X,B_1,\dots,B_h,B'_1,\dots B'_h\}$\; \label{line:pof_gs_all_buckets}

$B \gets \argmax_{Q \in \B} f(\eqstar_Q)$ \tcp*[l]{\textbf{$\eqstar_Q$ are the actions in $\eqstar$ performed by agent in $Q$}} \label{line:pof_gs_bucket}
$p_B \gets \max\{\alpstar_i \mid i \in B\}$\; \label{line:pof_gs_bucket_price}

for every $i\in B$, $\alpha'_i \gets (3/2) \cdot p_B$, and for every $j \notin B$, $\alpha'_i \gets 0$\;\label{line:pof_gs_contract_bucket}
Let $S' \gets \argmax_{S^\dagger \subseteq T}f(S^\dagger) - \sum_{i \in [n]} \sum_{j \in T_i \cap S^\dagger} \frac{c_j}{\alpha'_i}$ \tcp*[l]{where $\frac{0}{0}=0$ and $\frac{c}{0}=\infty$ for $c >0$}

\If{$\varphi(\contract', S') \ge \varphi(\contract^\dagger, S^\dagger)$}{
$(\contract,S) \gets (\contract', S')$
}\Else{
$(\contract,S) \gets (\contract^\dagger, S^\dagger)$
}

\Return $(\contract,S)$\; \label{line:pof_gs_return_bucket}
\end{algorithm}

\begin{proposition}
    The contract and equilibirium pair $(\contract,S)$ returned by Algorithm~\ref{alg:pof_gs} satisfies 
    \[
    \varphi(\contract,S) \ge \frac{\ln \ln n}{132\ln n} \varphi(\constar,\eqstar)
    \]
\end{proposition}
\begin{proof}
    Recall that the contract $(\contract^\dagger,S^\dagger)$ only incentivize a single agent $i \in A(\eqstar)$ and $S^\dagger \subseteq T_i$ is $i$th best-response to $\contract^\dagger = \noindexsubcontract{i}$.
    Thus, by Lemma~\ref{lem:bestresponsemonotonicity} $f(S^\dagger) \ge f(S^\star_i) \ge \varphi(\noindexsubcontract{i},S^\star_i)$, where $S^\star_i = \eqstar \cap T_i$ and the last inequality follows from Definition~\ref{def:goodobj_multi_multi}, property \ref{def:sandwichproperty}.

    Since $f$ is subadditive, the same analysis as in Lemma~\ref{lem:pof_num_buckets} yields
    $f(S') \ge \frac{\ln \ln n}{33 \ln n}f(S^\star_{-i})$.
    Since $A(S')$ only has small agents we have that $\sum_{i \in [n]}\alpha_i' = \sum_{i \in B}\alpha'_i \le 1/2$.
    Thus,
    \[
    \varphi(\alpha',S') \ge (1-\sum_{i \in [n]}\alpha_i')f(S') \ge \frac{\ln \ln n}{66 \ln n}f(S^\star_{-i})
    \]
    We conclude that 
    $$
    \varphi(\contract,S) \ge \max\{\varphi(\contract',S'), \varphi(\contract^\dagger,S^\dagger))\}
    \ge 
    \frac{\ln \ln n}{132 \ln n} (f(\eqstar_{-i}) + \varphi(\noindexsubcontract{i},\eqstar_i))
    \ge 
    \frac{\ln \ln n}{132 \ln n} \varphi(\constar,\eqstar),
    $$
    where the last inequality follows from Definition~\ref{def:goodobj_multi_multi}, property \ref{def:decompositionproperty}.
\end{proof}

\subsection{Unbounded \Price\ for Reward for Submodular $f$}\label{subsec:poe_submod_unbounded}

\begin{theorem}\label{thm:unbouded_poe_sm}
    Even with $n=2$ agents and $m=3$ actions, for submodular $f$, the \price\ for $f$ maximization is unbounded.
\end{theorem}
\newcommand{\bad}{\mathcal{B}}
\newcommand{\good}{\mathcal{G}}
Consider the following two agent example: agent $1$ has actions $T_1 = \{\bad, \good\}$, and agent $2$ has actions $T_2=\{2\}$ (a binary agent). 
We consider costs $c_\bad = \eps, c_{\good}=1+\eps/3$, $c_2 = \eps^4$, for $\eps < 0.4$ and a coverage\footnote{A strict subset of submodular functions.} reward function $f:2^{\{\good, \bad, 2\}}\rightarrow \reals_{\ge 0}$ depicted in Figure~\ref{fig:pof_gap_coverage}.
More precisely, we consider a universe $U = \{u_1, u_2, u_3, u_4\}$ where the weight of $u_1$ is 1 and the rest of the weights are $\eps$. Each action is associated with elements in $U$ via $\tau: T \to U$.
$\tau(\good) = \{u_1,u_2\}$, $\tau(\bad) = \{u_2,u_3\}$, and $\tau(2) = \{u_3,u_4\}$.
For any set $S \subseteq T$, let $\tau(S) = \cup_{j \in S} \tau (j)$, then
$f(S) = \sum_{u \in \tau(S)} w(u)$, where $w$ denotes the weight of the element.

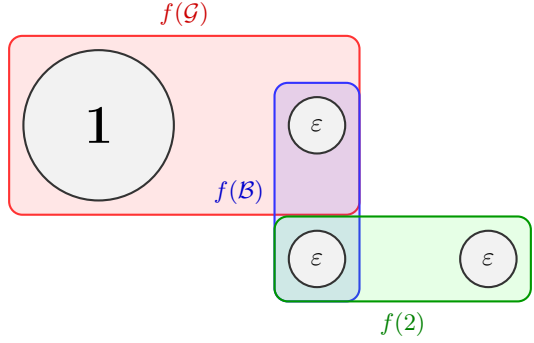
\begin{figure}
    \centering
    \begin{tikzpicture}[
        node distance = 1.0cm and 1.5cm, 
        main node/.style={
            circle, 
            draw=black!80, 
            thick, 
            fill=gray!10,
            minimum size=2.0cm, 
            font=\huge\bfseries 
        },
        small node/.style={
            circle, 
            draw=black!80, 
            thick, 
            fill=gray!10,
            minimum size=0.75cm, 
            font=\normalsize
        },
        group box/.style={
            rounded corners=5pt,
            thick,
            fill opacity=0.1
        },
        label text/.style={ font=\small\sffamily\bfseries }
    ]

    
    \node[main node] (n1) {1};

    \node[small node, right=of n1] (e1) {$\eps$};

    \node[small node, below=of e1] (e2) {$\eps$};

    \node[small node, right=of e2] (e3) {$\eps$};

    \begin{pgfonlayer}{background}
        
        \node[group box, draw=red!80, fill=red,
              fit=(n1) (e1), inner sep=5pt] (box_good) {};

        \node[group box, draw=blue!80, fill=blue,
              fit=(e1) (e2), inner sep=5pt] (box_bad) {};

        \node[group box, draw=green!60!black, fill=green,
              fit=(e2) (e3), inner sep=5pt] (box_two) {};
              
    \end{pgfonlayer}


    \node[label text, red!80!black, above] at (box_good.north) {$f(\good)$};

    \node[label text, blue!80!black, left] at (box_bad.west) {$f(\bad)$};

    \node[label text, green!50!black, below] at (box_two.south) {$f(2)$};

    \end{tikzpicture}
    \caption{The coverage function $f$ that exemplifies an unbounded \price.}
    \label{fig:pof_gap_coverage}
\end{figure}

\begin{lemma}
    The (unequal-pay) contract $\alpha_1 = 1-\eps^2, \alpha_2 = \eps^1$ yields $f(S)=1+3\eps$ from every equilibrium.
\end{lemma}
\begin{proof}
    Observe that for any $S_1\subseteq T_1$ it holds that 
    \[
    f(2\mid S_1) \ge \eps > \frac{c_2}{\alpha_2},
    \]
    thus in every equilibrium $S$ it holds that $S_2 = \{2\}$.

    Observe that for every $S\subseteq T$ with $2\in S$
    \[
    f(\bad \mid S\setminus \{\bad\}) \le f(\bad \mid \{2\}) = \eps < \frac{c_{\bad}}{\alpha_1},
    \]
    and thus in every equilibrium $S$ it holds that $\bad \notin S$.
    Finally, observe that
    \[
    f(\good \mid \{2\}) = 1+\eps > \frac{1+\eps/3}{1-\eps^2} =\frac{c_{\good}}{\alpha_1},
    \]
    So all in all the only equilibrium is $S=\{\good, 2\}$, and $f(S)=1+3\eps$
\end{proof}
\begin{lemma}
    No equal-pay contract $\contract$ with $\sum_{i=1}^n \alpha_i \le 1$ admits an equilibrium with $f(S)\ge 3\eps$.
\end{lemma}
\begin{proof}
    Let $\alpha$ be an equal-pay contract, and $S$ be an equilibrium. 
    If $S_2 \ne \emptyset$, then $\alpha_2 > 0$, and $\alpha_1 \le \frac{1}{2}$.
    Since $f(\good\mid \{\bad, 2\}) = 1 < \frac{c_{\good}}{\frac{1}{2}}$, this implies $\good \notin S$ and $f(S)\le 3\eps$.
    Otherwise, if $S_2 =\emptyset$, then, even if $\alpha_1=1$ (which maximizes $f$ subject to $S_2=\emptyset$) we have
    \[
    f(\{\good\}) - c_{\good} = \frac{2\eps}{3} < \eps = f(\{\bad\}) - c_{\bad},
    \]
    which implies that agent $1$ strictly prefers $\{\bad\}$ over $\{\good\}$,
    furthermore
    \[
    f(\{\good, \bad\})-c_{\good}-c_{\bad} =1+2\eps - 1-\frac{4\eps}{3} = \frac{2\eps}{3} < f(\{\bad\}) - c_{\bad},
    \]
    thus the only equilibrium in this case is $S=\{\bad\}$ with $f(S)=2\eps$.
\end{proof}

\section{Single-Agent Contracts Are $O(n)$-Approximately Optimal for Subadditive $f$}
\begin{theorem}
    When $f$ is subadditive, the optimal single-agent contract is an $O(n)$-approximation of the optimal unconstrained contract.
\end{theorem}
\begin{proof}
    Let $\alpha^\star, S^\star$ be a contract and equilibrium. If there exists an agent $j\in [n]$ such that $\alpha^\star_{j} > \frac{1}{2}$ and $(1-\alpha_{j}^\star) f(S_j^\star) \ge 4 f(S_{-j}^\star)$, the contract $\alpha_{j}=\frac{\alpha_{j}^\star+1}{2}$ and $\alpha_{i}=0$ for all $i\ne j$ has a principal's utility of at least $\frac{2}{17}(1-\sum_{i=1}^n \alpha_i^\star)f(S^\star)$, by Lemma~\ref{lem:pof_large_agent_approx}.
    Otherwise,
    let $Z=\{i\in [n]\mid \alpha^\star_i > \frac{1}{2}\}$. It holds that $f(S^\star_{[n]\setminus Z}) \ge \frac{1}{5}(1-\sum_{i=1}^n \alpha_i^\star)f(S^\star)$, since if $Z=\emptyset$ this is trivial, and if $Z=\{j\}$ then $(1-\alpha^\star_j)f(S^\star_j)< 4f(S^\star_{-j})$, and thus 
    \[
    (1-\sum_{i=1}^n \alpha_i^\star)f(S^\star) \le (1-\sum_{i=1}^n \alpha_i^\star)(f(S^\star_{j})+f(S^\star_{-j}) \le (1-\alpha_{j}^\star)f(S^\star_{j})+f(S^\star_{-j})\le 5f(S^\star_{-j}).
    \]
    Observe that, by subadditivity, there exists an agent $k\in [n]\setminus Z$ such that $f(S^\star_k) \ge \frac{1}{n}f(S^\star_{[n]\setminus Z}) \ge \frac{1}{5n}(1-\sum_{i=1}^n \alpha_i^\star)f(S^\star)$. Consider the contract $\alpha_k= \frac{3}{4}$ and $\alpha_i=0$ for all $i\ne k$. Let $S\subseteq T_k$ be an equilibrium of this contract where only agent $k$ acts. It holds that
    \[
    \begin{aligned}
    f(S) &\ge \alpha_k f(S) - c(S) \ge \alpha_k f(S^\star_k) - c(S^\star_k) \ge (\alpha_k - \alpha^\star_k) f(S^\star_k) + \alpha^\star_k f(S^\star_k) - c(S^\star_k) \ge (\alpha_k - \alpha^\star_k) f(S^\star_k) \\
    &\ge  \left(\frac{3}{4}-\frac{1}{2}\right)f(S^\star_k) \ge \frac{1}{4}f(S^\star_k) \ge \frac{1}{20n}\left(1-\sum_{i=1}^n \alpha_i^\star\right)f(S^\star).
    \end{aligned}
    \]
    So the principal's utility is
    \[
    \left(1-\alpha_k\right)f(S) =\frac{1}{4}f(S) \ge \frac{1}{80n}\left(1-\sum_{i=1}^n \alpha_i^\star\right)f(S^\star),
    \]
    as needed.
\end{proof}
\section{Constant-Approximation of the Optimal Equal-Pay Contract under Binary-Actions and XOS Rewards}
In this appendix we provide a constant-factor approximation to the optimal equal-pay contract in the binary-action setting under XOS $f$.
Our guarantees hold for any BEST objective (see Definition~\ref{def:goodobj_multi_multi}).
\begin{theorem}\label{thm:xos_binary_approx}
    Let $\varphi$ be a BEST objective. For XOS $f$, Algorithm~\ref{alg:xos_approx} runs in polynomial time given demand oracle access to $f$ and value oracle access to $\varphi$, and outputs an equal-pay contract $\contract$ and equilibrium $S$ such that $\varphi(\contract, S) \ge \Omega(1) \varphi(\contract^\star, S^\star)$ for any equal-pay contract $\contract^\star$ and equilibrium $S^\star\subseteq [n]$ such that $\sum_{i=1}^n \alpha^\star_i \le 1$.
\end{theorem}

\begin{algorithm} 
\caption{An $O(1)$-approximation for the optimal equal-pay contract under binary actions}\label{alg:xos_approx}
\KwIn{$n\in \mathbb{N}$, costs $c_1,\dots,c_n \ge 0$, and demand oracle access to an XOS set function $f:2^{[n]}\rightarrow \reals_{\ge 0}$, value oracle access to a BEST objective $\varphi$.}
\KwOut{An equal pay contract $\alpha$ and equilibrium $S$ such that $\varphi(\alpha, S) \ge \Omega(1)\varphi(\alpha^\star, S^\star)$ for any equal-pay contract $\alpha^\star$ and equilibrium $S^\star \subseteq [n]$.}
$A'\gets \{i\in [n]\mid \frac{c_i}{f(\{i\})} \le \frac{1}{2}\}$ \label{line:def_of_a_prime}\;
$\mathcal{C}\gets \left\{ \left(\eqcontract{\frac{c_i}{f(\{i\})}}{\{i\}}, \{i\}\right)\mid i\in [n]\right\} \cup \{(\eqcontract{0}{[n]}, \{i\in [n]\mid c_i = 0)\}$\;
\For{$k\gets 8,9,\dots n$}{
\For{$j\gets 0,\dots, \lfloor\log n\rfloor$} {
    $y \gets 2^j \max_{i\in A'} f(\{i\})$\;
    $p_i \gets \max\left(\frac{c_ik}{2}, \frac{y}{2k}\right)$\;
    $T\gets $ a demand bundle with respect to the price vector $p$\;
    $U\gets$ an arbitrary subset of $T$ of size $\min\left(|T|, \lfloor\frac{k}{8}\rfloor\right)$\;
    $V\gets$ the result of applying Lemma~\ref{lem:scaling_for_exist} on action profile $T$, subset of agents $U$, and $\gamma=2$\;
    $C\gets C\cup \left\{\left(\eqcontract{\frac{4}{k}}{V}, V\right)\right\}$\;
}
}
\Return an element of $\argmax_{(\alpha, S)\in \mathcal{C}} \varphi(\alpha, S)$\;
\end{algorithm}
\begin{lemma}\label{lem:approx_A'_f}
    Let $\contract^\star=\eqcontract{t}{S^\star}$ be an equal-pay contract such that $t\le \frac{1}{|S^\star|}$, $S^\star$ is dropout stable with respect to $\contract^\star$ and $S^\star \subseteq A' = \{ i \in [n] \mid c_i/f(\{i\}) \leq 1/2\}$.
    When $f$ is XOS, there exists an equal-pay contract and equilibrium pair $(\contract, S)\in \mathcal{C}$ (defined in Algorithm~\ref{alg:xos_approx}), such that $\varphi(\contract, S) \ge \frac{1}{256}f(S^\star)$. 
\end{lemma}
\begin{proof}
    Let $i^\star \in \argmax_{i\in A'} f(\{i\})$. Observe that $\left(\eqcontract{\frac{c_{i^\star}}{f(\{i^\star\})}}{\{i^\star\}}, \{i^\star\}\right) \in \mathcal{C}$, and indeed $\{i^\star\}$ is an equilibrium of $\alpha=\eqcontract{\frac{c_{i^\star}}{f(\{i^\star\})}}{\{i^\star\}}$. Furthermore, $\varphi(\alpha, \{i^\star\})\ge\left(1-\frac{c_{i^\star}}{f(\{i^\star\})}\right)f(\{i^\star\})\ge \frac{1}{2}f(\{i^\star\})$, where the first inequality is since $\varphi$ is a BEST objective and thus lower-bounded by the principal's utility, and the second is since $i^\star\in A'$.
    
    If $f(S^\star) \le 2f(\{i^\star\})$ then $\varphi(\alpha, \{i^\star\}) \ge \frac{1}{2}f(\{i^\star\}) \ge \frac{1}{4}f(S^\star)$, and we are done.
    
    If $|S^\star| < 8$, from subadditivity $f(\{i^\star\})\ge \frac{1}{7}f(S^\star)$, and thus $\varphi(\alpha, \{i^\star\}) \ge \frac{1}{2}f(\{i^\star\})\ge \frac{1}{14}f(S^\star)$, as needed.
    
    Otherwise, we have that $|S^\star|\ge 8$ and $f(S^\star)>2f(\{i^\star\}) = 2\max_{i\in A'} f(\{i\})$. Additionally, due to subadditivity, $f(S^\star) \le n\max_{i\in A'} f(\{i\})$. Thus, in some iteration $k=|S^\star|$ and $y$ satisfies $\frac{1}{2}f(S^\star) \le y \le f(S^\star)$. Let $V$ be as defined in this iteration, and $\alpha=\eqcontract{\frac{4}{k}}{V}$. We claim that $\varphi(\alpha, V)\ge \frac{1}{256}f(S^\star)$, concluding the proof.
    Firstly,
    \[
    \varphi(\alpha, V) \ge \left(1-\sum_{i\in [n]}\alpha_i\right) f(V) = \left(1-\frac{4}{k}|V|\right) f(V) \ge \left(1-\frac{4}{k}\left\lfloor\frac{k}{8}\right\rfloor\right)f(V) \ge \frac{1}{2}f(V),
    \]
    so it is left to show that $V$ is an equilibrium of $\alpha$ and $f(V)\ge \frac{1}{128}f(S^\star)$.

    Let $p$ be the price vector defined by $p_i = \max\left(\frac{c_ik}{2}, \frac{y}{2k}\right)$, and let $A_0 = \{i\in [n]\mid p_i = c_ik/2\}$ and $A_1=[n]\setminus A_0$. Since $T$ is a demand set with respect to the price vector $p$, we have
    \begin{equation}\label{eq:f_T_bound_1}
    f(T) \ge f(T)-\sum_{i\in T} p_i \ge f(S^\star \cap A_0)-\sum_{i\in S^\star \cap A_0} p_i =f(S^\star \cap A_0) - \sum_{i\in S^\star \cap A_0} \frac{c_i\cdot k}{2}.
    \end{equation}
    Since $S^\star$ dropout stable with respect to $\alpha^\star=\eqcontract{t}{S^\star}$, it holds that, for any $i\in S^\star$, $\frac{1}{k} \cdot f(i\mid S^\star\setminus \{i\}) \ge t\cdot f(i\mid S^\star \setminus \{i\}) \ge c_i$, i.e., $c_i\cdot k \le f(i\mid S^\star \setminus \{i\})$. Plugging this into Equation~\ref{eq:f_T_bound_1} we get
    \[
    f(T) \ge f(S^\star \cap A_0) - \sum_{i\in S^\star \cap A_0}\frac{1}{2}f(i\mid S^\star \setminus \{i\})\ge f(S^\star \cap A_0) - \frac{1}{2}f(S^\star \cap A_0) = \frac{1}{2}f(S^\star \cap A_0),
    \]
    where the second inequality is by Lemma~\ref{lem:xos_margs}.
    It also holds that
    \begin{align*}
    f(T) &\ge f(T)-\sum_{i\in T} p_i \ge f(S^\star \cap A_1) - \sum_{i\in S^\star \cap A_1} p_i=f(S^\star \cap A_1) - \sum_{i\in S^\star \cap A_1} y/2k\\
    &= f(S^\star \cap A_1) - |S^\star \cap A_1| \frac{y}{2k} 
    \ge f(S^\star \cap A_1) -\frac{y}{2} \ge f(S^\star \cap A_1) - \frac{1}{2}f(S^\star).
    \end{align*}
    On the one hand we have $f(T) \ge \frac{1}{2}f(S^\star \cap A_0)$, and on the other $f(T) \ge f(S^\star \cap A_1)-\frac{1}{2}f(S^\star)$, which yields
    \[
    3f(T) =2f(T)+f(T) \ge f(S^\star \cap A_0) + f(S^\star \cap A_1)-\frac{1}{2}f(S^\star) \ge f(S^\star) - \frac{1}{2}f(S^\star) =\frac{1}{2}f(S^\star), 
    \]
    i.e., $f(T) \ge \frac{1}{6} f(S^\star)$. We also note that $T$ is dropout stable with respect to $\eqcontract{\frac{2}{k}}{T}$, since for each $i\in T$ it holds that $\frac{2}{k} f(i\mid T\setminus \{i\}) \ge \frac{2}{k} p_i \ge \frac{2}{k} \frac{c_i \cdot k}{2} = c_i$.
    We next claim that $f(U) \ge \frac{1}{64} f(S^\star)$. Observe that if $U=T$, there's nothing to prove. Otherwise, it holds that $|U| = \left\lfloor \frac{k}{8}\right\rfloor$, and thus, by Lemma~\ref{lem:xos_margs},
    \[
    f(U) \ge \sum_{i\in U} f(i\mid T\setminus \{i\}) \ge \sum_{i\in U} \frac{y}{2k} =\left\lfloor \frac{k}{8}\right\rfloor\frac{y}{2k} \ge \frac{y}{32} \ge \frac{1}{64}f(S^\star).
    \]
    We now have that $T$ is dropout stable with respect to $\eqcontract{\frac{2}{k}}{T}$, and $f(U) \ge \frac{1}{64} f(S^\star)$, thus, by the guarantees of Lemma~\ref{lem:scaling_for_exist}, $V$ is an equilibrium of $\eqcontract{\frac{4}{k}}{V}$ and satisfies $f(V) \ge \frac{1}{128} f(S^\star)$, as needed.
\end{proof}
\begin{proof}[Proof of Theorem~\ref{thm:xos_binary_approx}]
    Let $\contract^\star= \eqcontract{t}{A^\star}$ be an equal-pay contract and $S^\star \subseteq [n]$ be an equilibrium such that $\sum_{i=1}^n \alpha^\star_i \le 1$.
    Let $Z=\{i\in [n]\mid c_i = 0\}$, and observe that $Z$ is an equilibrium of $\eqcontract{0}{[n]}$, and $\left(1-\sum_{i=1}^n 0\right) f(Z)\le \varphi(\eqcontract{0}{[n]}\le f(Z)$, i.e., $\varphi(\eqcontract{0}{[n]}, Z)=f(Z)$. It also holds that $S^\star \setminus A^\star \subseteq Z$. 
    
    If $t>\frac{1}{2}$, then $A^\star=\{i^\star\}$.
    By 
Definition~\ref{def:goodobj_multi_multi}\ref{def:decompositionproperty} we have 
    \[
    \begin{aligned}
    \varphi(\contract^\star, S^\star) 
    &\le f(S^\star_{-i^\star})+\varphi(\contract^\star\mid_{i^\star}, \{i^\star\}) 
    \le f(Z) +\varphi(\contract^\star\mid_{i^\star}, \{i^\star\}) 
    \\
    &=\varphi(\eqcontract{0}{[n]}, Z) +\varphi\left(\eqcontract{\frac{c_{i^\star}}{f(\{i^\star\})}}{\{i^\star\}}, \{i^\star\}\right),
    \end{aligned}
    \]
    So at least one of $(\eqcontract{0}{[n]},Z)$ and $\left(\eqcontract{\frac{c_{i^\star}}{f(\{i^\star\})}}{\{i^\star\}}, \{i^\star\}\right)$ is a $\frac{1}{2}$-approximately optimal contract with respect to $\varphi$, and since both are in $\mathcal{C}$ (as defined in Algorithm~\ref{alg:xos_approx}) we are done.

    Otherwise, if $t\le \frac{1}{2}$, we have
    \[
    \varphi(\contract^\star, S^\star) \le f(S^\star) \le f(S^\star \cap A^\star)+ f(S^\star \setminus A^\star) \le f(S^\star \cap A^\star) + f(Z) = f(S^\star \cap A^\star) + \varphi(\eqcontract{0}{[n]}, Z).
    \]
    Since $(\eqcontract{0}{[n]}, Z)\in \mathcal{C}$, it remains to show the existence of a pair $(\contract, S)\in \mathcal{C}$ such that $\varphi(\contract, S)$ approximates $f(S^\star \cap A)$.
    If $|S^\star \cap A| \le 20$, then some $i\in S^\star \cap A$ has $f(\{i\}) \ge \frac{1}{20}f(S^\star \cap A)$, and since $\left(\eqcontract{\frac{c_i}{f(\{i\})}}{\{i\}}, \{i\}\right)\in \mathcal{C}$, we are done.
    Otherwise, partition $S^\star \cap A$ into three sets: $S_1\uplus S_2 \uplus S_3 = S^\star \cap A$. Without loss of generality $f(S_1) \ge \frac{1}{2}f(S^\star \cap A)$. By applying the
    scaling-for-existence lemma (Lemma~\ref{lem:scaling_for_exist}),
    we get a subset $S'\subseteq S_1$ such that $f(S') \ge \frac{1}{2}f(S_1) \ge \frac{1}{6}f(S^\star \cap A)$ and $f(S')$ is subset-stable with respect to the contract $\eqcontract{2t}{S'}$, and $2t\le \frac{2}{|S^\star|} \le \frac{1}{\lceil(S^\star \cap A) / 3\rceil} \le \frac{1}{|S'|}$. Furthermore, $S'\subseteq S^\star \subseteq A'$, which implies by
    Lemma~\ref{lem:approx_A'_f}  the existence of a pair $(\contract, S)\in \mathcal{C}$ such that 
    $\varphi(\contract, S) \ge \frac{1}{256}f(S')$, as needed.
\end{proof}

\section{Optimal Equal-Pay Contract for Additive Functions}

In this section, we show that when the success probability function $f$ is additive, the optimal equal-pay contract can be computed in polynomial time.

\begin{theorem}\label{thm:opt_additive}
    When $f$ is additive, the optimal equal-pay contract  can be computed in polynomial time.
\end{theorem}
\begin{proof}
Consider a multi-agent combinatorial-actions instance $\multiInstance$ with additive $f$.
Under a contract $\boldsymbol{\alpha}$, the utility of agent $i$ is
\begin{align*}
\alpha_i \cdot f(S \sqcup S_{-i}) - c(S)
= \sum_{j \in S} \left( \alpha_i \cdot f(\{j\}) - c_j \right) + \alpha_i \cdot f(S_{-i}).
\end{align*}
Hence, agent $i \in \agents$ chooses an action $j \in T_i$ if and only if $\alpha_i \geq c_j / f(\{j\})$.

Note that an optimal equal-pay contract must set its payment equal to $c_j / f(\{j\})$ for some action $j \in T_i$ and some agent $i \in \agents$. 
Otherwise, the principal could strictly decrease the payment without changing the induced equilibrium.
Since there are only polynomially many such candidate values, one can enumerate all of them.
For each candidate payment $p$, the problem reduces to selecting the optimal subset of agents $S \subseteq A$ to incentivize at payment $p$.

Fix a payment $p$.
For each agent $i$, define $w_i^p = \sum_{j \in T_i} f(\{j\}) \cdot \mathbbm{1}[c_j / f(\{j\}) \leq p]$ which represents the success probability contributed by agent $i$ at payment $p$.
Without loss of generality, assume that $A = \{1, 2, \ldots, n\}$  and $w_1^p \geq w_2^p \geq \ldots \geq w_n^p$.
The principal's utility from incentivizing a set of agents $S \subseteq A$ at payment $p$ is then $(1 - p \cdot |S|) \cdot \sum_{i \in S} w_i^p$.
This expression is maximized when $S$ is chosen as a prefix of $\{1, \ldots, n\}$.
Since there are polynomially many such prefixes, one can check each of them to identify the optimal set of agents for payment $p$.
This completes the proof.
\end{proof}

\begin{remark}
    The above can easily be adjusted to output the equal-pay contract which maximizes the reward or the social welfare.
\end{remark}
\section{Missing Proofs from Section 2}

\subsection{Proof of Lemma~\ref{lemma:modified_doubling}}\label{apx:modified_doubling}

\modifiedDoublingLemma*
\begin{proof}
As $\Seq{}$ is an equilibrium with respect to $2 \contract$, agent $i\in A(S)$ weakly prefers $\Seq{}_i$ over $\Seq{}_i \cup S_i$. That is,
\[
(2 \alpha_i) f(\Seq{}) - c(\Seq{}_i) \geq (2 \alpha_i) f(\Seq{}_i \cup S_i , \Seq{}_{-i}) - c(\Seq{}_i \cup S_i)
\]
and therefore
\[
(2 \alpha_i) \bigg( f(\Seq{}_i \cup S_i , \Seq{}_{-i}) - f(\Seq{}) \bigg) \leq c(\Seq{}_i \cup S_i) - c(\Seq{}_i) = c(S_i \setminus \Seq{}_i).
\]

Furthermore, since $S$ is subset stable with respect to $\contract$, agent $i\in A(S)$ weakly prefers $S_i$ over $\Seq{}_i \cap S_i \subseteq S_i$. So, we have
\[
\alpha_i f(S) - c(S_i) \geq \alpha_i f(\Seq{}_i \cap S_i , S_{-i}) - c(\Seq{}_i \cap S_i).
\]
This implies
\[
\alpha_i \bigg(f(S) - f(\Seq{}_i \cap S_i , S_{-i})\bigg) \geq c(S_i) - c(\Seq{}_i \cap S_i) = c(S_i \setminus \Seq{}_i). 
\]

In combination, we obtain
\begin{equation}
f(\Seq{}_i \cup S_i , \Seq{}_{-i}) - f(\Seq{}) \leq \frac{\alpha_i}{2\alpha_i} \left(f(S) - f(\Seq{}_i \cap S_i , S_{-i})\right) = \frac{1}{2} \left(f(S) - f(\Seq{}_i \cap S_i , S_{-i})\right),
\label{eq:dagger}
\end{equation}
where we use that the expression is non-negative by monotonicity of $f$.

Renumber the agents such that $A(S)=\{1,\dots, k=|A(S)|\}$, and
Let $S^j_i = \Seq{}_i \cup S_i$ if $i \leq j$ and $S^j_i = \Seq{}_i$ otherwise (for $j=1,\dots k$). By this definition
\[
f(S) \leq f(S^k) = f(S^0) + \sum_{i=1}^k \left( f(S^i) - f(S^{i-1}) \right).
\]
Note that $S^0 = \Seq{}$. Furthermore, by submodularity of $f$, we have for all $i\in [k]$
\[
f(S^i) - f(S^{i-1}) = f(S_i \setminus \Seq{}_i \mid S^{i-1}) \leq  f(S_i \setminus \Seq{}_i \mid \Seq{})  = f(\Seq{}_i \cup S_i , \Seq{}_{-i}) - f(\Seq{}).
\]
Therefore, we have
\begin{align*}
    f(S) 
    &\leq f(\Seq{}) + \sum_{i=1}^k \left(f(\Seq{}_i \cup S_i , \Seq{}_{-i}) - f(\Seq{}) \right) \\
    &\leq f(\Seq{}) + \sum_{i=1}^k \frac{1}{2} \left(f(S) - f(\Seq{}_i \cap S_i , S_{-i})\right) \\
    &\leq f(\Seq{}) + \sum_{i=1}^n \frac{1}{2} \left(f(S) - f(\Seq{}_i \cap S_i , S_{-i})\right) \\
    &\leq f(\Seq{}) + \frac{1}{2} f(S \setminus \Seq{}), && 
    (\text{submodularity of } f)
\end{align*}

where in the second inequality we used Equation~\eqref{eq:dagger}. 
So, we have $f(S) \leq 2 f(\Seq{})$.
\end{proof}

\section{Missing Proofs from Section 3}\label{sec:proofs_from_3}

\subsection{Proof of Lemma~\ref{lem:agent_constrained_approx_demand}}
\label{sec:proof_of_agent_constrained_approx_demand}

\approximateDemandQueryWtihConstraint*

Algorithm~\ref{alg:agent_constrained_approx_demand} produces a sequence $\emptyset = S_0 \subseteq S_1 \subseteq \ldots \subseteq S_k$. Note that the algorithm might potentially add actions of the same agent in more than one iteration of the algorithm since the distorted objective is changing.

 We use the following notation: 
\begin{align*}
    \Phi_i(U) &= \left( 1-1/k\right)^{k-i} \cdot  f(U) - p(U)  \\
    \Psi_i(U,A) &= \max\{ 0, \left( 1-1/k\right)^{k-(i+1)} \cdot f(A \mid U) - p(A)   \}
\end{align*}

We start with the first lemma, which is analogous to \cite[Lemma 1]{Harshaw2019SubmodularMB}.

\begin{lemma}\label{lem:harshaw_lemma_1}
    In each iteration it holds that:
    \begin{align*}
        \Phi_{i+1}(S_{i+1}) - \Phi_i(S_i) = \Psi_i(S_i, A_{i,j_i}) + (1/k) \cdot (1-1/k)^{k-(i+1)} \cdot f(S_i).
    \end{align*}
\end{lemma}
\begin{proof}
    We expand:
    \begin{align*}
       & \quad  \Phi_{i+1}(S_{i+1}) - \Phi_i(S_i) \\
        &= (1-1/k)^{k-(i+1)} \cdot f(S_{i+1})- p(S_{i+1}) - (1-1/k)^{k-i} \cdot f(S_i) + p(S_i) \\
         &= (1-1/k)^{k-(i+1)} \cdot f(S_{i+1})- p(S_{i+1}) - (1-1/k)^{k-(i+1)} \cdot (1-1/k) \cdot f(S_i) + p(S_i) \\
         &= (1-1/k)^{k-(i+1)} \cdot \left[ f(S_{i+1}) - f(S_i) \right] - \left[ p(S_{i+1})-p(S_i)\right] + (1/k) \cdot (1-1/k)^{k-(i+1)} \cdot f(S_i).
    \end{align*}
    Now, if the if-statement in Line~\ref{line:if_positiev} holds, then  $S_{i+1} = S_i \cup A_{i,j_i}$ and $S_i \cap A_{i,j_i} = \emptyset$, and:
    \begin{align*}
        & \quad (1-1/k)^{k-(i+1)} \cdot \left[ f(S_{i+1}) - f(S_i) \right] - \left[ p(S_{i+1})-p(S_i)\right]  \\
        &=   (1-1/k)^{k-(i+1)} \cdot f(A_{i,j_i} \mid S_i) - p(A_{i,j_i})  \\
        &= \Psi_i(S_i, A_{i,j_i}).
    \end{align*}
    Otherwise, the if-statement in Line~\ref{line:if_negative} holds, and we have $S_{i+1} = S_i$, and:
    \begin{align*}
        (1-1/k)^{k-(i+1)} \cdot \left[ f(S_{i+1}) - f(S_i) \right] - \left[ p(S_{i+1})-p(S_i)\right]  &= 0 =  \Psi_i(S_i, A_{i,j_i}). \qedhere
    \end{align*}
\end{proof}

We next prove the second lemma, which is analogous to \cite[Lemma 2]{Harshaw2019SubmodularMB}.

\begin{lemma}\label{lem:harshaw_lemma_2}
    In each iteration, for any $U \subseteq T$ with $|A(U)| \leq k$, it holds that:
    \begin{align*}
        \Psi_i(S_i, A_{i,j_i}) \geq (1-1/e) \cdot (1/k) \cdot (1-1/k)^{k-(i+1)} \cdot \left[  f(U) - f(S_i) \right] - (1/k) \cdot p(U).
    \end{align*}
\end{lemma}
\begin{proof}
Observe that:
\begin{align*}
    k \cdot \Psi_i(S_i, A_{i,j_i}) &= k \cdot \max\left\{ 0, \left( 1-1/k\right)^{k-(i+1)} \cdot f(A_{i,j_i} \mid S_i) - p(A_{i,j_i})  \right\} \\ 
    &\geq  k \cdot \max_{j \in [n]} \left\{ \left( 1-1/k\right)^{k-(i+1)} \cdot f(A_{i,j} \mid S_i) - p(A_{i,j})  \right\} \\ 
    &\geq |A(U)| \cdot \max_{j \in [n]} \left\{  \left( 1-1/k\right)^{k-(i+1)} \cdot f(A_{i,j} \mid S_i) - p(A_{i,j}) \right\} \\
    &\geq |A(U)| \cdot \max_{j \in A(U)} \left\{ \left( 1-1/k\right)^{k-(i+1)} \cdot f(A_{i,j} \mid S_i) - p(A_{i,j}) \right\} \\
    &\geq \sum_{j \in A(U)} \left[ \left( 1-1/k\right)^{k-(i+1)} \cdot f(A_{i,j} \mid S_i) - p(A_{i,j}) \right] \\
    &\geq \sum_{j \in A(U)} \left[ (1-1/e) \cdot \left( 1-1/k\right)^{k-(i+1)} \cdot f(U_j \mid S_i) - p(U_j) \right] \\
    &\geq (1-1/e) \cdot \left( 1-1/k\right)^{k-(i+1)} \cdot  \sum_{j \in A(U)} f(U_j \mid S_i) - p(U) \\
    &\geq (1-1/e) \cdot \left( 1-1/k\right)^{k-(i+1)} \cdot  \left[ f(U) - f(S_i)  \right]- p(U),  
\end{align*}
where the fifth inequality follows from Line \ref{line:greedy} in Algorithm~\ref{alg:agent_constrained_approx_demand}.
\end{proof}

We are now ready to prove Lemma~\ref{lem:agent_constrained_approx_demand}.

\begin{proof}[Proof of Lemma~\ref{lem:agent_constrained_approx_demand}]
    Observe that:
    \begin{align*}
        \Phi_0(S_0) &= (1-1/k)^k \cdot f(\emptyset) - p(\emptyset) = 0\\
        \Phi_k(S_k) &=  (1-1/k)^0 \cdot f(S_k) - p(S_k) = f(S_k) - p(S_k)
    \end{align*}
    We will use the following inequality:
    \begin{align*}
        (1/k) \cdot \sum_{i=0}^{k-1} (1-1/k)^{k-(i+1)} = (1-(1-1/k)^k) \geq 1-1/e.
    \end{align*}
    Fix any $U \subseteq T$ with $|A(U)| \leq k$. We get:
    \begin{align*}
        f(S_k) - p(S_k) &= \Phi_k(S_k) - \Phi_0(S_0) \\
        &= \sum_{i=0}^{k-1} \left[ \Phi_{i+1}(S_{i+1}) - \Phi_i(S_i) \right] \\
        &= \sum_{i=0}^{k-1} \left[ \Psi_i(S_i, A_{i,j_i}) + (1/k) \cdot (1-1/k)^{k-(i+1)} \cdot f(S_i) \right] \\
        &\geq \sum_{i=0}^{k-1} \left[ (1-1/e) \cdot (1/k) \cdot (1-1/k)^{k-(i+1)} \cdot \left[  f(U) - f(S_i) \right] - (1/k) \cdot p(U) \right] \\
        & \quad \quad \quad +\sum_{i=0}^{k-1} (1/k) \cdot (1-1/k)^{k-(i+1)} \cdot f(S_i)  \\
        &= (1-1/e) \cdot (1/k) \cdot \sum_{i=0}^{k-1} (1-1/k)^{k-(i+1)} \cdot f(U) - p(U)  
        \\
        & \quad \quad \quad + (1/e) \cdot \sum_{i=0}^{k-1} (1/k) \cdot (1-1/k)^{k-(i+1)} \cdot f(S_i)  \\
        &\geq (1-1/e)^2 \cdot f(U) - p(U) + 0 \\
        &= (1-1/e)^2 \cdot f(U) - p(U). \qedhere
        \end{align*}
\end{proof}

\end{document}